\documentclass[a4 paper, 11pt]{article}
\pdfoutput=1
\usepackage{comment}
\usepackage{jheppub}
\usepackage{subcaption}
\usepackage{amsmath}
\usepackage{mathrsfs}
\usepackage{amssymb}
\usepackage{amscd}
\usepackage{dsfont}
\usepackage{enumerate}
\usepackage{amsfonts}
\usepackage{epsfig}
\usepackage{mathtools}
\usepackage{yfonts}
\usepackage{bbold}
\usepackage{breqn}
\usepackage[utf8]{inputenc}
\usepackage[english]{babel}
\usepackage{graphicx}
\allowdisplaybreaks[4]

\usepackage{jheppub}
\usepackage{mathtools}
\usepackage{float}
\usepackage{tikz-cd}

\date{}

\newcommand{\bp}{\bar{\partial}}

\newcommand{\MJ}{{\mathcal{J}}}

\newcommand{\HL}{\mathcal{H}_L}
\newcommand{\HR}{\mathcal{H}_R}
\newcommand{\MH}{\mathcal{H}}
\newcommand{\MP}{\mathcal{P}}
\newcommand{\MD}{\mathcal{D}}
\newcommand{\MO}{\mathcal{O}}

\newcommand{\MOh}{\mathcal{O}_{h,\bar{h},q}}

\newcommand{\ML}{{\mathcal{L}}}

 \usepackage[normalem]{ulem}
\usepackage[textsize=scriptsize,textwidth=2.5cm]{todonotes}

\let\exporig\exp
\usepackage{physics}
\usepackage{enumitem}
\let\exp\exporig

\affiliation[\ensuremath{\gamma}]{Yau Mathematical Sciences Center, Tsinghua University, Beijing 100084, China}
\affiliation[\ensuremath{\tau}]{Department of Mathematical Sciences, Tsinghua University, Beijing 100084, China}
\affiliation[\ensuremath{\sigma}]{Center for the Fundamental Laws of Nature, Harvard University, Cambridge, MA, USA}

\usepackage{xspace}

\newcommand{\z}{\left}
\newcommand{\y}{\right}
\newcommand{\p}{\partial}

\title{Symmetries, operators and correlators in $J\bar{T}$ deformed CFTs}

\author{Liangyu Chen$^\gamma$,} 

\author{Zhengyuan Du$^{\gamma,\tau,\sigma}$,}

\author{Wei Song$^{\gamma,\tau}$}

\emailAdd{liangyu-chen@mail.tsinghua.edu.cn} 
\emailAdd{duzy21@mails.tsinghua.edu.cn}

\emailAdd{wsong2014@mail.tsinghua.edu.cn}

\abstract
{
We use symmetries to define a new class of operators and compute their correlation functions in the 
$J\bar T$-deformed conformal field theory on the plane, following the strategy developed in \cite{Chen:2025jzb}.
The symmetry algebra of the deformed theory consists of a local Virasoro–Kac–Moody algebra in the left-moving sector and a non-local counterpart in the right-moving sector. This algebraic structure guides the definition of distinct operator classes. In this paper, we focus on two types of operators:
{\it Dressed} operators transform as primaries under the symmetries and depend on non-local coordinates, which involve integration over a region.
{\it Physical} operators, introduced in this work, depend only on local quantities at the classical level and can be expressed in a manageable way in terms of dressed operators and currents.
At the quantum level, we assume the existence of dressed operators and the Ward identities that relate them. The physical operators are then defined via the dressed operators and currents, the expression of which is a natural uplifting of the classical version.
This formulation allows the powerful constraints of conformal symmetry to be leveraged for computing physical observables. 
Consequently, we employ conformal perturbation theory to compute the two-point and $N$-point functions of physical operators. In momentum space, we sum the most UV-sensitive contributions to all orders; the results show precise agreement with string theory predictions. In position space, the leading-log contributions to the two-point correlators are summed to all orders in the deformation parameter, revealing non-perturbative behavior.
}

\begin{document}
\maketitle
    \section{Introduction}
The seminal work of Strominger and Vafa \cite{Strominger:1996sh} offered a microscopic derivation of the Bekenstein-Hawking entropy for a specific class of supersymmetric black holes in IIB string theory, with a near-horizon geometry described by AdS$_3$. 
 The essential part of the matching was later found to work in a universal way and can be put into the framework of the AdS$_3$/CFT$_2$ correspondence \cite{Brown:1986nw,Strominger:1997eq}.
The AdS/CFT correspondence \cite{Maldacena:1997re,Gubser:1998bc,Witten:1998qj} represents the most successful realization of the holographic principle \cite{tHooft:1993dmi,Susskind:1994vu}, positing that quantum gravity in Anti-de Sitter (AdS) space is equivalent to a conformal field theory (CFT) in one lower dimension. This duality provides a powerful framework for understanding both the quantum theory of gravity in the bulk and quantum field theory on the boundary. For example, the deep connection between spacetime geometry and quantum entanglement \cite{Ryu:2006bv,Hubeny:2007xt} has played an important role for addressing the black hole information paradox \cite{Almheiri:2019psf,Penington:2019kki, Almheiri:2019qdq}. However, the central role of AdS spacetime is also a key limitation: generic black holes and the astrophysically relevant spacetimes of cosmology typically lack an isolated AdS region. This fact motivates the exploration of holographic dualities beyond the standard AdS/CFT paradigm.

A step toward understanding realistic black holes holographically was made by the Kerr/CFT correspondence \cite{Guica:2008mu,Bredberg:2009pv,Castro:2010fd,Haco:2018ske,Haco:2019ggi,Bredberg:2011hp,Compere:2012jk} and its three-dimensional avatar, holography for warped AdS$_3$ (WAdS$_3$) \cite{Anninos:2008fx,Compere:2009zj,Compere:2014bia,Song:2016gtd}. A key feature of these proposals is an $SL(2,\mathbb{R})_L\times U(1)_R$ symmetry, which constitutes the isometry of both WAdS$_3$ and the near-horizon geometry of the extremal Kerr black hole \cite{Bardeen:1999px}. Compared to the full $SL(2,\mathbb{R})_L\times SL(2,\mathbb{R})_R$ isometry of AdS$_3$/CFT$_2$, this reduced symmetry suggests that the holographic dual breaks the right-moving conformal symmetry while potentially preserving the left-moving one. Further intuition comes from a top-down approach that embeds WAdS$_3$ into string theory, treating the dual as a deformation of AdS$_3$/CFT$_2$. Supergravity analyses indicate this deformation is generally irrelevant, driven by an operator of conformal dimension $(1,2)$ \cite{Guica:2010ej,Compere:2010uk,El-Showk:2011euy, Song:2011sr}.

The $J\bar{T}$ deformation \cite{Guica:2017lia} provides a concrete example of a $(1,2)$ deformation that leaves the left-moving conformal symmetry intact. In this construction, $J$ is a $U(1)$ current and $\bar{T}$ is the current associated with right-moving translations. Similar to $T\bar{T}$ deformations \cite{Zamolodchikov:2004ce, Smirnov:2016lqw, Cavaglia:2016oda}, the $J\bar{T}$ deformation exhibits a solvable structure. Notably, its spectrum on a cylinder admits a universal expression in terms of the undeformed data, the deformation parameter, and the cylinder radius \cite{Guica:2017lia,Chakraborty:2018vja}, its torus partition function is modular covariant \cite{Aharony:2018ics}, and the scattering amplitude is modified by a CDD factor \cite{Anous:2019osb}.

In the context of holography, applying the standard (double-trace) $J\bar{T}$ deformation to a CFT$_2$ corresponds to imposing a mixed boundary condition in the bulk \cite{Bzowski:2018pcy}, as it is holographically dual to a change in boundary conditions \cite{Witten:2001ua,Gubser:2002zh,Gubser:2002vv}. 
To achieve a more intrinsic deformation of the bulk geometry---a key ingredient in non-AdS holography, one way to proceed is to consider the single-trace version of the $J\bar{T}$ deformation, which is currently well-defined only when the undeformed theory is a symmetric product CFT \cite{Chakraborty:2018vja,Apolo:2018qpq}. The untwisted sector of this single-trace theory is identical to its double-trace counterpart, while the twisted sector can be constructed from the untwisted sector in a manner analogous to standard symmetric product CFTs. A concrete realization of the Kerr/CFT correspondence in three dimensions—featuring a warped AdS$_3$ (WAdS$_3$) bulk geometry dual to a $(1,2)$-deformed boundary—is provided by the WAdS$_3$/single-trace $J\bar{T}$ correspondence \cite{Chakraborty:2018vja,Apolo:2018qpq}. This is constructed in IIB string theory with NS-NS flux, where the WAdS$_3$ background \cite{Apolo:2018qpq} can be generated by applying a TsT transformation, a solution-generating technique \cite{Lunin:2005jy,Maldacena:2008wh}, to an AdS$_3\times \mathcal{N}$ background.  
A more general discussion on the connection between $TsT$ transformation and solvable irrelevent deformations can be found in \cite{Araujo:2018rho, Borsato:2018spz, Apolo:2019zai,Apolo:2021wcn}.

The symmetry structure has been pivotal in developing the Kerr/CFT correspondence and related proposals, and its manifestation in $J\bar{T}$-deformed CFTs is particularly illuminating. As established through field-theoretic and holographic calculations \cite{Bzowski:2018pcy,Guica:2020eab,Chakraborty:2023wel,Georgescu:2024ppd,Georgescu:2025jlx}, the left-moving conformal symmetries remain intact. The right-moving sector presents a more different picture: although standard conformal symmetry is broken down to a global translation, a non-local, field-dependent right-moving stress tensor emerges. The conserved charges built from its associated current indeed satisfy the Virasoro algebra under the Poisson bracket.
This intriguing symmetry structure immediately raises a pressing question: how do these symmetries organize states and operators, and can they be used as a practical tool for calculating physical observables such as correlation functions? 

In this paper, we study the interplay between symmetries and operators in $J\bar{T}$-deformed CFTs, generalizing the strategy developed for $T\bar{T}$ deformations \cite{Chen:2025jzb} by introducing a new type of operators, dubbed {\it physical} operators. Using a Hamiltonian formalism \cite{Kruthoff:2020hsi,Georgescu:2024ppd},  we provide an explicit construction of physical operators in terms of CFT data. This allows us to treat correlation functions via conformal perturbation theory and compute the leading-log contributions to two-point correlation functions to all orders in the deformation parameter. In momentum space, the resummed result exhibits a CFT-like form but with a momentum- and charge-dependent effective conformal weight. In position space, our analysis reveals a non-perturbative effect in the position-space correlators, similar to that discovered in \cite{Hirano:2025alr}.
As we will comment in appendix \ref{sec:compare}, the physical operator we consider in this paper is different from those discussed in the earlier work \cite{Guica:2019vnb,Guica:2021fkv,Georgescu:2022iyx}. Consequently, the result of the correlation function is also different. 

In the Hamiltonian formalism \cite{Kruthoff:2020hsi,Georgescu:2024ppd}, $J\bar{T}$ deformation on the plane is described by a canonical flow
  \begin{equation}
      \MD_\lambda H=0,\quad D_\lambda \equiv\p_\lambda-i[\mathcal{X},\cdot]
  \end{equation}
  where $\mathcal{X}$ is a bilocal operator constructed from the stress tensor and the $U(1)$ current.  We first construct the symmetries in the deformed theory, which possesses (Virasoro-Kac-Moody)$^2$, while the left-moving part is local and the remaining one is non-local. 
    
Starting from a local operator $\mathcal{O}^{(0)}(0,y)$ of the undeformed CFT$_2$ on an initial slice $t=0$, we discuss the following three types of operators:
   \begin{itemize}
    \item {\it Undeformed} operators: Constructed via time evolution of local operators on the initial slice
    \begin{align}\label{undefO}
        \mathcal{O}^{(0)}(t,y) = e^{iHt}\mathcal{O}^{(0)}(0,y)e^{-iHt}
    \end{align}
The undeformed operators are defined pointwise in spacetime and therefore are the natural candidates of local operators. However, these operators do not have a clean symmetry structure.  
    \item {\it Dressed} operators: Solutions to the flow equation
    \begin{align}\label{flowedeq}
        \mathcal{D}_\lambda \tilde{\mathcal{O}} = 0, \quad \tilde{\mathcal{O}}|_{\lambda=0} = \mathcal{O}^{(0)}
    \end{align}
When the undeformed operator is a primary operator in the original CFT, the dressed operator remains primary under the dressed conformal symmetries. These operators are therefore well-organized by the symmetry, and their correlation functions retain their CFT form. However,  as can be seen explicitly in \eqref{Flowedoperator}, the dressed operators depend on non-local coordinates which involve integrals of conserved currents. This highly non-local nature obscures the direct physical significance of the dressed operators.
    \item {\it Physical} operators:
    \begin{equation}
       \MO_{h,\bar{h},q}\equiv \frac{\MO_{h,\bar{h},q}^{(0)}}{\z(1-\lambda \tilde{\MJ}_+\y)\z(1-\lambda \tilde{\MJ}_++\frac{\lambda^2k}{4 \pi} \HR\y)^{ \bar{h}-1}}
    \end{equation}
    We consider a primary operator in the undeformed CFT with conformal weights $(h,\bar{h})$ and chiral charge $q$. In the deformed theory, the physical operator constructed from operators that can be defined pointwise in spacetime, such as the chiral current $\tilde{\mathcal{J}}_{+}$ and the right-moving stress tensor component $\mathcal{H}_{R}$. This distinguishes it from the non-local dressed operators, which depend on integrated quantities.
    
\end{itemize}
The physical operators can also be rewritten in terms of dressed operators:
 \begin{equation}
\begin{aligned}
\MOh(u,v)=\int d\sigma^2\delta(U(\sigma^+,\sigma^-)-u,V(\sigma^+,\sigma^-)-v)e^{-i\lambda q\int^{\sigma^{-}} \tilde{\MH}_R(\sigma^{-\prime})d\sigma^{-\prime}}\tilde{\MO}_{h,\bar{h},q}(\sigma^+,\sigma^-)\label{}
\end{aligned}
\end{equation}
Here, $(u,v)$ are null coordinates and $(U,V)$ are field-dependent coordinates. The dressed primary operator $\tilde{\mathcal{O}}_{h,\bar{h},q}$ is defined as the solution to the flow equation \eqref{flowedeq} with the initial condition of being a primary operator in the undeformed CFT.  

At the quantum level, we assume the existence of the dressed operator, and define the quantum version of the physical operator through its relation to the dressed operator. In Euclidean momentum space, physical operators are defined as 
\begin{equation}
    \MO_{h,\bar{h},q}(p,\bar{p})=\int dx^2 e^{ipx+i\bar{p}\bar{x} - \frac{i\lambda}{2\pi} \z[\bar{p}\int^xdx'j(x')+\z(iq+\frac{\lambda k\bar{p}}{4\pi}\y)\int^{\bar{x}}d\bar{x}'\bar{T}(\bar{x}')\y] }  \tilde{\mathcal{O}}_{h, \bar{h}, q} \left(x, \bar{x} \right)\label{Ophys-m-E}
\end{equation}
where normal ordering is assumed. 
This expression organizes calculations via symmetries, as all operators on the right-hand side are covariant under the canonical flow and maintain the standard CFT OPE. 
In particular, the computation of the correlation function is now turned into a conformal perturbation problem. 
Consequently, we can compute correlation functions to all orders in the deformation parameter $\lambda$ in both momentum and position space.

In momentum space, we sum over the most dominant terms in UV to obtain the all-order perturbative result:
\begin{equation}
\begin{aligned}
\left\langle\mathcal{O}_{h, \bar{h}, q}(p, \bar{p}) \mathcal{O}_{h, \bar{h},-q}(-p,-\bar{p})\right\rangle & \sim \frac{\pi \Gamma\left(1-2 h_\lambda\right)}{\Gamma\left(2 h_\lambda\right)}\left(\frac{|p|}{2}\right)^{4 h_\lambda-2}
\end{aligned}
\end{equation}
where
\begin{equation}
h_\lambda=h -\frac{i\lambda q\bar{p}}{2\pi} -\frac{\lambda^2k\bar{p}^2}{8\pi^2},\quad \bar h_\lambda = \bar h -\frac{i\lambda q\bar{p}}{2\pi} -\frac{\lambda^2k\bar{p}^2}{8\pi^2}
\end{equation}
The two-point function we obtain maintains the structural form of a CFT correlator, albeit with a momentum- and charge-dependent shift in the conformal weight. This result agrees precisely with findings from the TsT/$J\bar{T}$ correspondence \cite{ACSSW}. Furthermore, it mirrors the behavior observed in momentum-space calculations for $T\bar{T}$-deformed theories \cite{Cui:2023jrb, Aharony:2023dod, Chen:2025jzb}, suggesting a universal feature of such solvable deformations.

We further analyze the two-point function in position space. Its leading logarithmic contributions can be organized into a formal power series in the variable
\begin{equation}
    K\equiv-\frac{\lambda^2k}{4\pi^2}\frac{\log|x|^2}{(\bar{x}-\frac{\lambda q}{\pi}\log|x|^2)^2},
\end{equation}
which we sum using Borel resummation. For generic values of $K$-corresponding to a broad range of spacetime separations-the series is directly Borel summable. However, for $K \in \mathbb{R}_+$, standard Borel resummation fails due to a branch point on the integration path. This is a common obstruction in resurgence theory, typically requiring a non-perturbative completion. In our case, taking the Cauchy principal value of the Borel integral reveals that the full result includes an instanton-like contribution that is exponentially suppressed at small $K$.

The final answer for the two-point function in position space exhibits distinct scaling regimes. At long distances (IR), it shows the standard CFT power-law decay, $x^{-2h}\bar{x}^{-2\bar{h}}$, with unshifted conformal weights. At short distances (UV), it becomes proportional to $x^{-2h}(\log|x|^2)^{-\bar{h}}$ in zero charge sector and $x^{-2h}(\log|x|^2)^{-2\bar{h}}$ in non-zero charge sector, preserving the left-moving power law but displaying a milder, logarithmic decay in the right-moving sector. Due to the intact left-moving conformal symmetry, the UV divergence persists—a key difference from $T\bar{T}$ deformation. In the crossover region, the correlator develops oscillatory and cusp structures, whose physical origin remains an open question.

The rest of this paper is organized as follows. In Section \ref{sec:JTbar}, we discuss the symmetry structure and explicitly construct the dressed stress tensor and $U(1)$ current that generate it. In Section \ref{sec:operators}, we construct physical operators and derive their Ward identities with the dressed symmetries. In Section \ref{sec:dress}, we solve for the dressed operators at the classical level and provide an inverse relation expressing physical operators in terms of dressed ones. In Section \ref{sec:PQ}, we lift the physical operator to the quantum level. In Section \ref{sec:CorrF}, we compute correlation functions in both momentum space and position space.
We conclude with a summary and discussion in Section \ref{sec:sum}.

\section{Nonlocal symmetries in $J\bar{T}$ deformation}\label{sec:JTbar}

In this section, we briefly review $J\bar T$ deformation \cite{Guica:2017lia} in the Hamiltonian formalism and derive the dressed symmetries on the plane. Symmetries for $J\bar T$ deformed CFTs have been discussed on the cylinder in \cite{Guica:2020eab,Guica:2020uhm,Georgescu:2024ppd}. In Appendix D, we will show that symmetry generators on the plane are consistent with the large radius limit of that on those on the cylinder. 

\subsection{$J\bar T$ deformation as a canonical transformation}
In this subsection, we establish the basic setup for $J\bar T$ deformation. 
For any two-dimensional quantum field theory with two commuting conserved currents, one can define a class of solvable deformations by adding a composite current-current operator to the action:
\begin{equation}\label{Sd}
    \frac{\partial S}{ \partial \lambda}= \int d^2x  \, \mathcal{O}_{d.f} ,\quad \mathcal{O}_{d.f}= - {1\over 2}\epsilon^{ab} \epsilon^{\mu\nu}  J_{\mu a } J_{\nu b},
\end{equation}
where $\lambda$ is the deformation parameter and $S$ denotes the deformed action at $\lambda$. The currents $J_{\mu a}$ are conserved: $\partial^\mu J_{\mu a} = 0$. Different choices of the currents lead to different deformations. In this work, we focus on the $J\bar{T}$ deformation, where the two currents are taken as:
\begin{equation}\label{JTbarcurrents}
    J_{\underline{1}} = J_u du + J_v dv, \qquad
    J_{\underline{2}} = T_{vu} du + T_{vv} dv,
\end{equation}
with $u = y + t$, $v = y - t$ being light-cone coordinates.

This type of deformation can be analyzed in the Hamiltonian formalism. On a constant-time slice, the flow of the Hamiltonian takes the form
\begin{equation}\label{dH}
    \partial_\lambda H = -\int dy\, \mathcal{O}_{d.f}(y).
\end{equation}
On a cylinder with a spatial circle of circumference $R$, we can introduce a 
Green's function $G(y-y')$ satisfying $\partial_y G(y-y') = \delta(y-y') - 1/R$,
using which and the conservation laws, the Hamiltonian flow can be rewritten as
\begin{equation}\label{dHxy}
    \partial_\lambda H = i[\mathcal{X}, H] + \mathcal{Y},
\end{equation}
with
\begin{equation}\label{XY}
    \begin{aligned}
    \mathcal{X} &= -\frac{1}{2}\int dy dy'\, G(y-y') J_{ta}(y) J_{tb}(y') \epsilon^{ab}, \\
    \mathcal{Y} &= -\frac{1}{2R}\int dy dy'\, \epsilon^{\mu\nu}\epsilon^{ab} J_{\mu a}(y) J_{\nu b}(y').
\end{aligned}
\end{equation}
When considering the theory defined on the plane ($R \to \infty$), the term $\mathcal{Y}$ vanishes, and the deformation reduces to a canonical transformation generated by $\mathcal{X}$, preserving the energy spectrum while mixing states in the Hilbert space.

We distinguish three types of operators in the deformed theory: \emph{undeformed}, \emph{dressed}, and \emph{physical} operators.
In this paper, we mainly focus on \emph{physical} operators, the notion of which was first introduced in the $T\bar T$ deformations \cite{Chen:2025jzb} and will be generalized to the $J\bar T$ case in Section \ref{sec:operators}. In the rest of this subsection, we review the definitions of the first two types of operators introduced in the literature \cite{Kruthoff:2020hsi,Guica:2020eab,Guica:2021fkv}. 
The first two were studied in earlier works in $T\bar{T}$ deformed CFTs \cite{Kruthoff:2020hsi}, and their definitions carry over analogously to $J\bar{T}$ due to the similar structure of the flow equation. {\it Undeformed} operators are evolved with the deformed Hamiltonian:
\begin{equation}\label{undeformedO}
    \mathcal{O}^{(0)}(t,y) = e^{iHt} \mathcal{O}^{(0)}(0,y) e^{-iHt}.
\end{equation}
They can be defined pointwise in spacetime but do not respect the symmetries of the undeformed theory along the flow, complicating the analysis of correlation functions.
\emph{Dressed} operators $\tilde{\mathcal{O}}$, also called flowed operators \cite{Guica:2021fkv}, are defined as covariantly constant under the deformation:
\begin{equation}\label{fO}
    \mathcal{D}_\lambda \tilde{\mathcal{O}} = 0, \quad \text{with} \quad \mathcal{D}_\lambda = \partial_\lambda - i[\mathcal{X}, \cdot], \quad \tilde{\mathcal{O}}|_{\lambda=0} = \mathcal{O}^{(0)}.
\end{equation}
Note that the derivative with respect to $\lambda$ commutes with spacetime derivatives.  Then a functional $f(\tilde{\mathcal{O}})$ is also covariantly constant, i.e. $\partial_\lambda f(\tilde{\mathcal{O}})=i[\mathcal{X}, f(\tilde{\mathcal{O}})]$.
If the operators satisfy the relation $f(\tilde{\mathcal{O}})|_{\lambda=0}=0$ before the deformation, then this relation will be satisfied by the dressed operator as well. 
Thus, the dressed operators preserve important properties such as chirality, conservation laws, and algebraic structures along the flow. Their correlation functions remain unchanged compared to the undeformed theory, mimicking those of a CFT$_2$. In section 4, we will explicitly construct the dressed operators in $J\bar T$ deformed CFTs starting from the undeformed operators and stress tensor. As is evident from the expression \eqref{Flowedoperator}, the dressed operators depend on integrated quantities and hence are manifestly non-local. The non-local nature of the dressed operators makes their physical interpretation less transparent.

In summary, undeformed operators are defined pointwise but symmetry-unorganized, while dressed operators preserve symmetry structure at the cost of locality. In Sec. \ref{sec:operators}, we will introduce a new class—\emph{physical} operators—that aim to combine the advantages of both: an expression only depending on a given spacetime point and compatibility with symmetry organization. In the remainder of this section, we focus on the stress tensor, a special type of operator in $J\bar T$ deformed CFTs.

\subsection{The $J\bar T$ deformed stress tensor and currents}
In this section, we review the derivation of the deformed stress tensor in terms of the undeformed ones following \cite{Guica:2020uhm,Guica:2020eab,Georgescu:2024ppd}, and set up conventions along the way. Readers who are familiar with this part can skip this section. 

We denote the action of the $J\bar{T}$ deformed CFT$_2$ schematically by
    \begin{equation}
        S[\phi]=\int d^{2} x \, \mathcal{L}(\phi(x), \partial_{\mu} \phi(x)),
    \end{equation}
   where $\phi$ collectively denotes the fundamental fields. The translational invariance is preserved under the $J\bar{T}$ deformation, 
   generated by the stress tensor, which can be obtained via Noether's theorem \footnote{Note that
   we have assumed that 
   the stress tensor takes the standard form \eqref{Neother}, which works for a class of theories including free theories. These theories typically do not have classical central charges.
   For theories like Liouville theory, the stress tensor has to be modified, and our subsequent discussions do not apply directly.},
   \begin{equation}  \label {Neother}T_\mu\,^\nu=\eta_\mu\,^\nu\mathcal L- \partial_\mu\phi{\partial \mathcal{L}\over  \partial{\partial _\nu \phi}}
   \end{equation}
   We obtain the Hamiltonian density by a Legendre transformation,  \begin{equation}  
\mathcal{H} = \pi_\phi \partial_{t} \phi - \mathcal{L},\quad \pi_\phi={\partial\mathcal{L}\over \partial\partial_t \phi}
  \end{equation} 
It follows that the stress tensor can be expressed as
    \begin{equation}
        \begin{aligned}   \label{Tuv}
        & T_{t t}=\mathcal{H}, \quad T_{t y}=\frac{\partial \mathcal{H}}{\partial\left(\partial_y \phi\right)} \frac{\partial \mathcal{H}}{\partial \pi_\phi},\quad T_{y t}=\pi_\phi \partial_y \phi\equiv \mathcal{P} \\
        & T_{y y}= \hat {V}[ \mathcal{H}]-\mathcal{H}, \quad \hat {V}=\pi_\phi \frac{\partial}{\partial \pi_\phi}+\partial_y \phi\frac{\partial }{\partial\left(\partial_y \phi\right)} 
        \end{aligned}
    \end{equation}
Unlike the $T\bar{T}$
  deformation, which preserves Lorentz symmetry, the $J\bar{T}$ deformation explicitly breaks it, leading to a non-symmetric stress tensor. To discuss $J\bar{T}$ deformations, we assume that the original undeformed theory contains a compact scalar field $\psi$, so that  the undeformed Lagrangian has a decomposition as follows
\begin{equation}
	\ML^{(0)}=\frac{k}{2\pi}\Big((\p_t\psi)^2-(\p_y\psi)^2\Big)+\ML^{(0)}_m(\phi_m,\p_\mu\phi_m),
\end{equation}
where $k$ is the level of the $U (1)$ symmetry, and $\ML_m^{(0)}$ is a CFT which doesn't depend on $\psi$.
The undeformed Hamiltonian density is given by
\begin{equation}
	\MH^{(0)}=\frac{\pi}{2k}\pi_\psi^2+\frac{k}{2\pi}(\p_y\psi)^2+\MH^{(0)}_m(\pi_{\phi_m},\phi_m,\p_y\phi_m)
\end{equation}
 We assume a shift symmetry $\psi \to \psi + \epsilon$ persists after turning on the $J\bar{T}$ deformation. Then the Noether current can be viewed as the momentum along the {\it target} space direction $\psi$, with components given by:
\begin{equation}
    J_t\equiv\pi_\psi,\quad J_y\equiv \frac{\p\MH}{\p\p_y\psi}\overset{\lambda\to 0}{=}\frac{k}{\pi}\p_y\psi
\end{equation}
In addition to the Noether currents, a topological current exists following from the identity $dd\psi=0$, with components given by:
\begin{equation}
    K_t\equiv\frac{k}{\pi}\p_y\psi,\quad K_y\equiv \frac{k}{\pi}\frac{\p\MH}{\p\pi_\psi}\overset{\lambda =0}{=}\pi_\psi
\end{equation}
The charge of the topological current can be viewed as the winding number, which counts how many times the {\it worldsheet} parameterized by $y$ winds around the {\it target} spacetime coordinate $\psi$. The winding charges take discrete values and can be non-zero if the worldsheet has non-trivial topology.
The translational and topological currents can be linearly combined into two new currents
\begin{equation}
\begin{aligned}
    &\MJ_{\pm\alpha}\equiv \frac{J_\alpha \pm K_\alpha}{2},
    \end{aligned}\label{chiralJ}
\end{equation}
which in the undeformed limit are chiral and anti-chiral currents, respectively,
\begin{equation}\begin{aligned}
&\lim_{\lambda\to0} \MJ_{+u} \equiv \MJ_{+}^{(0)},\quad  \lim_{\lambda\to0} \MJ_{+v}=0,\\
& \lim_{\lambda\to0}\MJ_{-v}  \equiv-\MJ_{-}^{(0)},\quad\lim_{\lambda\to0}\MJ_{-u}=0, 
\end{aligned}\end{equation}
 As we will show later, the chirality property is not preserved after the  $J\bar T$ deformation, due to the mixing between left and right-moving sectors. For this reason, we will refer to these currents as quasi-chiral currents.
Nevertheless, the $ t $ components of these currents still have the same functional expression in terms of the phase space variables:
\begin{equation}\label{Relationcurrent}
\begin{aligned}
    &     \MJ_{\pm t}=\frac{\pi_\psi\pm\frac{k}{\pi}\p_y \psi}{2}=\MJ_{\pm }^{(0)}, 
\end{aligned}
\end{equation}
This means that the conserved charges do not flow with the deformation,  
\begin{equation}\label{unflowedcharge}
    Q_L\equiv\int \MJ_{+ t}dy,\quad Q_R\equiv \int \MJ_{-t}dy
\end{equation}
It is possible to define $J\bar{T}$ deformations by choosing any of the aforementioned currents as the $U(1)$ current in \eqref{JTbarcurrents}. 
In this paper, we consider $J\bar T$ deformation generated by the quasi-chiral current $\mathcal J_{+\mu}$:
\begin{equation} \label{flowH}
    \p_\lambda \MH=-\MO_{J\bar{T}}=\epsilon^{\alpha\beta}\MJ_{+\alpha}T_{v\beta}
\end{equation}
For convenience, we introduce the energy density in light-cone coordinates
    \begin{equation}
        \HL\equiv T_{ut}=\frac{\MH+\MP}{2},\quad \HR\equiv -T_{vt}=\frac{\MH-\MP}{2},\quad u=y+t,\quad v=y-t 
    \end{equation}
 Using the expressions of the stress tensor \eqref{Tuv} and current \eqref{chiralJ}, \eqref{Relationcurrent}, we can rewrite the flow equation \eqref{flowH} as a flow equation of $\HR$ 
 \begin{equation}\label{JTbarDeform}
     \p_\lambda \HL=\p_\lambda \HR=-\MJ_+^{(0)}\frac{\p \HR}{\p\p_y\phi}\frac{\p \HR}{\p\pi_{\phi}}+\frac{1}{2}\HR\z(\frac{\p\HR}{\p_y\psi}+\frac{k}{\pi}\frac{\p\HR}{\p \pi_\psi}\y)
 \end{equation} 
Now let us solve the above equation on the constant time slice $t=0$. Assuming the deformed right-moving energy $\mathcal{H}_{R} $ as a functional of its undeformed counterparts and $\mathcal{J}_{+}^{(0)}$, and use the initial condition that the undeformed theory is a CFT, we can find the following solution of the flow equation \cite{Guica:2020uhm,Georgescu:2024ppd}:
\begin{equation}\label{HRsolution}
\begin{aligned}
& \mathcal{H}_R=\frac{2\pi}{\lambda^2 k}\left(1-\lambda \mathcal{J}_{+}^{(0)}-\sqrt{\left(1-\lambda \mathcal{J}_{+}^{(0)}\right)^2-\frac{\lambda^2 k}{\pi} \mathcal{H}_R^{(0)}}\right),\quad \mathcal{H}_L=\mathcal{H}_R+\MP
\end{aligned}
\end{equation}
By substituting the solution \eqref{HRsolution} into \eqref{Tuv}, we obtain all the components of the stress tensor and the quasi-chiral current in null coordinates,
\begin{equation}\label{currentssol}
\begin{aligned}
& T_{uu}=\HL,\quad T_{uv}=0,
\quad T_{v u}=
\frac{\lambda \tilde{\mathcal{J}}_{+}\mathcal{H}_R}{1-\lambda \tilde{\mathcal{J}}_{+}}
\quad T_{v v}=
\frac{\mathcal{H}_R}{1-\lambda \tilde{\mathcal{J}}_{+}}\\
&\MJ_{+u}=\tilde{\MJ}_{+}+\frac{\lambda k}{2\pi}\frac{\lambda \tilde{\MJ}_+\HR}{1-\lambda\tilde{\MJ}_+},\quad \MJ_{+v}=\frac{\lambda k}{2\pi}\frac{\HR}{1-\lambda\tilde{\MJ}_+}
\end{aligned}
\end{equation}
where we have introduced the notation
\begin{align} \label{tJplus}\tilde{\mathcal{J}}_{+}=\MJ_+^{(0)}+\frac{\lambda k}{2\pi}\HR
\end{align}
From the above solution \eqref{currentssol}, it is evident that the current $\mathcal{J}_{+\mu}$ is no longer a chiral current once the deformation is turned on. However, as the $v$ component of the current is proportional to that of the right-moving stress tensor, we can construct a chiral current
\begin{align}\label{chiralcurrent}
\tilde{\mathcal{J}}_{+\alpha}=\mathcal{J}_{+\alpha}-{\lambda k\over 2\pi} T_{v\alpha}
\end{align}
The $v$ component is zero and the $u$ component satisfies $\tilde{\MJ}_{+u}=\tilde{\MJ}_+$. It is interesting to note that the following Sugawara-type combination is invariant under the flow, 
\begin{align}\partial_\lambda (\MH_L-{\pi \over k}{\tilde{\MJ}_+}^2 )=0 \end{align}
We can also invert the above relations between $(\MH_{R}, \tilde{\MJ}_{+})$ and $(\MH_{R}^{(0)}, \MJ_{+}^{(0)})$ for later convenience,
\begin{equation}\label{TransHJ}
 \begin{aligned}
       \MH_R^{(0)}&=\MH_R\z(1-\lambda \tilde{\MJ}_++\frac{\lambda^2 k }{4\pi }\MH_R\y)\\
        \MJ^{(0)}_+&=\tilde{\MJ}_+-\frac{\lambda k}{2\pi }\MH_R
    \end{aligned}
\end{equation}
Although initially derived on the $t=0$ slice by solving the flow equation, the relations \eqref{HRsolution} and \eqref{tJplus} between the deformed and undeformed currents can be extended to any constant-time slice through time evolution dictated by the deformed Hamiltonian. That is, these relations hold at any time $t$, provided the operators at different time slices are connected by the appropriate time evolution.
\begin{equation}\label{RSTt}
\begin{aligned}
    &\MH_{L/R}^{(0)}(t,y)=e^{iHt}\MH_{L/R}^{(0)}(0,y)e^{-iHt}, \quad \MH_{L/R}(t,y)=e^{iHt}\MH_{L/R}(0,y)e^{-iHt}\\
    &\MJ^{(0)}_+(t,y)=e^{iHt}\MJ^{(0)}_+(0,y)e^{-iHt}, \quad \quad \tilde{\MJ}_+(t,y)=e^{iHt}\tilde{\MJ}_+(0,y)e^{-iHt}
\end{aligned}
\end{equation} 
While the definition of the undeformed currents is consistent with the undeformed operator in \eqref{undeformedO}, it is crucial to note a fundamental distinction: these ``undeformed'' operators agree with those in the original $\mathrm{CFT}_2$ only at the initial time slice $t=0$. On a later time slice, the undeformed operator $\mathcal{O}^{(0)}$ within the $J\bar{T}$ deformed theory evolves according to the deformed Hamiltonian, whereas its counterpart in the original $\mathrm{CFT}_2$ evolves under the undeformed Hamiltonian.

\subsection{The dressed symmetry generators: stress tensor and   chiral currents}
In this subsection, we construct the dressed stress tensor and dressed chiral currents in the $J\bar{T}$ deformed $\mathrm{CFT}_2$ by solving the flow equation~\eqref{fO}.
The non-local symmetries have been previously constructed on the cylinder using topological current \cite{Georgescu:2024ppd}. In the following, our construction is carried out on the plane for the quasi-chiral current. While the resulting dressed current takes a similar form with the large radius result of \cite{Georgescu:2024ppd}, different choices of the current lead to different details in the explicit expression of the flow operator and the non-local coordinates.

From  \eqref{chiralcurrent}, we can see the deformation can be generated by $\tilde{\MJ}_{+\mu}$ equivalently
\begin{align} \label{dH}
    \p_\lambda \MH=-\MO_{J\bar{T}}=\epsilon^{\alpha\beta}\mathcal{J}_{+\alpha}T_{v\beta}=\epsilon^{\alpha\beta}\tilde{\mathcal{J}}_{+\alpha}T_{v\beta}
    =2\tilde{\MJ}_+T_{vv}
\end{align}
Throughout this paper, the range of integration  in $\int dy$ is from $-\infty$ to $\infty$.
The flow operator is then given by
\begin{align}
    {\mathcal{X}}_{J\bar{T}}=\int dydy'G(y-y')\tilde{\MJ}_+(y)\MH_R(y')
\end{align}
To understand the symmetry structure of the theory, we will solve the flow equation, which at the  classical level is given by \begin{align}\label{flowO}
&\mathcal{D}_{\lambda}\tilde \MO = 0,\quad 
\mathcal{D}_{\lambda}\equiv\partial_{\lambda} + \{\mathcal{X}_{J\bar{T}}, \cdot\},\quad  \tilde \MO|_{\lambda=0}=\MO^{(0)}.
\end{align}
In the following, we will solve the above equation for the dressed stress tensor $\tilde {\mathcal H}_L,\,\tilde{\mathcal{H}}_R$, and currents $\tilde{\mathcal{J}}_\pm$. 

\paragraph{The left-moving sector.} Using the Poisson brackets\footnote{The Poisson brackets here lack a classical central term, a consequence of our choice of stress tensor \eqref{Tuv}. It would be interesting to explicitly study cases where a classical central charge appears; we leave this for future work.} collected in Appendix \ref{app:poisson_brackets}, we find that the left-moving stress tensor and the chiral currents are both covariant, namely, 
\begin{align}\label{HLD}
&  \MD_\lambda \MH_L=0   ,\quad \tilde{\mathcal{H}}_{L} = \mathcal{H}_{L}\\
 &  \MD_\lambda \tilde{\MJ}_+=0, \quad  
\end{align}
Using the above covariantly constant currents, we can construct an infinite number of symmetry generators that also satisfy the flow equation \eqref{flowO},
\begin{equation}\label{Charges1}
    \begin{aligned}
        &\tilde{L}_m\equiv i^{m+1}\int dy \, u^{m+1}\MH_L,\qquad \tilde{\mathcal{J}}_{m}\equiv i^{m}\int dy \, u^{m}\tilde{\MJ}_+
    \end{aligned}
\end{equation}
The expression \eqref{Charges1} of the charge is formal. For $m>1$, the charges are finite for local excitations that fall off sufficiently enough at infinity. As is shown explicitly in appendix \ref{sec:compare}, these charges defined directly on the plane agree with those obtained from the plane limit of the cylinder generators \cite{Georgescu:2024ppd}.
In particular, the left-moving energy is given by $H_L=\tilde{L}_{-1}$, which is the same relation as in CFT$_2$ on the plane. 
From the relation between the chiral current and quasi-chiral current \eqref{TransHJ}, we learn the charges corresponding to the two currents are related by 
\begin{align}
    Q_L=\tilde{\MJ}_{0}-\frac{\lambda k}{2\pi}H_R\label{chargeflow}
\end{align}
which is in agreement with the charge flow  relation obtained in $J\bar T$ deformation \cite{Bzowski:2018pcy,Chakraborty:2018vja,Apolo:2018qpq}.
By construction, the stress tensor and $U(1)$ current in the left-moving sector are both local,
indicating that the left-moving Virasoro-Kac-Moody algebra is preserved under the $J\bar T$ deformation, as observed earlier in \cite{Guica:2020eab}. We will come back to the symmetry algebra and confirm this point later.

\paragraph{The right-moving sector.}
To determine the dressed stress $\tilde{\mathcal{H}}_{R}$ and $U(1)$ current $\tilde{\mathcal{J}}_{-}$ in the right-moving sector, let us first examine how $\mathcal{D}_{\lambda}$ acts on $\mathcal{H}_R$ and $\mathcal{J}_{-t}$; the results are
\begin{equation}
    \MD_\lambda \MH_R=\p_y\z(\MD_\lambda\bar{\eta}_0\y),\quad \MD_\lambda \MJ_{-t}=\p_y\z(\MD_\lambda \bar{\chi}_0\y)
\end{equation}
where 
\begin{equation}
	\begin{aligned}\label{etachi0}
	 {\bar \eta}_0\equiv\int dy'G(y-y') \mathcal{H}_{R}(y') ,\quad \bar{\chi}_0\equiv\int dy'G(y-y') \MJ_{-t}(y').
	\end{aligned}
\end{equation}
The above structure is reminiscent of the construction of $T\bar T$ deformation \cite{Chen:2025jzb}, in which case the non-local coordinate \cite{Dubovsky:2012wk} plays an essential role.
We expect the non-local coordinate  \cite{Azeyanagi:2012zd} observed in $J\bar T$ deformation is also useful in the construction of symmetry generators. On the plane,  let us define the non-local coordinate \footnote{Although the $J\bar{T}$ deformation from chiral, Noether, and topological currents all lead to the same deformed Hamiltonian, the resulting operators $\mathcal{X}$ are different. As a consequence, the non-local coordinates \eqref{nonlocalv} from the chiral current are different from those of \cite{Georgescu:2024ppd} from the topological current. We thank Monica Guica for discussions on this point.} 
\begin{align}   \label{nonlocalv}
\hat v\equiv v+\delta\hat v,\quad 
\delta \hat{v} \equiv -\lambda \chi_0+\frac{\lambda^2 k}{4 \pi}\bar{\eta}_0.
\end{align}
where 
\begin{equation}
    \chi_0 \equiv \int dy'\, G(y - y') \tilde{\mathcal{J}}_+(y').
\end{equation}
We observe the following useful relations $n\ge1 $
\begin{subequations} \label{SymInduction}
\begin{align}
  \MD_\lambda&\z(\frac{1}{n!}\MH_R\z( -\delta\hat{v}\y)^n\y) =\p_y\z(\frac{1}{n!}(\MD_\lambda\bar{\eta}_0) \z( -\delta\hat{v}\y)^{n}\y)-\frac{1}{(n-1)!}(\MD_\lambda \bar{\eta}_0)\z( -\delta\hat{v}\y)^{n-1}\\
       \MD_\lambda&\z(\frac{1}{n!}\MJ_{-t}\z( -\delta\hat{v}\y)^n\y)
    =\p_y\z(\frac{1}{n!} (\MD_\lambda\bar{\chi}_{0})\z( -\delta\hat{v}\y)^{n}\y)-\frac{1}{(n-1)!}(\MD_\lambda \bar{\chi}_{0})\z( -\delta\hat{v}\y)^{n-1}
\end{align}
\end{subequations}
where we have used
\begin{equation}
\begin{aligned}
& \MD_\lambda \bar{\eta}_0=-\frac{\z(\chi_0-\frac{\lambda k}{2\pi }\bar{\eta}_0\y)\MH_R}{1-\lambda \tilde{\MJ}_+},\quad \MD_\lambda \bar{\chi}_{0}=-\frac{\z(\chi_0-\frac{\lambda k}{2\pi }\bar{\eta}_0\y)\MJ_-^{(0)}}{1-\lambda \tilde{\MJ}_+}, \quad \MD_\lambda\chi_0=0,
\end{aligned}
\end{equation}
The relation \eqref{SymInduction} directly yields the solution to the flow equation \eqref{flowO}
\begin{equation}\label{dressedcureents}
    \begin{aligned}
        &\tilde{\MH}_R\equiv\sum_{n=0}^\infty\frac{(-1)^n}{n!}\p^n_y\z(\MH_R\delta \hat{v}^n\y),\quad \tilde{\MJ}_-\equiv\sum_{n=0}^\infty\frac{(-1)^n}{n!}\p^n_y\z(\MJ_{-t}\delta \hat{v}^n\y)
    \end{aligned}
\end{equation}
To check the flow equation is satisfied we have used the fact that \begin{align}\lim _{n\to \infty}\p_y^{n+1}\z(\frac{1}{n!}(\MD_\lambda\bar{\eta}_0) \z( -\delta\hat{v}\y)^{n}\y)= 0,  \quad \mathrm{for} \quad  |\lambda|<1 . \nonumber\end{align}
The above solution can be regarded as a functional relation on an arbitrary constant time slice.
In other words, currents on both sides of the equality, as well as the quantity $\delta \hat v$, should be regarded as a function of the coordinates $(t,y)$. 
Using the non-local coordinate transformation discussed in detail in Appendix \ref{symnonlocal}, we find a useful integral relation between the dressed currents and the deformed currents \eqref{fH}, which we reproduce here for convenience,
\begin{equation}\label{fH0}
	\begin{aligned}
    &\int dy'f(y')\tilde{\mathcal{H}}_R(y'-t)=\int dy' f(y'+\delta\hat{v}(t,y'))\HR(t,y')\\
    &\int dy' f(y')\tilde{\mathcal{J}}_-(y'-t)=\int dy' f(y'+\delta\hat{v}(t,y'))\MJ_{-t}(t,y')\\
	\end{aligned}
\end{equation}
where we have dropped total derivative terms, assuming evaluating on configurations that vanish sufficiently fast at infinity.
In particular, we find that the dressed currents evaluated on the non-local coordinates are related to the deformed and undeformed currents as 
\begin{align}\label{H0relation}
	\begin{aligned}
&\tilde{\mathcal{H}}_R(\hat{v})=\z({\partial \hat v\over \partial y}\y)^{-1}\HR(t,y)=\z({\partial \hat v\over \partial y}\y)^{-2} \HR^{(0)}(t,y),\\
		&\tilde{\mathcal{J}}_-(\hat{v})=\z({\partial \hat v\over \partial y}\y)^{-1}\MJ_{-t}(t,y)=\z({\partial \hat v\over \partial y}\y)^{-1} \MJ_{-}^{(0)}(t,y)
	\end{aligned}
    \end{align}
    where the nonlocal coordinate is defined in \eqref{nonlocalv} and  \begin{align}
       {\partial \hat v\over \partial y}= 1-\lambda \tilde{\MJ}_+(t,y)+\frac{\lambda^2 k}{4\pi }\MH_R(t,y)
    \end{align}
It is interesting to note that the relative factor between the dressed and undeformed currents in \eqref{H0relation} closely resembles the one resulting from a conformal transformation. However, the coordinate transformation applied here is non-local, field-dependent, and, furthermore, it mixes the left and right-moving sectors. These characteristics make a straightforward interpretation as a conformal transformation highly obscure. Nevertheless, this intriguing connection certainly warrants further investigation.

From the dressed currents \eqref{dressedcureents}, we obtain the following covariantly-constant symmetry generators in the right-moving sector, 
\begin{equation}\label{Charges2}
\begin{aligned}
    &    \tilde{\bar{L}}_m\equiv(-i)^{m+1}\int dyv^{m+1}\tilde{\MH}_R=(-i)^{m+1}\int dy(v+\delta \hat{v})^{m+1}\MH_R=(-i)^{m+1}\int dy \hat{v}^{m+1}\MH_R\\
    & \tilde{\bar{\mathcal{J}}}_{m}\equiv(-i)^{m}\int dyv^{m}\tilde{\MJ}_{-}=(-i)^{m}\int dy(v+\delta \hat{v})^{m}\MJ_{-t}=(-i)^{m}\int dy \hat{v}^{m}\MJ_{-t}
\end{aligned}
\end{equation}
where the second equality is obtained by applying integration by parts, and we have used the nonlocal coordinate introduced in \eqref{nonlocalv}. Similar to the left-moving sector, the right-moving charges constructed directly on the plane agree with the large radius limit of the charges on the cylinder. See appendix \ref{sec:compare} for more details. 
In particular, we find the expression for the right-moving energy and quasi-anti-chiral $U(1)$ charge,
\begin{equation}
   \begin{aligned}
       & H_R=\tilde{\bar L}_{-1}\qquad Q_R=\tilde{\bar{\mathcal{J}}}_{0}
   \end{aligned}
\end{equation}
Unlike the charge flow \eqref{chargeflow} appearing in the left-moving sector, the right-moving $U(1)$ charge corresponding to the quasi-anti-chiral current is just the zero mode of the anti-chiral current $\tilde{\bar{\mathcal{J}}}_-$.

\paragraph{The dressed symmetry algebra.}
As argued in \cite{Kruthoff:2020hsi,Georgescu:2024ppd,Chen:2025jzb}, the flow equation preserves conservation laws and chirality conditions.
As the dressed currents flow covariantly under the $J\bar T$ deformation and have a definite chirality property before the deformation, they remain conserved and satisfy the chirality condition along the $J\bar T$ flow. Now we verify this directly using the explicit form of the dressed currents. The total coordinate dependence comprises both explicit and implicit contributions, so the derivative expands as  
\begin{equation}
\partial_u\equiv\frac{1}{2}\z(\p_y+{\breve{\p}}_t+\left\{\, \cdot \, ,H\right\}\y) , \quad \partial_v\equiv\frac{1}{2}\z(\p_y-{\breve{\p}}_t-\left\{ \, \cdot \, ,H\right\}\y) \label{du}
\end{equation}
where $\p_y$ denotes the full derivative with respect to $y$, and  ${\breve{\p}}_{t}$ denotes taking the derivative with respect to the explicit time dependence.
We note that on any functionals of $\chi_0$, $\bar{\eta}_0$, and local operators, $\partial_y$ coincides with the Poisson bracket $\{\cdot, P\}$, while $\partial_u$ similarly acts as $\{\cdot, H_L\}$. Then we have
\begin{equation}
     \begin{aligned}
&\partial_v \MH_L=-\{\MH_L,H_R\}=-\{ \MH_L,{\tilde{\bar L}}_{-1}\}=0,\quad  
\partial_u \tilde{\MH}_R=\{\tilde{\MH}_R,H_L\}=\{ \tilde{\MH}_R, {\tilde{L}}_{-1}\}=0\\
&\partial_v \tilde{\MJ}_+=-\{\tilde{\MJ}_+,H_R\}=-\{ \tilde{\MJ}_+,{\tilde{\bar L}}_{-1}\}=0,\quad  
\partial_u \tilde{\MJ}_-=\{\tilde{\MJ}_-,H_L\}=\{ \tilde{\MJ}_-, {\tilde{L}}_{-1}\}=0
 \end{aligned}
 \end{equation}
Thus, the left-moving currents $\HL$ and $\tilde{\MJ}_+$  are holomorphic, while the right-moving dressed currents $\tilde{\MH}_R$ and $\tilde{\MJ}_-$ are antiholomorphic.

Finally, we obtain the algebra between the dressed symmetry generators, which reproduces the algebraic structure of their undeformed counterparts: two commuting copies of the Witt algebra, together with a $U(1)$ Kac-Moody algebra,
\begin{equation}\label{Vir-KM}
    \begin{aligned}
        &i\{\tilde{L}_n,\tilde{L}_m\}=(n-m)\tilde{L}_{n+m},\quad i\{\tilde{\MJ}_{n},\tilde{\MJ}_{m}\}=mk\delta_{m,-n},\quad i\{\tilde{L}_n,\tilde{\MJ}_{m}\}= -m\tilde{\MJ}_{m+n}\\
        &i\{\tilde{\bar{L}}_n,\tilde{\bar{L}}_m\}=(n-m)\tilde{\bar{L}}_{n+m},\quad i\{\tilde{\bar{\mathcal{J}}}_{n},\tilde{\bar{\mathcal{J}}}_{m}\}=mk\delta_{m,-n},\quad i\{\tilde{\bar{L}}_n,\tilde{\bar{\mathcal{J}}}_{m}\}= -m\tilde{\bar{\mathcal{J}}}_{m+n}\\
        &i\{\tilde{L}_n,\tilde{\bar{L}}_m\}=i\{\tilde{L}_n,\tilde{\bar{\mathcal{J}}}_m\}=i\{\tilde{\mathcal{J}}_n,\tilde{\bar{L}}_m\}=i\{\tilde{\mathcal{J}}_n,\tilde{\bar{\mathcal{J}}}_m\}=0
    \end{aligned}
\end{equation}
 
To summarize, we solve the deformed stress currents in terms of the undeformed ones \eqref{HRsolution}, and construct their dressed counterparts by solving the flow equation \eqref{flowO}.
In the left-moving sector, we find a Virasoro-Kac-Moody algebra generated by local currents:  The deformed stress tensor component $\HL$ directly solves the flow equation and remains chiral; The chiral current \eqref{tJplus} constructed from a linear combination of the quasi-chiral current and right-moving stress tensor is also covariantly constant. The effect of the $J\bar T$ deformation yields
a flowed chiral charge \eqref{chargeflow} and a deformed stress tensor \eqref{HRsolution}, but nevertheless still preserves local conformal symmetry. 
In the right-moving sector, however, the solution to the flow equation \eqref{flowO} is no longer local, but instead an infinite sum \eqref{dressedcureents} involving the non-local coordinate \eqref{nonlocalv}.
From the dressed currents we can construct infinitely many conserved charges, which under the Poisson bracket form two commuting sets of Virasoro-Kac-Moody algebra \eqref{Vir-KM}.

\section{The physical operators} \label{sec:operators}
In this section, we define {\it physical} operators and discuss their transformation under the symmetries. The basic logic is the same as what we did in \cite{Chen:2025jzb} for $T\bar{T}$ deformation. 

In the previous section, we worked out the symmetry generators. The natural next question is how these symmetries act on operators. Let us first consider the undeformed operators, which at $t=0$ are primary operators with conformal weights $(h,\bar h)$ and chiral charge $q$ under the chiral symmetry in the undeformed CFT$_2$, satisfying the following Poisson brackets \begin{equation}
\begin{aligned}\label{h0O0}
& \left\{\mathcal{H}_L^{(0)}\left(y_1\right), \mathcal{O}_{h,\bar{h},q}^{(0)}\left(y_2\right)\right\}=\left\{H_L^{(0)}, \mathcal{O}_{h,\bar{h},q}^{(0)}\left(y_2\right)\right\} \delta\left(y_1-y_2\right)+h \mathcal{O}_{h,\bar{h},q}^{(0)}\left(y_2\right) \partial_{y_1} \delta\left(y_1-y_2\right) \\
& \left\{\mathcal{H}_R^{(0)}\left(y_1\right), \mathcal{O}_{h,\bar{h},q}^{(0)}\left(y_2\right)\right\}=\left\{H_R^{(0)}, \mathcal{O}_{h,\bar{h},q}^{(0)}\left(y_2\right)\right\} \delta\left(y_1-y_2\right)-\bar{h} \mathcal{O}_{h,\bar{h},q}^{(0)}\left(y_2\right) \partial_{y_1} \delta\left(y_1-y_2\right)\\
&\left\{\MJ_{+}^{(0)}\left(y_1\right),\mathcal{O}_{h,\bar{h},q}^{(0)}\left(y_2\right)\right\}=-iq\mathcal{O}^{(0)}_{h,\bar{h},q}\delta(y_1-y_2)
\end{aligned}
\end{equation}
The Poisson brackets above are preserved under time evolution as given in \eqref{undeformedO} and therefore valid at any time.  However, the undeformed currents are no longer conserved under the deformed Hamiltonian and thus do not generate any symmetries. The physical relevance of the undeformed stress tensor remains unclear, despite its simple Ward identity.  To study the symmetry structure, we instead analyze the transformation of the operators under the dressed symmetries constructed in the previous section. The Poisson brackets between the dressed  
currents \eqref{dressedcureents} with the undeformed operators are lengthy and not very illuminating. Instead, it is simpler to present the Ward identities in terms of the conserved charges. For instance, the zero mode generators yield the following,
\begin{equation}\label{WI0}
\begin{aligned}
   i\{\tilde{L}_0,\MOh^{(0)}(y)\}&=  u \partial_{u} \MOh^{(0)}(y) + \z(h +\lambda \mathcal{B}_J^{(0)}\y)\MOh^{(0)}(y) \\
   i\{\tilde{\bar{L}}_0,\MOh^{(0)}(y)\}&= v\p_v\MOh^{(0)}(y)+\p_v \z((\delta\hat{v}+\frac{\lambda^2 k}{4\pi }\bar{\eta}_0)\MOh^{(0)}(y)\y)
   +\z(\bar{h}+\lambda \mathcal{B}_R^{(0)}-i\lambda q \bar{\eta}_0\y)\MOh^{(0)}(y) \\
    i \{\tilde{\mathcal{J}}_{0},\MOh^{(0)}(y)\}&=q\MOh^{(0)}(y)+\frac{i\lambda k}{2\pi}\p_v\MOh^{(0)}(y)
\end{aligned}
\end{equation}
where 
\begin{equation}
\mathcal{B}^{(0)}_J=-\frac{\bar{h}\tilde{\MJ}_+}{1-\lambda \tilde{\MJ}_+}
 , 
\quad \mathcal{B}_R^{(0)} =  \frac{\lambda k}{2\pi }\frac{(\bar{h}-1)\HR}{1-\lambda \tilde{\MJ}_+}
\end{equation}
The derivatives are defined in \eqref{du} and act as Poisson brackets, $\partial_u = \{ \cdot ,H_L \}$ and $\partial_v= -\{ \cdot ,H_R \}$, on any functional of $\delta \hat{v}$, $\bar{\eta}_0$, $\MOh^{(0)}$, $\MH_L$, and $\MH_R$. 
Compared to that of CFT, the above Ward identities receive two types of modifications: the appearance of  $\mathcal B_J^{(0)}$ and $\mathcal B_R^{(0)}$, which formally resemble a field-dependent shift of the conformal weights; and terms with 
$\delta\hat{v}$ or $\bar{\eta}_0$ $\partial_v$ acting on $\MOh^{(0)}$. The flow equations possess a similar structure
\begin{equation}
\begin{aligned}\label{dO0}
\mathcal{D}_\lambda \mathcal{O}_{h,\bar{h},q}^{(0)}(y) &=\p_v\z((-\chi_0+\frac{\lambda k}{2\pi}\bar{\eta}_0)\MO^{(0)}_{h,\bar{h},q}\y)-iq\bar{\eta}_0\MO^{(0)}_{h,\bar{h},q}+\Big( {\mathcal{B}}_J^{(0)}+\mathcal{B}_R^{(0)}\Big)\mathcal{O}_{h,\bar{h},q}^{(0)}
\end{aligned}
\end{equation}
As we have explicitly shown in the previous section, the conformal invariance is preserved in the left-moving sector, which implies the existence of local operators that transform as a primary operator under the left-moving Virasoro algebra. 
The extra $\mathcal{B}^{(0)}_{J}$ appearing in the first line of \eqref{WI0} must be canceled by a modification of the operator. 
Furthermore, the appearances of the two extra terms $\mathcal{B}^{(0)}_{J}$ and $\mathcal{B}^{(0)}_{R}$ are similar in both the Ward identities and flow equations. This motivates us to define a new operator $\MOh$ so that the two extra terms both disappear. 
 Inspired by the construction in $T\bar T$ deformations \cite{Chen:2025jzb}, we dress the undeformed operator in the following expression:
\begin{align}
   &\mathcal{O}_{h,\bar{h},q} \equiv \left(1-\lambda \tilde{\MJ}_+\right)^{a} \z(\frac{\p \hat{v}}{\p y}\y)^{b}  \mathcal{O}_{h,\bar{h},q}^{(0)}
\end{align}
where the exponents $a,b$ are to be determined. The factor $\frac{\p \hat{v}}{\p y}$ originates from the relation between the deformed and undeformed currents \eqref{H0relation} and the nonlocal coordinate transformation \eqref{nonlocal}, while the factor $(1-\lambda \tilde{\MJ}_+)$ comes from the Jacobian determinant \eqref{Jacobian}. Crucially, the flow equations for these operators and their Poisson brackets with the dressed symmetry generators retain the same structure as in 
\eqref{WI0} and \eqref{dO0}, but with the extra factors replaced by:
\begin{equation}
\mathcal{B}^{(0)}_J\to \mathcal{B}_J=-\frac{(\bar{h}+a+b)\tilde{\MJ}_+}{1-\lambda\tilde{\MJ}_+},\quad \mathcal{B}^{(0)}_R\to\mathcal{B}_R=\frac{\lambda k}{2\pi}\frac{(\bar{h}+b-1)\HR}{1-\lambda\tilde{\MJ}_+}
\end{equation}
As mentioned before, requiring the operator to be a primary operator under the left-moving conformal symmetry imposes the condition \begin{align}\label{BJ0}\mathcal{B}_J=0\Longrightarrow a+b=-\bar h
\end{align}
The similarity between $\mathcal{B}_J$
and $\mathcal{B}_R$ in the flow equation and the transformation law further motivates us to require  $\mathcal{B}_R=0$ as well.  
The two conditions can be satisfied simultaneously by choosing $ a=-1,\,  b= -\bar{h}+1$. This yields the following definition of {\it physical} operator in $J\bar T$ deformed CFT as:
\begin{equation}  \label{physicalOp}
    \mathcal{O}_{h,\bar{h},q} \equiv \frac{\mathcal{O}_{h,\bar{h},q}^{(0)}}{\left(1-\lambda \tilde{\MJ}_+\right)\left(1-\lambda \tilde{\mathcal{J}}_++\frac{\lambda^2 k}{4\pi}\HR\right)^{\bar{h}-1}}=\det\z[\frac{\p \hat{x}}{\p x}\y]\z(\frac{\p \hat{v}}{\p y}\y)^{-\bar{h}}\mathcal{O}_{h,\bar{h},q}^{(0)}
\end{equation}
Similar to the $T\bar T$ deformed CFTs \cite{Chen:2025jzb}, the relation between the physical operator and the undeformed operator is also determined by the nonlocal coordinate transformation, with the Jacobian determinant multiplied by a factor that resembles \footnote{ 
If we replace $y$ by $v$ in $({\partial \hat v\over \partial y })^{-h}$, it would be exactly the conformal factor that comes from a conformal transformation for a primary operator with right-moving conformal weight $\bar h$, ignoring the field dependence of the transformation. } a right-moving conformal transformation. 
Comparing with the $T\bar T$ case, however, we now have a stronger argument for the physical operator \eqref{physicalOp}: vanishing of one of the extra coefficients \eqref{BJ0} is a consequence of local conformal symmetry in the left-moving sector. 
While the motivation of requiring $\mathcal{B}_R=0$ is less compelling, this choice leads to a relation between the physical operator 
\eqref{physicalOp} and the undeformed operator that resembles a non-local conformal transformation. As we will show in the next section, the physical operator will be a key ingredient to build the dressed operators which solve the flow equation \eqref{flowO}.

So far we have motivated the definition of the physical operator \eqref{physicalOp} using the Ward identities of the zero mode generators  \eqref{WI0}. Let us represent the remaining Ward identities.

\paragraph{The left-moving Ward identities.}

The physical operator transforms as a primary under the left-moving conformal symmetry, 
\begin{equation}
\begin{aligned}
i\{\tilde{L}_m,\MOh\}&=i^mu^{m+1}\p_u\MOh+i^m(m+1)hu^m\MOh\\
   i\{\tilde{\MJ}_{m},\MOh\}&=i^mu^m\z(q\MOh+\frac{i\lambda k}{2\pi}\p_v\MOh\y)=i^mu^m\z(q\MOh+\frac{i\lambda k}{2\pi}\{H_R, \MOh\}\y)
\end{aligned}
\end{equation}Here $m\in \mathbb{Z}$. Note that the Poisson bracket with the chiral charge acquires an additional term which comes from the flow of the charge \eqref{chargeflow}.  
This is more evident in the momentum space, with the physical operators defined by the Fourier transformation 
\begin{align}
    \MOh(p_L,p_R)
&=\mathcal{F}_{p_L,p_R}[ {O}_{h,\bar{h},q}(u,v)]\equiv \int dudv e^{-ip_L u+ip_R v}{\MO}_{h,\bar{h},q}(u,v) 
\end{align}
Then the physical operator will transform as
\begin{equation}\label{leftWard}
    \begin{aligned}
        i\{\tilde{L}_m,\MOh(p_L,p_R)\}&=(-1)^m(m+1)(h-1)\frac{\p^m\!\MOh(p_L,p_R)}{\p{p_L^m}}\\
        &+(-1)^{m+1} p_L\frac{\p^{m+1}\!\MOh(p_L,p_R)}{\p p_L^{m+1}},\qquad m\geq -1\\
        i\{\tilde{\MJ}_m,\MOh(p_L,p_R)\}&=(-1)^mq_\lambda\frac{\p^m\MOh(p_L,p_R)}{\p p_L^{m}},\quad q_\lambda=q+\frac{\lambda k}{2\pi} p_R,\quad m\geq 0
    \end{aligned}
\end{equation}
which means that the physical operator $\MOh(p_L,p_R)$ transforms as a primary under the left-moving symmetry with chiral charge $q_\lambda$,  consistent with charge flow \eqref{chargeflow}.
\paragraph{The right-moving Ward identities.}The transformation of the physical operator under the right-moving Virasoro generator, however,  involves the non-local variables, 
\begin{equation}
\begin{aligned}
   i\{\tilde{\bar{L}}_m,\MOh\}&=\z((-i)^m\hat{v}^{m+1}+(m+1)\frac{\lambda^2k}{4\pi}\bar{\eta}_m\y)\p_v\MOh\\&+(m+1)\z((-i)^m\Big(\bar{h}+\frac{\lambda^2k}{2\pi}\frac{\HR}{1-\lambda \tilde{\MJ}_+}\Big)\hat{v}^m-i\lambda q\bar\eta_m \y)\MOh,\quad m\in \mathbb{Z}
\end{aligned}
\end{equation}
where $\bar\eta_m,\,\bar{\chi}_m$ is the extension of the zero mode version \eqref{etachi0}:
\begin{equation}
\begin{aligned}\label{etam}
   &\bar{\eta}_m(t,y)\equiv (-i)^m\int dy'G(y-y')(y'-t+\delta\hat{v})^m\HR(y')\\
   &\bar{\chi}_m(t,y)\equiv (-i)^m\int dy'G(y-y')(y'-t+\delta\hat{v})^m\MJ_{-t}(y')
\end{aligned}
\end{equation}
The Ward identity in momentum space is given by 
\begin{equation}\label{TTbarWard}
\begin{aligned}
   i\{\tilde{\bar{L}}_m,\MOh(p_L,p_R)\}&=(m+1)(\bar{h}-1)\sum_{k=0}^{m}\frac{(-1)^km!}{k!(m-k)!}\frac{\p^k\mathcal{F}_{p_L,p_R}[(-i\delta\hat{v})^{m-k}\MOh]}{\p p_R^k}\\
   &+p_R\sum_{k=0}^{m+1}\frac{(-1)^{k}(m+1)!}{k!(m+1-k)!}\frac{\p^k\mathcal{F}_{p_L,p_R}[(-i\delta\hat{v})^{m+1-k}\MOh]}{\p p_R^k}\\
   &-i(m+1)\lambda \z(q+\frac{\lambda k}{4\pi}p_R\y)\mathcal{F}_{p_L,p_R}[\bar{\eta}_m\MOh],\qquad\quad  m \geq -1
\end{aligned}
\end{equation}
Furthermore, if the operator $\MOh^{(0)}$ carries anti-chiral charge $\bar{q}$ under $\MJ_{-}^{(0)}$, as defined by the Poisson bracket:
\begin{equation}
\{\MJ_{-}^{(0)}(y_1),\MO^{(0)}_{h,\bar{h},q}(y_2)\}=-i\bar{q}\MO^{(0)}_{h,\bar{h},q}\delta(y_1-y_2),
\end{equation}
Then, after the $J\bar{T}$ deformation, the corresponding physical operator satisfies the following additional Ward identity:
\begin{equation}
    \begin{aligned}
    i\{\tilde{\bar{\mathcal{J}}}_{m},\MOh\}&=\z((-i)^m\hat{v}^m\bar{q}+(-i)^{m-1}m\z(\frac{\lambda^2k}{4\pi}\frac{\MJ_{-t}}{1-\lambda\tilde{\MJ}_+}\y)\hat{v}^{m-1}-im\bar{\chi}_{m-1}\lambda q\y)\MOh\\
   &+m\bar{\chi}_{m-1}\frac{\lambda^2k}{4\pi}\p_v\MOh,\quad m\in \mathbb{Z}
   \end{aligned}
   \end{equation}
   and the Ward identity in momentum space
   \begin{equation}
   \begin{aligned}\label{Jbarward}
        i\{\tilde{\bar{\mathcal{J}}}_m,\MOh(p_L,p_R)\} &=\bar{q}\sum^{m}_{k=0}\frac{(-1)^km!}{k!(m-k)!}\frac{\p^k\mathcal{F}_{p_L,p_R}[(-i\delta\hat{v})^{m-k}\MOh]}{\p p_R^k}\\
  &-im\lambda\z(q+\frac{\lambda k}{4\pi}p_R\y)\mathcal{F}_{p_L,p_R}[\bar{\chi}_{m-1}\MOh],\ \qquad\quad\quad   m \geq 0
    \end{aligned}
\end{equation}

\section{The dressed operators}\label{sec:dress}
In this section, we derive the dressed operators by explicitly solving the flow equation \eqref{fO} at the classical level, with the initial condition that they are primary operators in the undeformed theory. The physical operator \eqref{physicalOp} provides a convenient starting point as the flow equation is simplified significantly as compared to the undeformed ones. In the following, we will first solve the flow equation by expressing the dressed operators in terms of the physical operators, and then invert the relation.

\subsection{Construction of the dressed operators}
Acting the covariant derivative $\mathcal D_\lambda$ on the physical operator \eqref{physicalOp}, we get 
\begin{equation}
\begin{aligned}\label{dO}
\mathcal{D}_\lambda \mathcal{O}_{h,\bar{h},q}&=\p_v\z((-\chi_0+\frac{\lambda k}{2\pi}\bar{\eta}_0)\MO_{h,\bar{h},q}\y)-iq\bar{\eta}_0\MO_{h,\bar{h},q}
\end{aligned}
\end{equation}
To proceed, we first multiply the physical operator by a factor of $e^{i\lambda q\bar{\eta}_0}$ to absorb the charge term in the flow equation \eqref{dO}, so that the right-hand side is a total derivative.
\begin{equation}\begin{aligned}
& \mathcal{D}_\lambda e^{i\lambda q\bar{\eta}_0} \mathcal{O}_{h,\bar{h},q} =-\p_v\z((\chi_0-\frac{\lambda k}{2\pi}\bar{\eta}_0)e^{i\lambda q\bar{\eta}_0} \MOh\y)
\end{aligned}
\end{equation}
The resulting expression formally resembles the total differential of a function under a shift from $(u,v)$ to $(u,v+\delta \hat v)$, with the key distinction that the differential element $\delta \hat v$ depends on the coordinates and is placed inside the derivatives. We can define a similar structure at order $n$
\begin{equation}
  { \tilde {\mathcal O}}_{h,\bar{h},q}^{(n)} \equiv  \frac{1}{n!} \partial_v^{n}\left((-\delta\hat{v})^{n}e^{i\lambda q\bar{\eta}_0} \mathcal{O}_{h,\bar{h},q}\right)
\end{equation}
This leads to the following recursion relation:
\begin{equation}
\mathcal{D}_\lambda { \tilde {\mathcal O}}_{h,\bar{h},q}^{(n)}=M^{(n)}-M^{(n+1)}
\end{equation}
where 
\begin{equation}
\begin{aligned}
        M^{(n)}&\equiv \frac{n}{n!}\p_v^n\z((\chi_0-\frac{\lambda k}{2\pi}\bar{\eta}_0)(-\delta \hat{v})^{n-1}e^{i\lambda q\bar{\eta}_0}\MOh\y)
\end{aligned}
\end{equation}
The recursion relation thus yields the following solution for the dressed operator:
\begin{align}\label{Flowedoperator}
   & \tilde{\mathcal{O}}_{h,\bar{h},q} (u,v)\equiv \sum_{n=0}^\infty   \frac{1}{n!} \partial_v^{n}\left((-\delta\hat{v})^{n}e^{i\lambda q\bar{\eta}_0} \mathcal{O}_{h,\bar{h},q}\right)\\
&\mathcal D_\lambda    \tilde{\mathcal{O}}_{h,\bar{h},q} =0 \nonumber
\end{align}
From the above expression, we see that the dressed operators depend on the non-local coordinates and are manifestly non-local. Since the $J\bar{T}$ deformation on the plane is a canonical transformation, the algebra of dressed operators is preserved. Consequently, the dressed operators defined in \eqref{Flowedoperator} are primary operators with conformal weights $(h,\bar{h})$ and chiral charge $q$ under the dressed symmetry generators
\begin{equation}
\begin{aligned}
   &i\{\tilde{L}_m,\tilde{\MO}_{h,\bar{h},q}\}=(m+1)hx^m\tilde{\MO}_{h,\bar{h},q}+x^{m+1}\p_{x}\tilde{\MO}_{h,\bar{h},q}\\
   &i\{\tilde{\bar{L}}_m,\tilde{\MO}_{h,\bar{h},q}\}=(m+1)\bar{h}\bar{x}^m\tilde{\MO}_{h,\bar{h},q}+\bar{x}^{m+1}\p_{\bar{x}}\tilde{\MO}_{h,\bar{h},q}\\
   &i\{\tilde{\mathcal{J}}_m,\tilde{\MO}_{h,\bar{h},q}\}=qx^m\tilde{\MO}_{h,\bar{h},q}
\end{aligned}
\end{equation}
 where $x=iu$ and $\bar{x}=-iv$ are complex coordinates in Euclidean space. Furthermore, if the primary operator $\MO^{(0)}_{h,\bar{h},q}$ is charged under $\mathcal{J}_{-}^{(0)}$ with $\bar{q}$ in the undeformed CFT, it follows that the corresponding dressed operator satisfies:
\begin{equation}
i\{\tilde{\bar{\mathcal{J}}}_m,\tilde{\MO}_{h,\bar{h},q}\}=\bar{q}\bar{x}^m \tilde{\MO}_{h,\bar{h},q}
\end{equation}

\subsection{Alternative expressions of the dressed operators}
For an arbitrary function $ f(u, v) $ that decays sufficiently rapidly, integration by parts on
\eqref{Flowedoperator} reveals that:
\begin{equation}\label{inteO}
\int dudvf(u,v)\tilde{\MO}_{h,\bar{h},q}(u,v)=\int dudv f(u,\hat{v})e^{i\lambda q\bar{\eta}_0}    \MO_{h,\bar{h},q}(u,v)
\end{equation}
where the nonlocal coordinates are defined in \eqref{nonlocalv}. This formula allows us, similar to the case of dressed symmetry generators, to derive a set of useful relations between the dressed and physical operators through appropriate choices of the function $f(u,v)$.
\begin{itemize}
    \item 
Applying the choice $f(u,v)=e^{-ip_Lu+ip_Rv}$, we have the operator in  momentum space 
\begin{equation}
\begin{aligned}
   \tilde{\MO}_{h,\bar{h},q}(p_L,p_R)&\equiv \mathcal{F}_{p_L,p_R}[\mathcal{\tilde O}_{h,\bar{h},q}(u,v)]  =\int du dv e^{-ip_Lu+ip_Rv}\tilde{\MO}_{h,\bar{h},q}(u,v)\\
   &=\int dudv e^{-ip_Lu+ip_R\hat{v}}e^{i\lambda q\bar{\eta}_0}\MOh (u,v)
\end{aligned}
\end{equation}
This relation establishes two equivalent expressions of the dressed momentum-space operator: by definition, it is the Fourier transform of the dressed operator in $(u,v)$ coordinates; equivalently, it is obtained by decomposing the composite operator $e^{i\lambda q\bar{\eta}_0}\MOh (u,v)$ into a plane-wave basis with respect to the nonlocal coordinates $(u, \hat{v})$.
\item Applying the function 
\begin{equation}
f(u',v')=\delta(u'-u,v'-\hat{v}(v)),
\end{equation}
we establish the relation between the dressed and the physical operator:
\begin{equation}\label{relation}
\tilde{\MO}_{h,\bar{h},q}(u,\hat{v})={e^{i\lambda q\bar{\eta}_0}(1-\lambda \tilde{\MJ}_+)\MOh(u,v)\over (1-\lambda\tilde{\MJ}_++\frac{\lambda^2k}{4\pi}\HR)}, \end{equation}
The left-hand side of the equation is obtained by replacing the coordinate $v$ with the nonlocal coordinate $\hat{v}$ in the dressed operator \eqref{Flowedoperator}. This yields a closed form relation: the dressed operator in the auxiliary space $(u, \hat{v})$ is related to a local functional of the physical operator at $(u,v)$.
\item Furthermore, combining this with the relation between the physical and undeformed operators, we find 
\begin{equation}\label{O0relation}
\tilde{\MO}_{h,\bar{h},q}(u,\hat{v})=e^{i\lambda q\bar{\eta}_0}\z(1-\lambda \tilde{\MJ}_++\frac{\lambda^2k}{4\pi}\HR\y)^{-\bar{h}}\mathcal{O}_{h,\bar{h},q}^{(0)}(u,v)
\end{equation}
To give the dressed operator living in the original coordinate $(u,v)$, we employs the inverse nonlocal coordinate transformation $U(\sigma^+,\sigma^-)$ and $V(\sigma^+,\sigma^-)$, which are field-dependent functions of arbitrary variables $\sigma^+, \sigma^-$ defined as
\begin{equation}\begin{aligned}
&U(\sigma^+,\sigma^-)\equiv\sigma^+\\
 &  V(\sigma^+,\sigma^-) \equiv \sigma^-+\lambda\int G(\sigma^+-u')\tilde{\MJ}_+(u') du'-\frac{\lambda^2k }{4\pi}\int G(\sigma^--v') \tilde{\MH}_R(v')dv' 
\end{aligned}
\end{equation}
Plugging the non-local coordinates $(u,\hat{v})$ in we get back the $(u,v)$ coordinates
\begin{equation}\label{sigmauv}
    U(u,\hat{v})=u,\quad V(u,\hat{v})=v
\end{equation}
This gives another expression for the dressed operator,
\begin{equation}\label{O0relation2}
\tilde{\MO}_{h,\bar{h},q}({u},{v})={e^{i\lambda q\int G(v-v')\tilde{\MH}_R(v')} (1-\lambda \tilde{\MJ}_+(u))^{\bar{h}}\over \left(1-\frac{\lambda^2k}{4\pi}\tilde{\MH}_R (v)\right)^{\bar{h}}} \,  \mathcal{O}_{h,\bar{h},q}^{(0)}(U(u,v),V(u,v))
\end{equation}

Therefore we express the dressed operator on the physical space $(u,v)$ in terms of undeformed operators that depend on the nonlocal variables $(U, V)$.
\end{itemize}

\subsection{Physical operators from dressed operators}
Finally, applying 
\begin{equation}
    f(u,v)=e^{-i\lambda q\int^v\tilde{\MH}_R(v')dv'}\delta(U(u,v)-u_0,V(u,v)-v_0)
\end{equation}
We get\begin{align}
\MOh(u_0,v_0)=\int dudv\delta(U(u,v)-u_0,V(u,v)-v_0)e^{-i\lambda q\int^v\tilde{\MH}_R(v')dv'}\tilde{\MO}_{h,\bar{h},q}(u,v)\label{Ophys1}
\end{align}
We may rename the integration variables $u$ and $v$ on the right-hand side as $\sigma^\pm$, since they are dummy variables, and omit the subscripts of $(u_0, v_0)$, which are arbitrarily chosen. This yields a formal expression for the physical operator in terms of the dressed ones.
\begin{align}
\MOh(u,v)=\int d\sigma^2\delta(U(\sigma ^+,\sigma^-)-u,V(\sigma^+,\sigma^-)-v)e^{-i\lambda q\int^{\sigma^{-}} \tilde{\MH}_R(\sigma^{-\prime})d\sigma^{-\prime}}\tilde{\MO}_{h,\bar{h},q}(\sigma^+,\sigma^-)\label{Ophys}
\end{align}
Note that the delta function on the right-hand side localizes the integral to the points satisfying $V(\sigma^+,\sigma^-) = v$. As we show in \eqref{sigmauv}, the solution is \footnote{ We assume that there are no other solutions to $ V(\sigma^+,\sigma^-)=v$. } \begin{equation}\sigma^+=u, \quad \sigma^-=\hat v(u,v), \label{onshell}\end{equation}
We can also perform a Fourier transform by substituting
\begin{align}
&f(u,v) =e^{-i\lambda q\int^v\tilde{\MH}_R(v')dv'}e^{-ip_L U(u,v)+ip_R V(u,v)},   \quad f(\hat{u}, \hat{v}) = e^{-i\lambda q\bar{\eta}_0(u,v)}e^{-i p_L u+ i p_R v}
\end{align}
 into \eqref{inteO}. It follows that the physical operator in momentum space takes the form
 \begin{equation}\label{TTbardress}
\begin{aligned}
\MOh(p_L,p_R)
&=\int dudv e^{-i\lambda q\int^v\tilde{\MH}_R(v')dv'} e^{ -ip_L U(u,v)+ip_R V(u,v)} \tilde{\MO}_{h,\bar{h},q}(u,v) 
\\
&=\mathcal{F}_{p_L,p_R}[e^{ip_R\lambda \int^u\tilde{\MJ}_+(u')du'-i\lambda\z(q+\frac{\lambda kp_R}{4\pi}\y)\int^v \tilde{\MH}_R(v')dv'}\tilde{O}_{h,\bar{h},q}(u,v)]
\end{aligned}
\end{equation}

To summarize, we have explicitly constructed the dressed operators by solving the flow equation. These are given either by the infinite sum in \eqref{Flowedoperator}, or by the closed expressions in \eqref{O0relation2} written in terms of operators on nonlocal variables. We also give an expression for the physical operators in terms of dressed operators, which allows us to lift that to quantum level and calculate the correlation function.

\section{Physical operators at the quantum level}\label{sec:PQ}

In this section we will first discuss the definition of the physical operators at the quantum level, following the construction of \cite{Chen:2025jzb}. Then we discuss the Ward identities.

\subsection{A definition of the physical operator}

In the previous two sections, we explicitly constructed the dressed operators at the classical level in terms of the undeformed and physical operators. Conversely, the physical operators can also be expressed in terms of the dressed operators, both in position space \eqref{Ophys} and momentum space \eqref{TTbardress}. In these expressions—particularly the momentum-space one \eqref{TTbardress}—all operators on the right-hand side are organized according to conformal symmetry. This allows us to reverse the logic and define the physical operator at the quantum level from the dressed operators, following the same prescription as in \cite{Chen:2025jzb}:
We assume the existence of the dressed operators and dressed currents at the quantum level, defined as solutions to the flow equation \eqref{flowedeq}, and take them as the starting point. The physical operators are then defined through their relation to the dressed operators, which takes the same formal form as in the classical case, supplemented by normal ordering. Explicitly, the position-space operator is defined as:
\begin{align}\label{Quantumposion}
\MOh(u,v)=\int d\sigma^2\delta(U(\sigma^+,\sigma^-)-u,V(\sigma^+,\sigma^-)-v)e^{-i\lambda q\int^{\sigma^{-}} \tilde{\MH}_R(\sigma^{-\prime})d\sigma^{-\prime}}\tilde{\MO}_{h,\bar{h},q}(\sigma^+,\sigma^-)
\end{align}
which is the same as \eqref{Ophys} but assuming normal ordering on the right-hand side. 
Written in the momentum space, the physical operator is defined as 
 \begin{equation}
\begin{aligned}\label{Ophys-m-L}
\MOh(p_L,p_R)\equiv\int dudv e^{-ip_Lu+ip_Rv}\z[e^{ip_R\lambda \int^u\tilde{\MJ}_+(u')du'-i\lambda\z(q+\frac{\lambda kp_R}{4\pi}\y)\int^v \tilde{\MH}_R(v')dv'}\tilde{O}_{h,\bar{h},q}(u,v)\y]
\end{aligned}
\end{equation}
which is identical to \eqref{TTbardress} but with the right-hand side understood as normal ordered. 
It is also convenient to consider operators in  Euclidean space, adopting the following Wick rotation:
\begin{equation}
\begin{aligned}
        &u\to -ix,\ v\to i\bar{x},\quad p_L\to -i p,\ p_R\to -i\bar{p}\\
    &\tilde{\MH}_{L}(u)\to\frac{T(x)}{2\pi},\ \tilde{\MH}_{R}(v)\to\frac{\bar{T}(\bar{x})}{2\pi},\ \MJ_+(u)\to \frac{j(x)}{2\pi}
\end{aligned}
\end{equation}
and $it\to t_E$ becomes the Euclidean time. This yields the following expression for the physical operator in momentum space:
\begin{equation}
    \MO_{h,\bar{h},q}(p,\bar{p})=\int dx^2 e^{ipx+i\bar{p}\bar{x} - \frac{i\lambda}{2\pi} \z[\bar{p}\int^xdx'j(x')+\z(iq+\frac{\lambda k\bar{p}}{4\pi}\y)\int^{\bar{x}}d\bar{x}'\bar{T}(\bar{x}')\y] }  \tilde{\mathcal{O}}_{h, \bar{h}, q} \left(x, \bar{x} \right)\label{Ophys-m-E}
\end{equation}
The operator in square brackets takes the form of a Wilson line, like dressing of the dressed operator.

The definition of physical operators given in \eqref{Ophys-m-E} follows the same strategy used in the $T\bar T$ deformation \cite{Chen:2025jzb}, where it was shown to coincide at the classical level with the operator discussed in \cite{Aharony:2023dod}. We note, however, that our physical operator \eqref{Ophys-m-E} in the $J\bar T$ deformation differs from the one defined in \cite{Guica:2021fkv}; a detailed comparison is provided in Appendix \ref{sec:compare}.
Here we offer further remarks on the quantum definition of physical operators. In what follows, we adopt the following perspective:
 \begin{itemize}
    \item[i)] 
    We assume the existence of the dressed symmetry generators, defined as solutions to the flow equation \eqref{flowedeq} with appropriate initial conditions. By construction, both the Ward identities among these operators and their correlation functions remain identical to those in the CFT.
    \item[ii)] 
    We define the physical operator from the dressed operators using \eqref{Quantumposion} in position space, and \eqref{Ophys-m-L} \eqref{Ophys-m-E} in momentum space with Lorentzian and Euclidean signature. 
In the following, we will mainly focus on the momentum space definition in Euclidean space \eqref{Ophys-m-E}.
\item[iii)] 
At the classical level, the physical operators can be expressed in terms of the undeformed operators via \eqref{physicalOp}, where the right-hand side involves only quantities defined at a single point.  At the quantum level, however, \eqref{physicalOp} must be normal ordered, and its precise form is not yet known. In particular, it remains unclear whether the physical operators can still be written purely in terms of pointlike operators at the quantum level. Nevertheless, the subsequent calculations rely only on the definition \eqref{Ophys-m-E}, especially the calculation of correlation functions.
\end{itemize}
In summary, from this section onward, we take the dressed variables as the fundamental building blocks. The physical operators are then defined in momentum space via \eqref{Ophys-m-E}. With this perspective in mind, we briefly comment on the status of the undeformed operators. As seen from the classical relation \eqref{physicalOp}, there is an additional factor involving ${\mathcal H}_R$ that relates the physical and undeformed operators. Consequently, the expression for the undeformed operator in terms of the dressed operators becomes more involved, and their precise treatment remains unclear.

\subsection{Ward identities}

We work in momentum space, with fixed left and right-moving energies $(p,\bar p)$. 
From the classical Ward identities, we learn that 
the three types of operators have the following symmetry properties. 

\paragraph{At the classical level} The dressed operator $\tilde{\mathcal{O}}_{h,\bar{h},q}(p,\bar p)$ is a primary operator with conformal weights $(h,\bar h)$ and U(1) charges $q,\bar q$
under the dressed  Virasoro-Kac-Moody symmetry \eqref{Charges1} \eqref{Charges2}.
The physical operator ${\MOh(p, \bar p)}$ behaves as a primary operator under the left-moving Virasoro-Kac-Moody symmetry \eqref{Charges1}, but does not transform in a nice way under the right-moving Virasoro-Kac-Moody symmetry \eqref{Charges2}.
The undeformed operator $\MOh^{(0)}(p,\bar p)$
is not a primary operator of either the left or the right moving Virasoro generators, but it has definite charges. 
These classical properties are summarized in Table \ref{tab:operator_charges}.
\begin{table}[ht]
\centering
\caption{Charges of selected operators: Here we summarize the charges of 3 kinds of operators at the classical level. }
\label{tab:operator_charges}
\begin{tabular}{l c c c c c c c}
\hline
 & $Q_L$ & $\tilde{\mathcal{J}}_0$ & $h$& $\bar{h}$&$H_L=\tilde{{L}}_{-1}$ &$H_R=\tilde{\bar{L}}_{-1}$ & $\tilde{\bar {\mathcal J}}_0=Q_R$\\
\hline
$\MOh^{(0)}(p_L,p_R)$ & $q$ &$q_\lambda$  &  &  & $p_L$&$ p_R$&  $\bar q$\\
$\tilde{\mathcal{O}}_{h,\bar{h},q}(p_L,p_R)$ & $q-\frac{\lambda k }{2\pi}p_R$ & $q$ & $h$ & $\bar h$&$p_L$& $p_R$ & $\bar q$\\
$\MOh(p_L,p_R)$ & $q$ & $q_\lambda$ & $h$&  & $p_L$& $p_R$ &$\bar q$ \\
\hline
\end{tabular}
\end{table}
\paragraph{At the quantum level}
As explained earlier, one advantage of the physical operator defined in \eqref{Ophys-m-E} is that it is expressed purely in terms of the dressed operators and symmetry generators, which obey the same Ward identities as in a CFT, i.e.
\begin{equation}
\begin{aligned}
    & \bar{T}\left(\bar{w}_1\right)  \tilde{\mathcal{O}} \left(x\right)\sim\left[\frac{h}{\left(\bar{w}_1-\bar{x}\right)^2}+ \frac{1}{\bar{w}_1-\bar{x}}\partial_{\bar{x}}\right]\tilde{\mathcal{O}}\left(x\right)\\
& j\left(w_1\right) \tilde{\mathcal{O}} \left(x\right)= \frac{ q } {w_1-x} \tilde{\mathcal{O}}\left(x\right)
\end{aligned}
\end{equation}
We assume the chiral currents have level $k$ with the OPE,
\begin{equation}
    j(w) j(x)\sim {k \over(w-x)^2 }
\end{equation}
 For simplicity, let us assume a Sugawara decomposition \footnote{We leave the discussion for the most general cases where this assumption can not be made to future study. } of the stress tensor \begin{align}
    T(x)=\hat T(x)+{:jj(x):\over 2k},\quad  
\end{align} 
where $\hat T$ and $j$ are decoupled. 
Using the ingredients above, we can compute the OPE between the physical operator \eqref{Ophys-m-E} and both the stress tensor and the chiral current. The resulting Ward identities elevate the classical relations \eqref{leftWard}, \eqref{TTbarWard}, and \eqref{Jbarward} to the quantum level. As observed in the classical analysis, the physical operator does not transform as a primary operator in the right-moving sector. Nonetheless, it carries a definite right-moving charge $\bar q$ under the anti-chiral current $\tilde{\bar J}_0$, and a fixed right-moving momentum $-i\bar{p}$. In momentum space, we have the following Ward identities,
\begin{equation}
\begin{aligned}\label{WIQ}
    [{\tilde L}_n ,  \mathcal{O}_{h,\bar h,q}(p,\bar p)]&=(-i)^n(n+1)\z(h-\frac{i\lambda q\bar{p}}{2\pi}-\frac{\lambda^2k\bar{p}^2}{8\pi^2}-1\y)\p_{p}^n\MOh(p,\bar{p})-(-i)^np\p_p^{n+1}\MOh(p,\bar{p})\\
  [{\tilde {\mathcal J}}_n,  \mathcal{O}_{h,\bar h,q}(p,\bar p)]&=  (-i)^n(q-{i\lambda k\over 2\pi} \bar p)\p_{p}^n\mathcal{O}_{h,\bar h,q}(p,\bar p)\\
[\tilde{\bar{\mathcal{J}}}_0,\MOh(p,\bar{p})]&=\bar{q}\MOh(p,\bar{p})
\end{aligned}
\end{equation}
 In the left moving sector, the above Ward identities indicate that the physical operator \eqref{Ophys-m-E} is a primary operator under the left-moving Virasoro-Kac-Moody algebra with a shifted conformal weight, and shifted charge under the chiral current,
\begin{align}
    h_\lambda=h-\frac{i\lambda q\bar{p}}{2\pi}-\frac{\lambda^2k\bar{p}^2}{8\pi^2},\quad q_\lambda=q-{i\lambda k\bar p\over 2\pi} \label{shiftL}
\end{align}
Note that the quasi-chiral charge and Noether charge of the operator $\mathcal{O}_{h,\bar h,q}(p,\bar p)$ remain the same, taking values as $q$ and  $q+\bar q$, respectively. These charges are consistent with the charge flow in the spectrum \cite{Chakraborty:2018vja,Apolo:2018qpq}.

We summarize the quantum numbers of dressed and physical operators in table \ref{tab:quantum}.
\begin{table}[ht]
\centering
\caption{Quantum numbers of the dressed and physical operators }
\label{tab:quantum}
\begin{tabular}{|l| c| c| c| c| c| c|}
\hline
 & $Q_L$ & $\tilde{\mathcal{J}}_0$ & $h$& $\bar{h}$ &$\mathcal P_0$ & $\tilde{\bar {\mathcal J}}_0=Q_R$\\
\hline
$\tilde{\mathcal{O}}_{h,\bar{h},q}(p,\bar{p})$ & $q+\frac{i\lambda k }{2\pi}\bar{p}$ & $q$ & $h$ & $\bar h$& $q+\bar q+\frac{i\lambda k }{2\pi}\bar{p}$ & $\bar q$\\
\hline 
$\MOh(p,\bar{p})$  & $q$ &$q_\lambda$ & $h_\lambda$ & & $q+\bar q$ &$\bar q$ \\
\hline
\end{tabular}
\end{table}

\section{Correlation functions} \label{sec:CorrF}

Having defined the physical operators in $J\bar{T}$-deformed CFTs in terms of dressed operators, we now calculate their correlation functions by leveraging the Virasoro-Kac-Moody symmetry.

\subsection*{The flow equation in position space}
 Let us consider the flow equations of the correlation functions under $J\bar{T}$ deformation. Flow equations for correlation functions in $T\bar{T}$ deformation were discussed in the literature \cite{Cardy:2019qao,Kruthoff:2020hsi,Chen:2025jzb}.
Using the relation \eqref{Quantumposion} between physical operators and dressed operators, we have 
\begin{equation}
\begin{aligned}
        \p_\lambda\left\langle \MO_{h,\bar{h},q}(u,v)... \right\rangle&=\int d\sigma^2\Big\langle \Big(\z(\int^{\sigma^+}du'\tilde{\MJ}_+(u')-\frac{\lambda k}{2\pi}\int^{\sigma^-}dv'\tilde{\MH}_R(v')\y)\\
        &\p_v\delta(U(\sigma^+,\sigma^-)-u,V(\sigma^+,\sigma^-)-v)\tilde{\MO}_{h,\bar{h},q}(\sigma^+,\sigma^-)\\
        &-iq\int^{\sigma^-}dv'\tilde{\MH}_R(v')\delta(U(\sigma^+,\sigma^-)-u,V(\sigma^+,\sigma^-)-v)\tilde{\MO}_{h,\bar{h},q}(\sigma^+,\sigma^-)\Big)...\Big\rangle
\end{aligned}
\end{equation}
where we have used the invariance of dressed operator correlation functions under the canonical flow, so that the derivative acts only on the explicit $\lambda$-dependence. Substituting the relation between the dressed and deformed stress tensors from \eqref{Ginverse}, the flow equation becomes:
\begin{equation}
\begin{aligned}\label{floweq}
        \p_\lambda\left\langle \MO_{h,\bar{h},q}(u,v)... \right\rangle&=\Big\langle \epsilon^{ij}\Big(-\p_v\z(\int^{(u,v)} dx'_{j}\MJ_{+i}(x')\MO_{h,\bar{h},q}(u,v)\y)\\
        &+iq\int^{(u,v)}dx_j'T_{vi}(x')\MOh(u,v)\Big)... \Big\rangle
\end{aligned}
\end{equation}

\subsection*{Series expansion}
To compute the correlation functions, we consider
the physical operator in momentum space \eqref{Ophys-m-E} which we reproduce here for convenience:
\begin{equation}
    \MO_{h,\bar{h},q}(p,\bar{p})=\int dx^2 e^{ipx+i\bar{p}\bar{x} - \frac{i\lambda}{2\pi} \z[\bar{p}\int^xdx'j(x')+\z(iq+\frac{\lambda k\bar{p}}{4\pi}\y)\int^{\bar{x}}d\bar{x}'\bar{T}(\bar{x}')\y] }  \tilde{\mathcal{O}}_{h, \bar{h}, q} \left(x, \bar{x} \right)\label{Ophys-m-E1}
\end{equation}
As discussed in Section \ref{sec:JTbar}, the Ward identities involving the dressed currents $j(x)$, $\bar{T}(\bar{x})$ and the dressed operators $\tilde{\MO}_{h,\bar{h},q}(x,\bar{x})$ are preserved along the flow and retain the same form as in the undeformed CFT. This allows us to compute correlation functions between physical operators by expanding the Wilson line like dressing factor as a power series, and thus evaluate the correlation function order by order using the Ward identities. For the two-point function, we obtain the following momentum space expansion:
\begin{align}\label{OpOp}
&    \langle \MOh(p,\bar{p})\MO_{h,\bar{h},-q}(-p,-\bar{p}) \rangle= \sum_{n_1,\bar{n}_1, n_2,{\bar n}_2}  [n_1,\bar{n}_1;n_2,{\bar n}_2] 
   \end{align}
where
\begin{equation}
\begin{aligned}
 &[n_1, \bar{n}_1 ; n_2, \bar{n}_2]\\
 &\quad = \frac{\z(-\frac{i\lambda}{2\pi}\y)^{n_1+\bar{n}_1}\z(\frac{i\lambda}{2\pi}\y)^{n_2+\bar{n}_2}\bar{p}^{n_1+n_2}\z(iq+\frac{\lambda k\bar{p}}{4\pi}\y)^{\bar{n}_1+\bar{n}_2} }{n_1! \bar{n}_1! n_{2}! \bar{n}_{2}!} \int d x d \bar{x} \,\ e^{i p x+i \bar{p} \bar{x}} \,\  G_{n_1, \bar{n}_{1} ; n_2, \bar{n}_2}(x, 0) \\
 &G_{n_1, \bar{n}_1 ; n_2, \bar{n}_2}\left(x_1, x_2\right) =\left\langle: \chi_j^{n_1}\bar{\chi}_{\bar{T}}^{\bar{n}_1} \tilde{\mathcal{O}}_{h, \bar{h}}\left(x_1\right)::\chi_j^{n_2}\bar{\chi}_{\bar{T}}^{\bar{n}_2}\tilde{\mathcal{O}}_{h, \bar{h}}\left(x_2\right):\right\rangle \\
&\chi_j(x)=\int^xdx'j(x'),\quad  \bar{\chi}_{\bar{T}}(\bar{x})=\int^{\bar{x}}d\bar{x}'\bar{T}(\bar{x}')
\end{aligned}
\end{equation}
Here, we adopt a point splitting regularization 
\begin{equation}\label{pointsplitting}
    :\chi^n(x)\tilde{\MO}(x):=\oint_{x+\epsilon}\frac{dw}{2\pi i} \frac{1}{w-x}\chi^n(w)\tilde{\MO}(x)-(\log(\epsilon)\text{-divergence})
\end{equation}
Our procedure is as follows: we first perform the perturbative calculation up to second order. We then demonstrate that the leading UV contributions can be resummed into a closed form, which takes the same form as the result derived from string theory in \cite{ACSSW}.
 
\subsection{Correlation functions in momentum space}
The zeroth-order two-point function is identical to the undeformed CFT correlator, expressed in position space and momentum space as:
\begin{align}
G(x,\bar x)_{(0)}=\frac{1}{|x|^{4h}}, \quad  G(p,\bar p)_{(0)}= \frac{\pi\Gamma(1-2h)}{\Gamma(2h)}\z(\frac{|p|}{2}\y)^{4h-2}
\end{align}
where we have set $\bar{h}=h$ for convenience. Applying the Ward identity,
\begin{equation}
\begin{aligned}
    &\langle \bar{T}\left(\bar{w}_1\right) \prod_{i=1}^N \tilde{\mathcal{O}}_i \left(x_i\right)\rangle=\left\{\sum_{i=1}^N\left[\frac{h_i}{\left(\bar{w}_1-\bar{x}_i\right)^2}+ \frac{1}{\bar{w}_1-\bar{x}_i}\partial_{\bar{x}_i}\right]\right\}\langle\prod_{i=1}^N \tilde{\mathcal{O}}_i\left(x_i\right)\rangle\\
&\langle j\left(w_1\right) \prod_{i=1}^N \tilde{\mathcal{O}}_i \left(x_i\right)\rangle=\sum_{i=1}^N \frac{ q_{i} } {w_1-x_i} \langle\prod_{j=1}^N \tilde{\mathcal{O}}_j\left(x_j\right)\rangle
\end{aligned}
\end{equation}
and the correlation functions between chiral currents:
\begin{equation}
    \langle j\left(w_1\right) j\z(w_2\y)\rangle=\frac{k}{(w_1-w_2)^2}
\end{equation}
we get
\begin{equation}
    \begin{aligned}
        &\langle  \bar{\chi}_{\bar{T}}(w_1)\prod_{i=1}^{N}\tilde\MO_i(x_i)\rangle=\z\{\sum_{i=1}^N\z[-\frac{h_i}{\bar{w}_1-\bar{x}_i}+\log(\bar{w}_1-\bar{x}_i)\p_{\bar{x}_i}\y]\y\}\langle \prod_{i=1}^{N}\tilde\MO_i(x_i)\rangle\\
        &\langle  \chi_{j}(w_1)\prod_{i=1}^{N}\tilde\MO_i(x_i)\rangle=\z\{\sum_{i=1}^Nq_i\log(w_1-x_i)\y\}\langle \prod_{i=1}^{N}\tilde\MO_i(x_i)\rangle\\
        &\langle \chi_j(w_1)\chi_j(w_2)\rangle=k\log(w_1-w_2)
        \end{aligned}\label{ope}
\end{equation}
The above formulas will serve as the building block for our perturbative calculation. 
As the physical operator \eqref{Ophys-m-E} has been written purely in terms of currents and primary operators in a CFT, we can discuss the left-moving and right-moving properties separately.
From the second and third lines of \eqref{ope}, we can see that the left-moving contribution can be computed exactly, yields a factor of the form in the two-point function:
\begin{equation}
    x^{\frac{i\lambda \bar{p}q}{\pi}+\frac{\lambda^2k}{4\pi^2}\bar{p}^2}
\end{equation}
For the right-moving part, at $n$-th order it will contribute a factor 
\begin{equation}
    \frac{1}{n!}\z(\frac{i\lambda q\bar{p}}{\pi}+\frac{\lambda^2 k\bar{p}^2}{4\pi^2}\y)^n(\log^n\bar{x}+a_{n,n-1}\log^{n-1}\bar{x}+...)
\end{equation}
As a final result, in momentum space, we obtain a compact result:
\begin{equation}
   \begin{aligned}
   &G(p,\bar{p})=\sum_{n=0}^\infty G(p,\bar{p})_{(n)}\\
       & G(p,\bar{p})_{(n)}=\frac{1}{n!}\int dx^2 e^{ipx+i\bar{p}\bar{x}}\z(\frac{i\lambda q\bar{p}}{\pi}+\frac{\lambda^2k\bar{p}^2}{4\pi^2}\y)^nA_n(\log \bar{x})x^{\frac{i\lambda q\bar{p}}{\pi}+\frac{\lambda^2k\bar{p}^2}{4\pi^2}}G(x,\bar{x})_{(0)}
   \end{aligned}
\end{equation}
where $G(x,\bar{x})_{(0)} = |x|^{-4h}$ is the leading-order two-point function, and $A_n(\theta)$ is an $n$-th order polynomial in $\theta$ with a unit leading coefficient:
\begin{equation}
    A_n(\theta)=\theta^n+a_{n,n-1}\theta^{n-1}+...+a_{n,0},\quad \theta=\log\bar{x}
\end{equation}
We explicitly list the first two polynomials:
\begin{equation}
    \begin{aligned}
        A_1(\theta)&=\theta-\frac{h-1}{2h},\\
    A_2(\theta)&=\theta^2+\frac{6h^2+11h+4}{2h(2h+1)}\theta-\frac{c}{48h(2h+1)}-\frac{4h^4+4h^3-3h^2-21h-12}{4h^2(2h+1)^2}
    \end{aligned}
\end{equation}
Further details are provided in Appendix \ref{app:2ndOrder}. Furthermore, we can perform Fourier transformation with respect to $p$, yielding:
\begin{equation}\label{pleading}
\begin{aligned}
    &
G(x, \bar{p})=\sum^{\infty}_{n=0}G(x,\bar{p})_{(n)}\\
&G(x,\bar{p})_{(n)}=\frac{1}{n!}\int d\bar{x}e^{i\bar{p}\bar{x}}\z(\frac{i\lambda q\bar{p}}{\pi}+\frac{\lambda^2 k\bar{p}^2}{4\pi^2}\y)^nA_n(\log \bar{x})x^{\frac{i\lambda q\bar{p}}{\pi}+\frac{\lambda^2k\bar{p}^2}{4\pi^2}}G(x,\bar{x})_{(0)}
\end{aligned}
\end{equation}
The result demonstrates an exact shift of the left-moving conformal weight in the $x$-direction by $-\frac{i\lambda q\bar{p}}{2\pi}-\frac{\lambda^2k\bar{p}^2}{8\pi^2}$.

At each order of $G(p,\bar{p})$, the dominant UV contribution comes from the leading log term in $A_n(\log x)$. Summing these contributions gives a formal expression in momentum space:
\begin{equation}\label{Gn}
    \begin{aligned}
        \langle \MOh(p,\bar{p})\MO_{h,\bar{h},-q}(-p,-\bar{p})\rangle&\sim\int dx^2 e^{ipx+i\bar{p}\bar{x}}|x|^{\frac{2i\lambda q\bar{p}}{\pi}+\frac{\lambda^2 k\bar{p}^2}{2\pi^2}}\frac{1}{|x|^{4h}}\\
        &=\frac{\pi \Gamma(1-2h_\lambda)}{\Gamma(2h_\lambda)}\z(\frac{|p|}{2}\y)^{4h_\lambda-2},\quad h_\lambda=h-\frac{i\lambda q\bar{p}}{2\pi}-\frac{\lambda^2k\bar{p}^2}{8\pi^2}
    \end{aligned}
\end{equation}
This produces a function analogous to a CFT two-point function with shifted conformal weight $h_{\lambda}$. We can take a close look at \eqref{Gn} in the Lorentzian signature. After the analytic continuation to time direction introduced in \cite{Bautista:2019qxj} to \eqref{Gn}, we get the Wightman 2-point function in momentum space
\begin{equation}
    \begin{aligned}
        &\langle \MO_{h,\bar{h},q}(p_L,p_R) \MO_{h,\bar{h},-q}(-p_L,-p_R)\rangle\sim \frac{\pi^2\theta(E-|P|)}{2\Gamma(2h_\lambda)^2}\z(\frac{|p|}{2}\y)^{4h_\lambda-2}
    \end{aligned}
\end{equation}
where $E \equiv p_L + p_R$ and $P \equiv p_L - p_R$ represent the energy and momentum of the operator. The resulting shift in the conformal weight is given by:
\begin{equation}
    h_\lambda =h +\frac{\lambda qp_R}{2\pi}+\frac{\lambda^2 kp_R^2}{8\pi^2}
\end{equation}

Finally, we can perform a similar calculation to the $N$-point correlation functions, and we can get the following result,
\begin{equation}\label{npf}
    \langle\prod^{N}_{i=1}\MO_{h_i,\bar{h}_i,q_i}(p_i,\bar{p}_i)\rangle\sim\mathcal{F}_{p_i,\bar{p}_i}\z[\prod_{j=1}^N\prod_{k\neq j,k=1}^N|x_j-x_k|^{-\frac{i\lambda q_j\bar{p}_k}{\pi}-\frac{\lambda^2k\bar{p}_j\bar{p}_k}{4\pi^2}}\langle\prod^N_{i=1}\tilde{\MO}_{h_i,\bar{h}_i,q_i}(x_i,\bar{x}_i)\rangle\y]
\end{equation}

The leading log result \eqref{Gn} takes a similar structure of a momentum-space two-point function in CFTs with a shifted conformal weight $h_\lambda=\bar h_\lambda$. This agrees with the result obtained from holographic calculation in the TsT/$J\bar T$ correspondence \cite{ACSSW}. The shifted conformal weight also agrees with that obtained in \cite{Guica:2021fkv} in the left-moving sector, but differs in the right-moving sector. This is mainly due to the fact that we are considering different operators from those of \cite{Guica:2021fkv}. We provide a more detailed comparison in appendix \ref{sec:compare}.

\subsection{Correlation functions in position space}

To get the correlation function in position space, we can expand $\z(\frac{i\lambda q\bar{p}}{\pi}+\frac{\lambda^2k\bar{p}^2}{4\pi^2}\y)$ in \eqref{Gn} and replace each $\bar{p}$ by the derivative acting on $e^{i\bar{p}x}$. Then we perform  integration by parts. The $1/\bar{x}$ resulting from acting $\bp$ on $A_n$ will not contribute to the leading log term. As a result we have:
\begin{equation}\label{Position}
    G(x,\bar{x})\sim\int d\theta \delta(\theta-\log |x|^2)e^{-\theta\z(\frac{\lambda q}{\pi}\p_{\bar{x}}+\frac{\lambda^2k}{4\pi^2}\p_{\bar{x}}^2\y) }G(x,\bar{x})_{(0)}
\end{equation}
Here we have introduced an auxiliary variable $\theta$ to avoid the terms coming from the commutator between $\log|x|^2$ and $\bp$. 
The integral over $\theta$ is taken as the last step of the computation. This is reminiscent of the star product in non-commutative field theory \cite{Seiberg:1999vs}.

\subsection*{Borel resummation of the correlation function}
We now perform a Borel resummation of the correlation function to analyze its non-perturbative effects (see \cite{Marino:2012zq} for a review). A similar analysis was carried out for $T\bar{T}$-deformed CFTs in \cite{Hirano:2025alr}. We initiate the procedure by expanding the expression \eqref{Position}, thereby generating a formal power series,
\begin{equation} \label{twoSumSeries}
    \begin{aligned}
        G(x, \bar{x}) &=\frac{1}{|x|^{4h}} \sum_{r=0}^{\infty} \sum_{m=0}^{\infty} \frac{ 1}{r! m!}  \z(-\frac{\lambda q\log(|x|^2)}{\pi}\y)^r\z(-\frac{\lambda^2k\log(|x|^2)}{4\pi^2}\y)^m\bp^{r+2m}G(x,\bar{x})_{(0)} \\
        &= \frac{1}{|x|^{4h}} \sum_{r=0}^{\infty} \sum_{m=0}^{\infty} \frac{(2h)_{2m+r}}{r!m!} Z^{r} W^{m}
    \end{aligned}
\end{equation}
where $(2h)_{2m+r}\equiv \Gamma(2h+2m+r)/\Gamma(2h)$ is the Pochhammer symbol and $Z, W$ are defined as
\begin{equation}
    Z  = \frac{  \lambda q\log (|x|^2 )}{\pi \bar{x}}, \quad W  = -\frac{ \lambda^{2} k\log (|x|^2 )}{4 \pi^{2} \bar{x}^{2}} 
\end{equation}
Splitting the Pochhammer symbol as $(2h)_{2m+r}=(2h)_{2m}(2h+2m)_r$, we can sum over the power series in $Z$ and obtain
\begin{equation}
 \begin{aligned}\label{2pfsum}
        G(x,\bar{x})
     &=\frac{1}{\z((1-Z)x\bar{x}\y)^{2h}}Y(K),\qquad Y(K)\equiv \sum_{m=0}^{\infty}\frac{(2h)_{2m}}{m!}K^m
 \end{aligned}
\end{equation}
where $Y(K)$ is a series in the new variable $K$ defined as
\begin{equation}\label{YK}
    \begin{aligned}
        &K\equiv \frac{W}{(1-Z)^2}=-\frac{\lambda^2k}{4\pi^2}\frac{\log(|x|^2 )}{(\bar{x}-\frac{\lambda q}{\pi}\log(|x|^2 ))^2}
    \end{aligned}
\end{equation}
The series $Y(K)$ is divergent as the coefficient grows factorially. Similar to $T\bar{T}$ case \cite{Hirano:2025alr}, a Borel resummation can be carried out. To do so, we first perform the Borel transformation to $Y(K)$ by dividing the original expansion coefficient by a factorial factor,
\begin{equation} \label{BorelTrans}
    \mathcal{B}[Y](\zeta)=\sum_{m=0}^\infty\frac{(2h)_m}{m!m!}\zeta^m={}_2F_1\z(h,h+\frac{1}{2},1,4\zeta\y)
\end{equation}
and then obtain the result of the original series formally by performing the following integral
\begin{align}\label{Brm}
    \mathscr{S}Y(K) 
    =\frac{1}{K}\int^{e^{i\arg K}\infty}_0 \mathcal{B}[Y](\zeta)e^{-\frac{\zeta}{K}}d\zeta
\end{align}
where the integration path is taken along a ray with the same argument as $K$.
The hypergeometric function possesses a branch point at $\zeta = \frac{1}{4}$, indicating the presence of non-perturbative effects. We now evaluate the integral \eqref{Brm}, distinguishing between the case where the integration path passes through the branch point and the case where it does not.

\paragraph{Case I: } $K \notin \mathbb{R}_+$.
In this case, the branch point $\zeta={1\over 4}$ does not lie on the integration path, and the series is directly Borel resummable, yielding
\begin{equation}\label{noninstanton}
    G(x,\bar{x}) = \frac{1}{\left((1-Z)x\bar{x}\right)^{2h}} \mathscr{S}Y(K) = \frac{U\left(h,\frac{1}{2},-\frac{1}{4K}\right)}{\left(\frac{\lambda^2 k}{\pi^2}x^2\log(|x|^2)\right)^{h}}, \quad \arg(-K) \in (-\pi,\pi).
\end{equation}
As a sanity check, we can expand the above result as a power series of $K$. This corresponds to performing an asymptotic expansion of the hypergeometric $U$ function, which indeed reproduces the original series \eqref{2pfsum}, without introducing any non-perturbative effect.

\paragraph{Case II: }$K \in \mathbb{R}_+$.
In this case, the integration path goes through the aforementioned branch point at $\zeta={1\over4}$, which leads to a Stokes discontinuity in the integral \eqref{Brm}. Let $\mathscr{S}^{(\pm)} Y(K) $ denote the integral \eqref{Brm} performed above and below the branch cut respectively. The discontinuity is then given by  
\begin{align}\label{DiscY}
   \text{Disc}  Y(K) &=\mathscr{S}^{(+)}Y(K)-\mathscr{S}^{(-)}Y(K)\\
  & =\frac{i\pi 2^{-2h}}{\Gamma(h)\Gamma(h+\frac{1}{2})}e^{-\frac{1}{4K}}K^{-\frac{1}{2}-h}U\z(1-h,\frac{3}{2},\frac{1}{4K}\y)
\end{align}
where we have used identities of the hypergeometric function \eqref{Hyper} to single out the contribution of the multi-value property. The correlation function is then proportional to the principal value of the integral
\begin{equation}\label{instanton}
    \begin{aligned}
        G(x,\bar{x})&=\frac{1}{((1-Z)x\bar{x})^{2h}}\z(\mathscr{S}^{(\pm)}Y(K)\mp\frac{1}{2}\text{Disc} Y(K)\y)\\
        &=\z(-\frac{1}{\frac{\lambda^2k}{4\pi^2}x^2\log(|x|^2 )}\y)^{h}\frac{1}{2\Gamma(2h)}\Big[\cos(h\pi)\Gamma(h){}_1F_1\z(h,\frac{1}{2},-\frac{1}{4K}\y)\\
        &+\sin(h\pi)\Gamma(h+\frac{1}{2})K^{-\frac{1}{2}}{}_1F_1\z(h+\frac{1}{2},\frac{3}{2},-\frac{1}{4K}\y)\Big],\quad K\in \mathbb{R}_+
    \end{aligned}
\end{equation}
For small $K$, the correlation function possesses a trans-series expansion:
\begin{equation}
    \begin{aligned}
        G(x,\bar{x})&\sim\frac{1}{((1-Z)x\bar{x})^{2h}}\sum_{n=0}^\infty\frac{(2h)_{2n}}{n!}K^{n}\\
    &-\frac{\sqrt{\pi}}{2^{2h+1}\Gamma(2h)((1-Z)x\bar{x})^{2h}}e^{-\frac{1}{4K}}K^{-2h}\sum_{n=0}^\infty\frac{(1-h)_n(\frac{1}{2}-h)_n}{n!}(-4K)^{n+\frac{1}{2}}
    \end{aligned}
\end{equation}
The first line recovers the perturbative series \eqref{YK}, and the second line comes from the non-perturbative instanton contribution, which is exponentially suppressed as $K \to 0^+$. The two series are comparable as $\Re(K)\sim \mathcal {O}(1)$, and therefore generate an effective UV cutoff at the scale of $\lambda \sqrt{k}$.

Now we perform the analysis of the behavior of the correlation function in both the $q=0$ and $q\neq 0$ sectors.

\begin{figure}[htbp]
    \centering
    \begin{subfigure}[b]{0.45\textwidth}
        \centering
        \includegraphics[width=\textwidth]{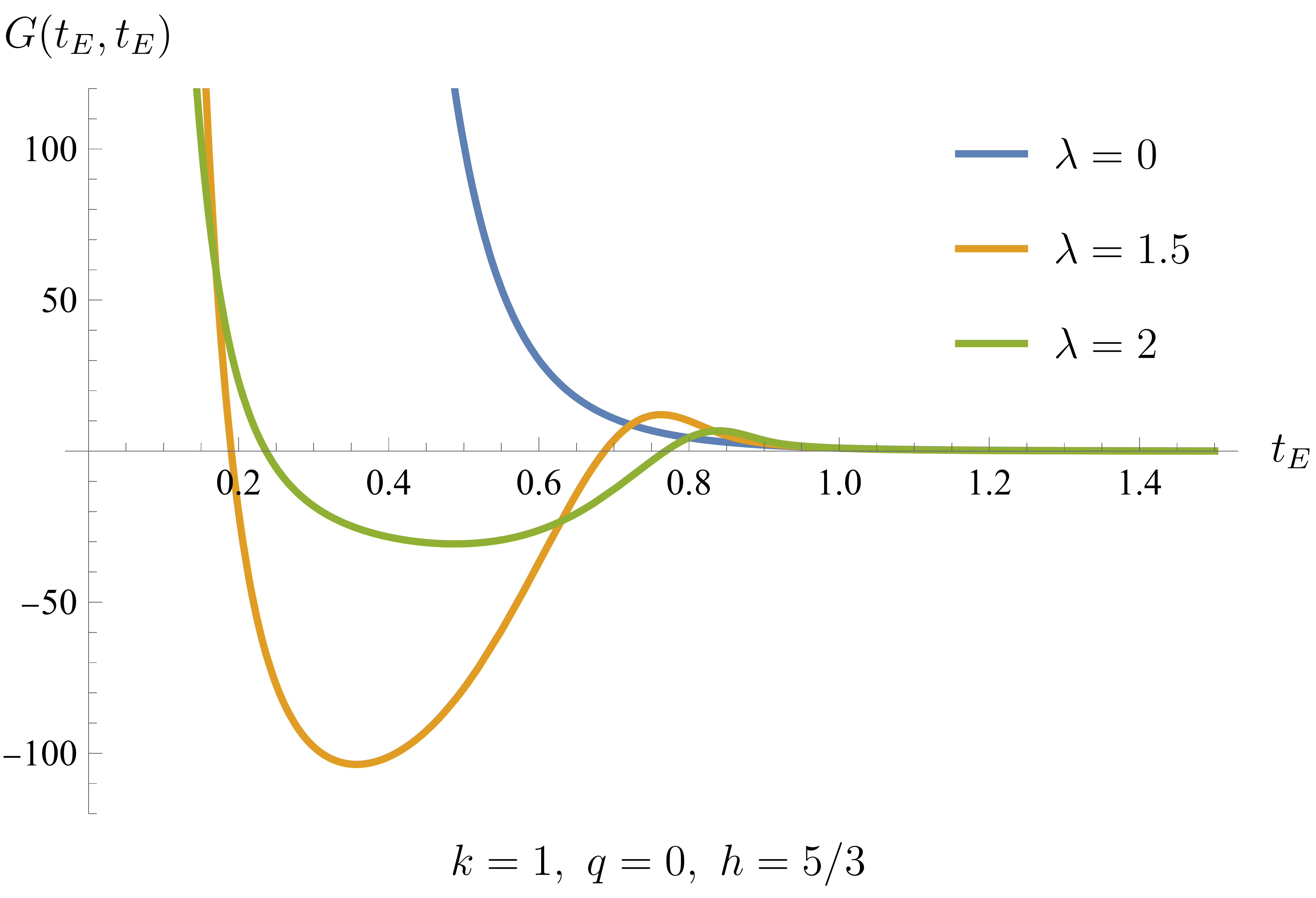}
        \caption{}
        \label{fig:sub1}
    \end{subfigure}
    \hfill
    \begin{subfigure}[b]{0.45\textwidth}
        \centering
        \includegraphics[width=\textwidth]{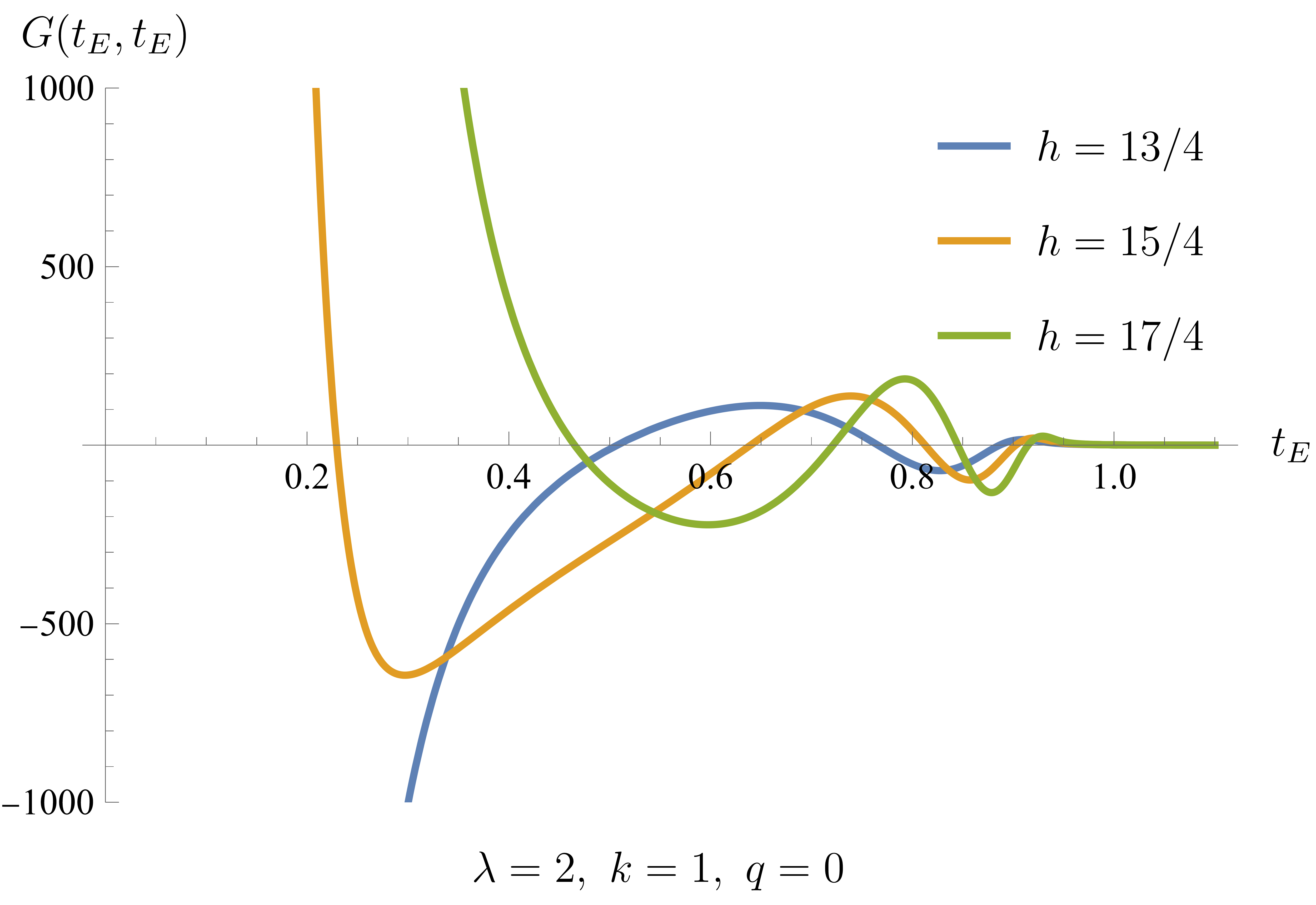}
        \caption{}
        \label{fig:sub2}
    \end{subfigure}
    \caption{The zero charge sector: we consider the case $x = \bar{x} = t_E$ with fixed $k = 1$. The UV behavior of the two-point function is controlled by  \eqref{smallx}. Figure 1(a) shows different values of $\lambda$, and  Figure 1(b) is for different $h$.}
    \label{fig:q=0 t}
\end{figure}

\paragraph{Zero charge sector.}When the chiral charge vanishes, the condition for instanton contributions, $K \in \mathbb{R}_+$, is realized as follows:
\begin{subequations}
    \begin{align}
        &\text{For } |x| > 1,\quad \Re(\bar{x}) = 0 \implies t_E = 0,\quad x = iy,\ \bar{x} = -iy, \\
        &\text{For } |x| < 1,\quad \Im(\bar{x}) = 0 \implies y = 0,\quad x = \bar{x} = t_E.
    \end{align}
\end{subequations}
The IR- and UV-behaviors are summarized below,
\begin{itemize} 
\item In the long-distance regime $|x| \gg 1$, we have $Z \to 0$ and $K \to 0$, so the correlation function reduces to the CFT result:
\begin{equation}\label{farregion}
    G(x,\bar{x}) \sim \frac{1}{|x|^{4h}}, \quad |x| \gg 1.
\end{equation}
\item In the very short-distance limit $|x| \ll 1$, the asymptotic behavior is given by
\begin{equation}\label{smallx}
    G(x,\bar{x})\sim \frac{\sqrt{\pi}C(h)}{\Gamma(h+\frac{1}{2})x^{2h}}\z(\frac{1}{\frac{\lambda^2k}{\pi^2}\log(|x|^2)}\y)^h,
\end{equation}
where 
\begin{equation}
    C(h)=\begin{cases}
        e^{-i\pi h}\cos(h\pi)&\text{Instanton sector}\\
        1&\text{Non-instanton sector}
    \end{cases}
\end{equation}
In both cases,  the very-short distance behavior looks like replacing $\bar x$ in the CFT result \eqref{farregion} by $\sqrt{\log|x|^2}$, while keeping the power-law dependence in $x$. 
The $x^{-2h}$ behavior is as expected because the left-moving conformal invariance is still kept. On the other hand, the replacement ${\bar x}^{-2h}\to \z({\log|x|^2}\y)^{-h}$  modifies the UV behavior drastically, from being power-law divergent to vanishing. 
A similar replacement structure has been observed in $T\bar{T}$-deformed CFTs~\cite{Hirano:2025alr}, which happens in both the left and right moving sectors so that the entire two-point function vanishes in the deep UV. 
In the $J\bar T$ deformation, however, the two-point function is still divergent due to the preserved conformal symmetry in the left-moving sector. This is more evident in the polar coordinate, $|G(r,\theta)|\to (r^2\log r)^{-h}$ in the polar coordinate. We learn that the  UV behaviour in the $J\bar{T}$ case is milder than a CFT, but more divergent than the $T\bar{T}$ case.  
\end{itemize}
For illustration, we present numerical evaluations of the two-point function. For simplicity, we consider the case of pure time-direction separation with $ y = 0 $. In the region $ t_E \geq 1 $, the behavior is governed by \eqref{noninstanton}, which exhibits monotonic decay. Instanton effects occur for $ t_E < 1 $, where the two-point function is given by \eqref{instanton}. Thus, $ G(t_E, t_E) $ is described by the piecewise function
\begin{equation}
        G(t_E,t_E) =
    \begin{cases}
        &\z(-\frac{\lambda^2k}{4\pi^2}t_E^2\log(t_E^2 )\y)^{-h}\frac{1}{2\Gamma(2h)}\Big[\cos(h\pi)\Gamma(h){}_1F_1\z(h,\frac{1}{2},\frac{\pi^2}{\lambda^2k}\frac{t_E^2}{\log(t_E^2)}\y)\\
        &+\sin(h\pi)\Gamma(h+\frac{1}{2})\z(-\frac{\lambda^2k}{4\pi^2}\frac{\log(t_E^2)}{t_E^2}\y)^{-\frac{1}{2}}{}_1F_1\z(h+\frac{1}{2},\frac{3}{2},\frac{\pi^2}{\lambda^2k}\frac{t_E^2}{\log(t_E^2)}\y)\Big],\ t_E<1\\
        &\\
        &{\left(\frac{\lambda^2 k}{\pi^2}t_E^2\log(t_E^2)\right)^{-h}}U\left(h,\frac{1}{2},\frac{\pi^2}{\lambda^2k}\frac{t_E^2}{\log(t_E^2)}\right), \quad t_E>1
    \end{cases}
\end{equation}
As shown in Figure \ref{fig:q=0 t}, the two-point function exhibits oscillatory behavior for $ t_E < 1 $. These oscillations are not periodic in nature. They arise from the interplay between the prefactor and the extremal structures of the confluent hypergeometric functions ${}_1F_1\left(h; \frac{1}{2}; z\right)$ and ${}_1F_1\left(h+\frac{1}{2}; \frac{3}{2}; z\right)$ as $ z = \frac{\pi^2}{\lambda^2 k} \frac{t_E^2}{\log(t_E^2)} $ varies over the negative real axis. Within the interval $ t_E \in (0,1) $, the argument $ z $ is monotonic functions of $ t_E $. The observed oscillations arise from the interplay of  the ${}_1F_1$ functions and the prefactors.
This results in a non-monotonic profile with a finite number of extrema depending on the parameters $ h $ and $ \lambda $. As $ \lambda $ increases, the position of the oscillation moves toward $ t_E = 1 $, while its amplitude is suppressed. Larger values of $ h $ amplify the magnitude of these oscillatory behavior, and the UV behavior acquires alternating signs due to the $ \cos(h\pi) $ factor.

\begin{figure}[htbp]
    \centering    
    \begin{subfigure}[b]{0.45\textwidth}
        \centering
        \includegraphics[width=\textwidth]{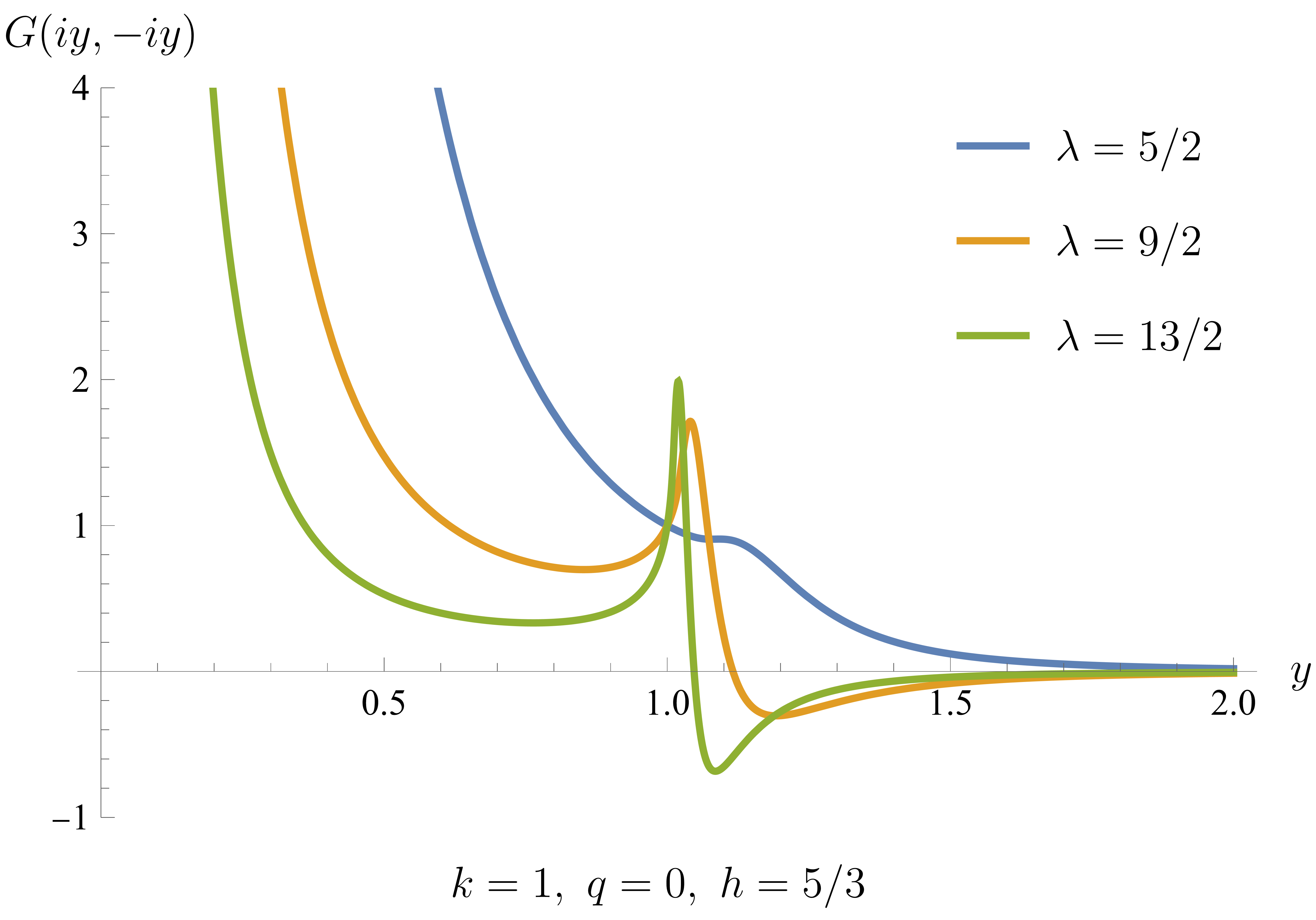}
        \caption{}
        \label{fig:sub3}
    \end{subfigure}
    \hfill
    \begin{subfigure}[b]{0.45\textwidth}
        \centering
        \includegraphics[width=\textwidth]{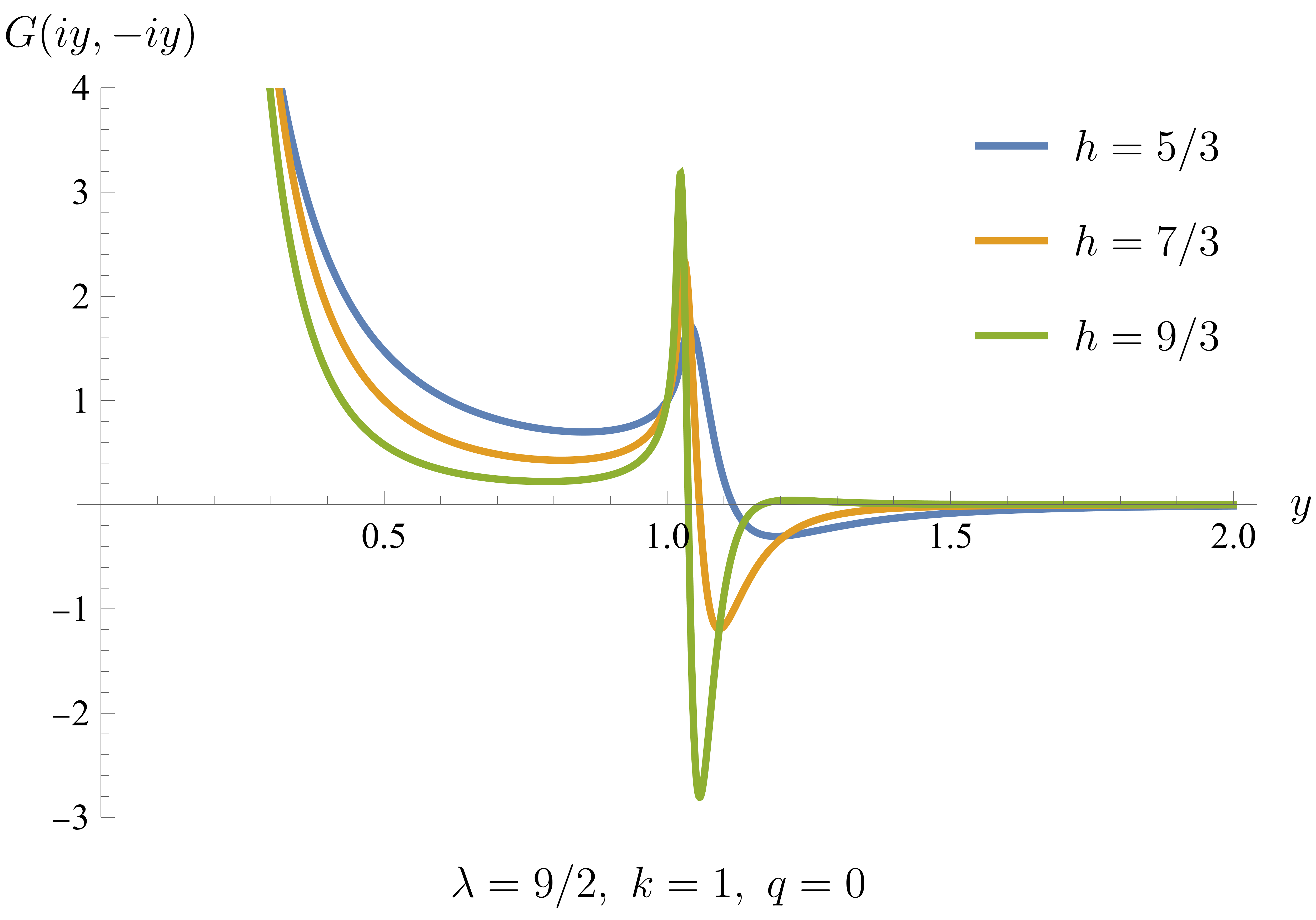}
        \caption{}
        \label{fig:sub4}
    \end{subfigure}
    
    \caption{The zero charge sector: we consider the case $x = -\bar{x} =iy$ with fixed $k = 1$. The UV behavior is controlled by \eqref{smallx}. Again, the behavior of the two-point function is shown for different values of $\lambda$ and $h$. }
    \label{fig:q=0 y}
\end{figure}

We also consider the two-point function with pure spatial separation, setting $ t_E = 0 $. In this case, the region $ y < 1 $ is governed by \eqref{noninstanton}, while instanton effects occur for $ y > 1 $, where the correlator is given by \eqref{instanton}. The full dependence of $ G(iy, -iy) $ is thus described by the piecewise function
\begin{equation}
        G(iy,-iy) =
    \begin{cases}
    &{\left(-\frac{\lambda^2 k}{\pi^2}y^2\log(y^2)\right)^{-h}}U\left(h,\frac{1}{2},-\frac{\pi^2}{\lambda^2k}\frac{y^2}{\log(y^2)}\right), \quad y<1\\
    &\\
        &\z(\frac{\lambda^2k}{4\pi^2}y^2\log(y^2 )\y)^{-h}\frac{1}{2\Gamma(2h)}\Big[\cos(h\pi)\Gamma(h){}_1F_1\z(h,\frac{1}{2},-\frac{\pi^2}{\lambda^2k}\frac{y^2}{\log(y^2)}\y)\\
        &+\sin(h\pi)\Gamma(h+\frac{1}{2})\z(\frac{\lambda^2k}{4\pi^2}\frac{\log(y^2)}{y^2}\y)^{-\frac{1}{2}}{}_1F_1\z(h+\frac{1}{2},\frac{3}{2},-\frac{\pi^2}{\lambda^2k}\frac{y^2}{\log(y^2)}\y)\Big],\ y>1
    \end{cases}
\end{equation}
As shown in Figure~\ref{fig:q=0 y}, oscillatory features emerge near $ y = 1 $. The magnitudes of these oscillations increase monotonically with both $ \lambda $ and $ h $. Notably, the argument of the confluent hypergeometric functions, $\frac{\pi^2}{\lambda^2 k} \frac{y^2}{\log(y^2)} $, is no longer monotonic for $ y > 1 $. The function ${}_1F_1$ is also non-monotonic, leading to a non-monotonic variation of the ${}_1F_1$ functions across this regime.

\paragraph{Non-zero charge sector.}Now we analyze the behavior of the two-point function in the presence of non-zero chiral charge. In this case, instanton contributions arise when the following conditions are satisfied: \begin{subequations}
\begin{align}
       &\text{For } |x| > 1,\quad \Im(\bar{x}) \neq 0,\quad \Re(\bar{x}) = \frac{\lambda q}{\pi} \log(|x|^2) \implies y^2 = e^{\frac{\pi}{\lambda q} t_E} - t_E^2 \label{yinstanton}, \\
        &\text{For } |x| < 1,\quad \Im(\bar{x}) = 0 \implies y = 0,\quad x = \bar{x} = t_E.
\end{align}
    \end{subequations}
The \eqref{yinstanton} defines a curve $y(t_E)$ along which the instanton solution is realized. Given that $|x| > 1$, the condition  
\begin{equation}
    \mathrm{sign}(\lambda q) \cdot t_E > 0
\end{equation}  
must be satisfied. However, even when this holds, for $|\lambda q| > \frac{e\pi}{2}$, the right-hand side of \eqref{yinstanton} can become negative for certain ranges of $t_E$,  causing the curve to split into two disjoint curves.

\begin{figure}[htbp]
    \centering    
    \begin{subfigure}[b]{0.45\textwidth}
        \centering
        \includegraphics[width=\textwidth]{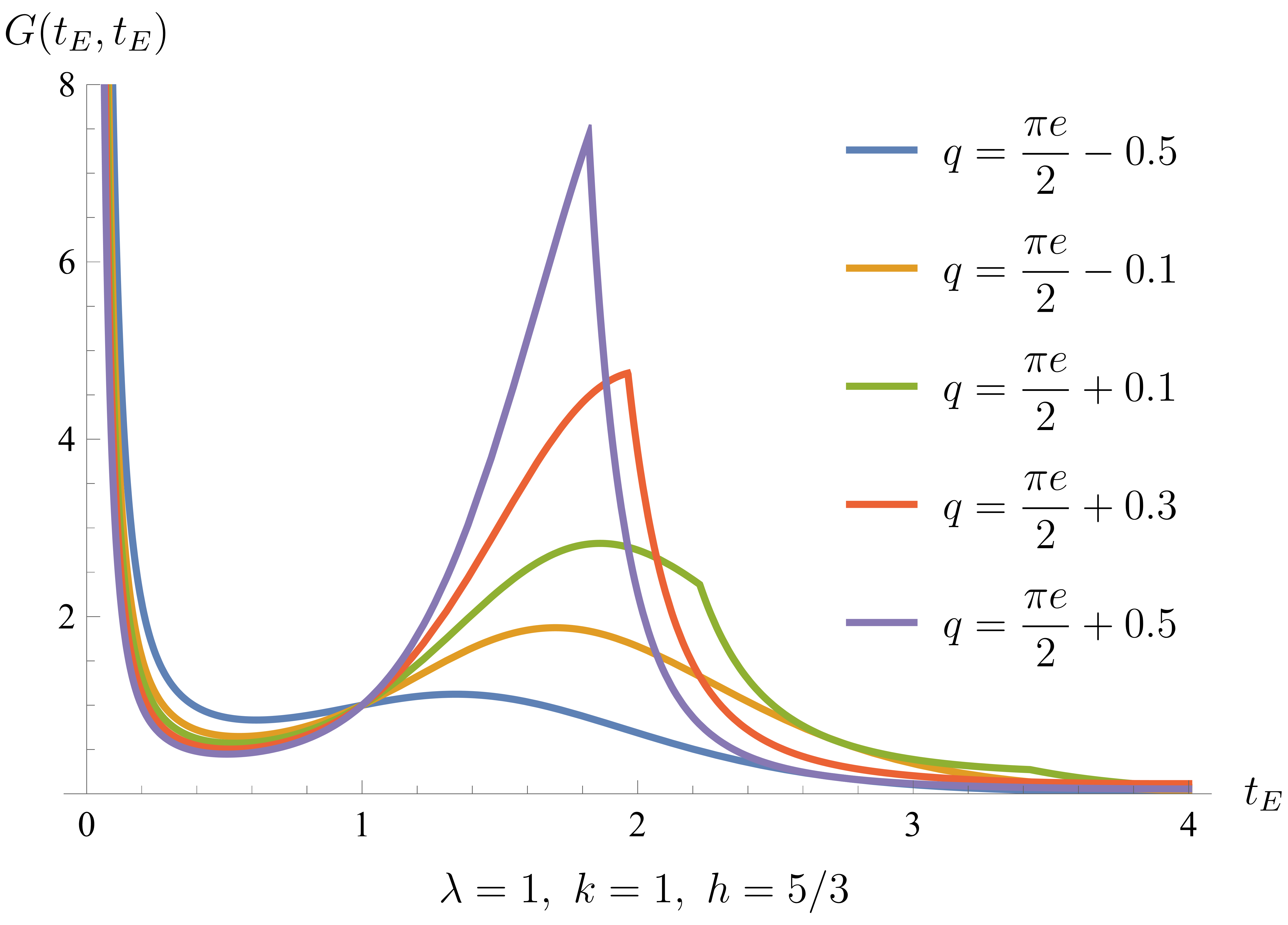}
        \caption{}
        \label{fig:sub5}
    \end{subfigure}
    \hfill
    \begin{subfigure}[b]{0.45\textwidth}
        \centering
        \includegraphics[width=\textwidth]{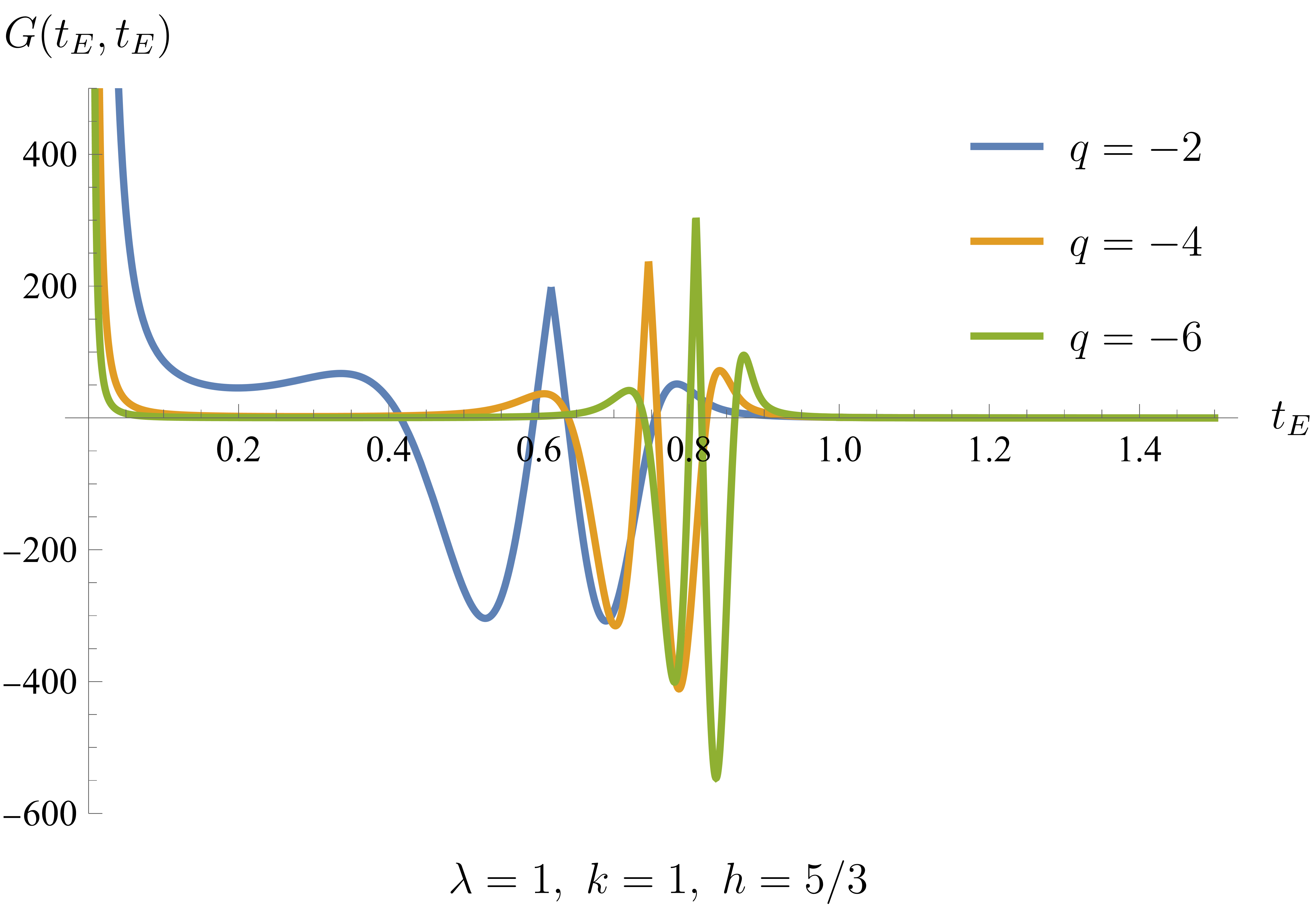}
        \caption{}
        \label{fig:sub6}
    \end{subfigure}
    \caption{Non-zero charge sector: the two-point function for different chiral charges $q$, with $x = \bar{x} = t_E$ and fixed parameters $\lambda = k = 1$, $h = 5/3$.}
    \label{qneq0}
\end{figure}

In addition to the instanton effects, we also observe that a cusp appears in the two-point function when $Z = 1$, corresponding to
\begin{equation}\label{cusp}
    \bar{x} = \frac{\lambda q}{\pi} \log |x|^2.
\end{equation}  
The 2-point function is continuous at this point, but its derivative exhibits a discontinuity. Since the right-hand side of \eqref{cusp} is real, a solution exists only when $y = 0$. For $|\lambda q| > \frac{e\pi}{2}$, one solution lies in the region $|t_E| > 1$, while another solution always exists within $|t_E| < 1$. At the cusp, the correlation function takes a finite value in both the instanton and non-instanton sectors. The IR- and UV-behaviors are summarized below:
\begin{itemize}
    \item In the long-distance region $|x| \gg 1$, the parameter $K$ becomes small, the correlation function reduces to the CFT result,  
\begin{equation}
    G(x,\bar{x}) \sim \frac{1}{|x|^{4h}}, \quad |x| \gg 1.
\end{equation}  
\item In the very short-distance region $|x| \ll 1$, $Z$ becomes large while $K$ remains small. In this limit, the asymptotic behavior is given by  
\begin{equation}
    G(x,\bar{x}) \sim \frac{1}{x^{2h} \left( \bar{x} - \frac{\lambda q}{\pi} \log |x|^2 \right)^{2h}}, \quad |x| \ll 1.
\end{equation}  
Thus, in the short-distance regime, the $\bar{x}$-dependence of the correlation function is effectively shifted to $\bar{x} - \frac{\lambda q}{\pi} \log |x|^2$.
\end{itemize}
To demonstrate the behavior of the two-point function for different chiral charges, we present numerical computations in Figure~\ref{qneq0}. We set $x = \bar{x}=t_E$, in which case a cusp appears. The two-point function is invariant under the transformation $(\bar{x}, \lambda q) \to (-\bar{x}, -\lambda q)$, so it suffices to consider the region $t_E > 0$ with both positive and negative values of $\lambda q$. The region $t_E\ge 1$ is described by \eqref{noninstanton}, and the region $t_E<1$ is given by \eqref{instanton}. The two-point function $G(t_E,t_E)$ is described by a piecewise function,
\begin{equation}
\begin{aligned}\label{G2tq}
    &        G(t_E,t_E)=\\
    &
    \begin{cases}
        &\z(-\frac{\lambda^2k}{4\pi^2}t_E^2\log(t_E^2 )\y)^{-h}\frac{1}{2\Gamma(2h)}\Big[\cos(h\pi)\Gamma(h){}_1F_1\z(h,\frac{1}{2},\frac{\pi^2}{\lambda^2k}\frac{(t_E-\frac{\lambda q}{\pi}\log(t_E^2))^2}{\log(t_E^2)}\y)\\
        &\!\!\!\!\!\!+\sin(h\pi)\Gamma(h+\frac{1}{2})\z(\!-\frac{\lambda^2k}{4\pi^2}\frac{\log(t_E^2)}{(t_E-\frac{\lambda q}{\pi}\log(t_E^2))^2}\!\y)^{-\frac{1}{2}}\!\!{}_1F_1\z(h+\frac{1}{2},\frac{3}{2},\frac{\pi^2}{\lambda^2k}\frac{(t_E-\frac{\lambda q}{\pi}\log(t_E^2))^2}{\log(t_E^2)}\y)\Big],\ t_E<1\\
        &\\
        &{\left(\frac{\lambda^2 k}{\pi^2}t_E^2\log(t_E^2)\right)^{-h}}U\left(h,\frac{1}{2},\frac{\pi^2}{\lambda^2k}\frac{(t_E-\frac{\lambda q}{\pi}\log(t_E^2))^2}{\log(t_E^2)}\right), \quad t_E>1
    \end{cases}
\end{aligned}
\end{equation}
Figure \ref{fig:sub5} describes the 2-point function with different positive charges. When $q$ is small, there is an oscillation in the $t_E\sim 1$ region. As $q$ increases through the critical value $\frac{e\pi}{2\lambda}$, the oscillation transforms into a cusp in the region $t_E>1$. The magnitude increases monotonically with $q$. When $q$ is very large, the oscillation disappears and the magnitude of the cusp becomes very large. In contrast, Figure \ref{fig:sub6} describes the 2-point function with different negative chiral charges. There will always be a cusp surrounded by oscillations. The magnitude increases with the charge $|q|$.

\section{Summary}\label{sec:sum}
In this paper, we study symmetries, operators, and correlation functions of $J\bar T$-deformed CFTs on the plane. The deformation is generated by a quasi-chiral current $\MJ_{+\alpha}$ and the right-moving stress tensor $T_{v\alpha}$. The $J\bar T $ operator can also be equivalently written in terms of current $\tilde{\mathcal{J}}_{+\alpha}$, which remains chiral even after the deformation \eqref{dH}, the charge of associated with which flows with the deformation parameter $\lambda$ \eqref{leftWard}. 

In the Hamiltonian formalism,  
we construct the symmetry generators by solving the flow equation \eqref{flowedeq}, the resulting symmetry algebra is two copies of the Virasoro-Kac-Moody algebra \eqref{Vir-KM}. In the left-moving sector, the deformed stress tensor $\mathcal H_L$ and the chiral $U(1)$ current $\tilde{\mathcal{J}}_{+}$  are already solutions to the flow equation. Therefore the Virasoro-Kac-Moody acts locally in the left-moving sector. In the right-moving sector, however, the dressed stress tensor and the anti-chiral $U(1)$ current \eqref{dressedcureents} are constructed in terms of the non-local right-moving coordinate $\hat v.$ 
By using these dressed symmetries, we have explicitly constructed the physical operator \eqref{physicalOp} by generalizing the approach we established for $T\bar{T}$ deformations in \cite{Chen:2025jzb}. Then, we studied the behavior of this physical operator under the covariant derivative $\mathcal{D}_{\lambda}$ and found that it satisfies a nice recursive relation. We therefore obtained an explicit solution \eqref{Flowedoperator} for the dressed operator. 

We then proceeded to compute the correlation function of the physical operators. To do so, we lifted our operator to the quantum level \eqref{Quantumposion}, and converted the computation as a conformal perturbation in terms of the deformation parameter $\lambda$. 
The leading-log contributions can be singled out in both momentum space and position space. The result in momentum space is given by \eqref{Gn} for the two-point function and \eqref{npf} for the $N$-point function, which exhibits a CFT structure but shifted conformal weights. 

When turning to position space, the series \eqref{twoSumSeries} is asymptotic with zero radius of convergence, and we performed a Borel transform on the leading logarithmic series. In the generic case, the Borel transform takes the form of a hypergeometric function \eqref{BorelTrans} and exhibits no singularities along the integration path. Consequently, the corresponding Laplace integral can be evaluated directly, yielding a result \eqref{noninstanton} expressed in terms of the confluent hypergeometric function of the second kind, $ U $. In contrast, when $K\in \mathbb{R}_+$, a branch point lies on the integration path. In this case, standard Borel resummation failed, due to the singularity obstructing the integration path. To address this, we employed lateral Borel resummation, computing the Borel resummation just above and below the singularity. The difference between these two resummations gave rise to a Stokes discontinuity \eqref{DiscY}, which is governed by an instanton-like contribution that is exponentially suppressed and determined by the location of the branch point. To eliminate the ambiguity associated with the lateral sums, we adopted the symmetric average of the two lateral resummations. This prescription corresponded to taking the Cauchy principal value \eqref{instanton} of the Borel integral, ensuring a real-valued and unambiguous result.
As the Borel resummation is performed in terms of the variable $K$, a complicated quantity depending both on the spacetime separation, and other parameters including the deformation parameter $\lambda$,  the charge, and the conformal weights, we still need to extract the information about how correlations change with positions.  We detailed the analysis with figures in two special cases: when the separation is purely in time and purely in space. The two-point function exhibits a CFT-like decay in the far infrared, a growth that sits between CFT and $T\bar T$ deformation in the deep UV, and an intermediate region with interesting oscillating and cusp structures.

We conclude by outlining several interesting questions for future work. First, the relationship between our momentum-space and position-space results requires deeper investigation. These are not related by a simple Fourier transformation, differing instead by an infinite series of sub-leading logarithmic terms. It is therefore unclear whether they provide a complete physical picture, raising the question of whether analogous non-perturbative effects exist in momentum space and how they might be characterized.
Second, our use of point-splitting regularization, which discards all cutoff dependence, effectively imposes a rigid UV cutoff. However, indications of a dynamical cutoff of order $\lambda\sqrt{k}$ suggest a more subtle interplay between the deformation and the UV regulator that merits a sharper understanding.
Third, extending our Euclidean correlation functions to the Lorentzian signature is a subtle but crucial task. This analytical continuation could unveil richer physics, such as in the lightcone limit—the Lorentzian counterpart of the short-distance Euclidean limit—which requires careful treatment.
Finally, the non-perturbative effect we observe is reminiscent of phenomena in the partition functions of $T\bar{T}$-deformed CFTs \cite{Benjamin:2023nts,Dei:2024sct,Gu:2025tpy}.  It is interesting to further explore the physical meanings of these non-perturbative behaviors.

    \section*{Acknowledgements}
We thank Luis Apolo, Jie Gu, Monica Guica, Yunfeng Jiang, Rubén Monten, Huajia Wang, Juntao Wang and Qiang Wen for useful discussions. 
The work is supported by the NSFC special fund for theoretical physics No. 12447108 and the national key research and development program of China No. 2020YFA0713000. ZD would like to thank the support of the PhD student fund for short-term overseas visits and the Simons Foundation Collaboration on Celestial Holography. LC would also like to thank the support of the Shuimu Tsinghua Scholar Program of
Tsinghua University. ZD and WS would like to thank the conference Exploring TTbar Deformations, Integrable Models, and String Theory in Ningbo University for stimulating discussions. WS is grateful to the Simons Collaboration on Celestial Holography Annual Meeting, New Frontiers in Supergravity, Strings, and Related Topics, and ICBS 2025 for the opportunities to present related work and receive valuable feedback. 

\appendix
\section{Details of calculations}
In this appendix, we collect some details of the calculations of this paper.
\subsection{Some basic Poisson brackets} \label{app:poisson_brackets}

Here, we present the Poisson brackets between the stress tensor and currents after $J\bar{T}$ deformation:
\begin{align}\label{HLRp}
		&\{\HL(y_1),\HL(y_2)\}=-\p_{y_1}\HL(y_1)\delta(y_1-y_2)+2\HL(y_2)\p_{y_1}\delta(y_1-y_2)\nonumber\\
		&\{\HR(y_1),\HR(y_2)\}=\p_{y_1}\z(\frac{\HR(y_1)}{1-\lambda\tilde{\MJ}_+}\y)\delta(y_1-y_2)-\frac{2\HR(y_2)}{1-\lambda\tilde{\MJ}_+}\p_{y_1}\delta(y_1-y_2)\nonumber\\
		&\{\HL(y_1),\HR(y_2)\}=\p_{y_1}\z(\frac{\lambda\tilde{\MJ}_+\HR(y_1)}{1-\lambda\tilde{\MJ}_+}\y)\delta(y_1-y_2)-\frac{\lambda\tilde{\MJ}_+\HR(y_2)}{1-\lambda\tilde{\MJ}_+}\p_{y_1}\delta(y_1-y_2)\nonumber\\
		&\{\HL(y_1),\tilde{\MJ}_+(y_2)\}=\tilde{\MJ}_+(y_1)\p_{y_1}\delta(y_1-y_2)\nonumber\\
		&\{\HR(y_1),\tilde{\MJ}_+(y_2)\}=-\frac{\lambda k}{2\pi}\frac{\HR(y_2)}{1-\lambda\tilde{\MJ}_+}\p_{y_1}\delta(y_1-y_2)\\
		&\{\tilde{\MJ}_+(y_1),\tilde{\MJ}_+(y_2)\}=\frac{k}{2\pi}\p_{y_1}\delta(y_1-y_2)\nonumber\\
                        &\{\HL(y_1),\MJ_{-t}(y_2)\}=-\frac{\lambda\tilde{\MJ}_+\MJ_{-t}(y_1)}{1-\lambda \tilde{\MJ}_+}\p_{y_1}\delta(y_1-y_2)\nonumber\\
		&\{\HR(y_1),\MJ_{-t}(y_2)\}=-\frac{\MJ_{-t}(y_1)}{1-\lambda\tilde{\MJ}_+}\p_{y_1}\delta(y_1-y_2)\nonumber\\
        &\{\MJ_{-t}(y_1),\MJ_{-t}(y_2)\}=-\frac{k}{2\pi }\p_{y_1}\delta(y_1-y_2)\nonumber\\
		&\{\tilde{\MJ}_+(y_1),\MJ_{-t}(y_2)\}=-\frac{\lambda k }{2\pi
        }\frac{\MJ_{-t}(y_1)}{1-\lambda \tilde{\MJ}_+}\p_{y_1}\delta(y_1-y_2)\nonumber
\end{align}
Here, we use the Poisson brackets of the undeformed theory and the relation:
\begin{equation}\label{currentrelation}
    	\begin{aligned}
		&\frac{\p \HL}{\p \HL^{(0)}}=1,\quad \frac{\p \HL}{\p \HR^{(0)}}=\frac{\lambda\tilde{\MJ}_+}{1-\lambda\tilde{\MJ}_+},\quad \frac{\p \HL}{\p \MJ_+^{(0)}}=\frac{\lambda\HR}{1-\lambda\tilde{\MJ}_+}\\
		&\frac{\p \MH_R}{\p \MH_L^{(0)}}=0,\quad \frac{\p \MH_R}{\p\MH_R^{(0)}}=\frac{1}{1-\lambda\tilde{\MJ}_+},\quad  \frac{\p \MH_R}{\p\MJ_+^{(0)}}=\frac{\lambda\MH_R}{1-\lambda \tilde{\MJ}_+}\\
		&\frac{\p \tilde{\MJ}_+}{\p \MH_L^{(0)}}=0,\quad \frac{\p \tilde{\MJ}_+}{\p\MH_R^{(0)}}=\frac{\lambda k}{2\pi}\frac{1}{1-\lambda\tilde{\MJ}_+} ,\quad \frac{\p \tilde{\MJ}_+}{\p\MJ_+^{(0)}}=1+\frac{\lambda^2k}{2\pi}\frac{\MH_R}{1-\lambda\tilde{\MJ}_+}
	\end{aligned}
\end{equation}

\subsection{The second order perturbative calculations of correlation functions}\label{app:2ndOrder}

In this appendix, we will collect the details of the second-order perturbative calculations of the two-point functions. By using the Ward identities and  considering the normal ordering, we have:
    \begin{align}
                &\langle \bar{\chi}_T({\bar{x}}_1) \bar{\chi}_T({\bar{x}}_2)\tilde{\MO}({\bar{x}}_1)\tilde{\MO}({\bar{x}}_2)\rangle=\Big[-\frac{h^2}{({\bar{x}}_1-{\bar{x}}_2)^2}+\frac{4h\log({\bar{x}}_1-{\bar{x}}_2)}{({\bar{x}}_1-{\bar{x}}_2)^2}-\frac{c}{12(\bar{w}_1-\bar{w}_2)^2}\\
        &\qquad -\frac{(1+h)\log({\bar{x}}_1-{\bar{x}}_2)}{{\bar{x}}_1-{\bar{x}}_2}\left(\p_{{\bar{x}}_1}-\p_{{\bar{x}}_2}\right) +\log({\bar{x}}_2-{\bar{x}}_1)^2\p_{{\bar{x}}_1}\p_{{\bar{x}}_2}\Big]\langle \tilde{\MO}({\bar{x}}_1)\tilde{\MO}({\bar{x}}_2)\rangle, \\
                &\langle \bar{\chi}_T({\bar{x}}_1) \bar{\chi}_T({\bar{x}}_1)\tilde{\MO}({\bar{x}}_1)\tilde{\MO}({\bar{x}}_2)\rangle=\Big[\frac{h(1+h)}{({\bar{x}}_1-{\bar{x}}_2)^2}-\frac{2h\log({\bar{x}}_1-{\bar{x}}_2)}{({\bar{x}}_1-{\bar{x}}_2)^2}  -\frac{2(1+h)\log({\bar{x}}_1-{\bar{x}}_2)}{{\bar{x}}_1-{\bar{x}}_2}\p_{{\bar{x}}_2}\\
        &\qquad +\log({\bar{x}}_1-{\bar{x}}_2)^2\p_{{\bar{x}}_2}^2\Big]\langle \tilde{\MO}({\bar{x}}_1)\tilde{\MO}({\bar{x}}_2)\rangle
    \end{align}
Summing them up, we can obtain the second-order contribution:
\begin{equation}
    \begin{aligned}
        & G(p,\bar{p})_{(2)}  =\z(\frac{\lambda q}{2\pi}-\frac{i\lambda^2 k\bar{p}}{8\pi^2}\y)^2 \int dx^2 \, e^{ipx +i\bar{p}\bar{x}} \Bigg\{\frac{(2h+1)h}{\bar{x}^2}-\frac{6h\log\bar{x}}{\bar{x}^2}+\frac{c}{12\bar{x}^2}\\
        &+\frac{4(1+h)\log\bar{x}}{\bar{x}}\p_{\bar{x}}+2(\log\bar{x})^2\p_{\bar{x}}^2  \Bigg\} \, \left\langle \tilde{\MO}(x)\tilde{\MO}(0) \right\rangle
    \end{aligned}
\end{equation}
We perform the integration by parts to move all of derivatives to the left, and the result can be obtained as:
\begin{equation}
\begin{aligned} G(p, \bar{p})_{(2)} & = {1\over 2!}\int dx^2 e^{ipx+i\bar p\bar x} \, \left(\frac{i\lambda q}{2\pi}+\frac{\lambda^2k\bar{p}^2}{4\pi^2}\right)^2 A_2(\log \bar{x})  \, x^{\frac{i\lambda q\bar{p}}{\pi}+\frac{\lambda^2k\bar{p}^2}{4\pi^2}}G(x,\bar{x})_{(0)}\\
A_2(\theta)&=\theta^2+\frac{6h^2+11h+4}{2h(2h+1)}\theta-\frac{c}{48h(2h+1)}-\frac{4h^4+4h^3-3h^2-21h-12}{4h^2(2h+1)^2}
\end{aligned}
\end{equation}

\subsection{Useful identities}
To calculate the Borel resummation in the presence of instantons, we make use of the following identity for the hypergeometric function:
    \begin{align}\label{Hyper}
        {}_2F_1(h,h+\frac{1}{2},1,4\zeta)&=\frac{\Gamma(\frac{1}{2}-2h)}{\Gamma(1-h)\Gamma(\frac{1}{2}-h)}{}_2F_1(h,h+\frac{1}{2},2h+\frac{1}{2},1-4\zeta)\nonumber\\
        &+\frac{\Gamma(2h-\frac{1}{2})}{\Gamma(h)\Gamma(h+\frac{1}{2})}(1-4\zeta)^{\frac{1}{2}-2h}{}_2F_1(1-h,\frac{1}{2}-h,\frac{3}{2}-2h,1-4\zeta)
    \end{align}
The multi-value property is encoded in the factor $(1-4\zeta)^{\frac{1}{2}-2h}$. The Stokes discontinuity can be obtained by integrating over the second line in \eqref{Hyper},
\begin{equation}
    \begin{aligned}
        \text{Disc} &
        Y(K)\\
        &=-\frac{i2^{2-4h}\Gamma(2h-\frac{1}{2})\cos(2h\pi)}{K\Gamma(h)\Gamma(h+\frac{1}{2})}e^{-\frac{1}{4K}}\!\int_{0}^\infty dtt^{\frac{1}{2}-2h}e^{-\frac{t}{K}} {}_2F_1\z(1-h,\frac{1}{2}-h,\frac{3}{2}-2h,-4t\y)\\
        &=\frac{i\pi 2^{-2h}}{\Gamma(h)\Gamma(h+\frac{1}{2})}e^{-\frac{1}{4K}}K^{-\frac{1}{2}-h}U\z(1-h,\frac{3}{2},\frac{1}{4K}\y)
    \end{aligned}
\end{equation}

\section{Symmetries and the non-local coordinates}\label{symnonlocal}
Having established the dressed symmetries, we proceed to a detailed discussion of the associated nonlocal coordinates.
During the process of constructing the symmetry generators, we observe the left-moving sector remains local, whereas a non-local coordinate in the right-moving  \eqref{nonlocalv} naturally appears. 
This motivates us to consider the following coordinate transformation,
\begin{equation}\label{nonlocal}
    \begin{aligned}
        & \hat{u}\equiv u,\quad \hat{v}\equiv v +\delta\hat{v},\quad  \delta\hat{v}\equiv-\lambda \chi_0+\frac{\lambda^2k}{4\pi}\bar{\eta}_0
    \end{aligned}
\end{equation}
This definition immediately yields the following non-trivial components in the Jacobian matrix, 
\begin{equation}
    \begin{aligned}
        &\p_u\hat{v}=-\lambda \tilde{\MJ}_+\z(1+\frac{\lambda^2 k}{4\pi}\frac{\MH_R}{1-\lambda \tilde{\MJ}_+}\y),\quad \p_v\hat{v}=1+\frac{\lambda^2 k}{4\pi}\frac{\MH_R}{1-\lambda \tilde{\MJ}_+}
    \end{aligned}
\end{equation}
so that the determinant is given by
\begin{align}\label{Jacobian}
    \det\z[\frac{\p\hat{x}}{\p x}\y]=1+\frac{\lambda^2k}{4\pi}\frac{\HR}{1-\lambda \tilde{\MJ}_+}
\end{align}
Inverting this relation gives
\begin{equation}\label{Inverse}
    \begin{aligned}
        &\p_{\hat{u}} v=\lambda \MJ_+,\quad \p_{\hat{v}} v=1-\frac{\lambda^2 k}{4\pi }\frac{\MH_R}{1-\lambda \tilde{\MJ}_++\frac{\lambda^2k}{4\pi}\MH_R}
    \end{aligned}
\end{equation}
These can be integrated so that the inverse transformation is given by
 \begin{align}
&   u=U( \hat{u},\hat v),   \quad  v=V(\hat{u},\hat v)  \label{U2u}
\end{align}
where $U,\,V$ are field-dependent functions of arbitrary variables $\sigma^+, \sigma^-$ defined as
\begin{equation}\label{U2s}\begin{aligned}
&U(\sigma^+,\sigma^-)\equiv\sigma^+\\
 &  V(\sigma^+,\sigma^-) \equiv \sigma^-+\lambda\int G(\sigma^+-u')\tilde{\MJ}_+(u') du'-\frac{\lambda^2k }{4\pi}\int G(\sigma^--v') \tilde{\MH}_R(v')dv' 
\end{aligned}
\end{equation}

Applying integration by parts to the expression for the dressed currents \eqref{dressedcureents} leads to the following useful relation:
\begin{equation}\label{fH}
	\begin{aligned}
    &\int dy'f(y')\tilde{\mathcal{H}}_R(y'-t)=\int dy' f(y'+\delta\hat{v}(t,y'))\HR(t,y')\\
    &\int dy' f(y')\tilde{\mathcal{J}}_-(y'-t')=\int dy' f(y'+\delta\hat{v}(t,y'))\MJ_{-t}(t,y')\\
	\end{aligned}
\end{equation}
where all quantities are evaluated on a fixed time slice, and $f(y)$ is an arbitrary function. Notable choices for $f(y)$ include:
\begin{itemize}
    \item 
Applying $f(y') = \delta(y' - (y + \delta\hat{v}))$ for $\HR$ and $\MJ_{-t}$ in \eqref{fH}, we obtain the equal-time relation between the dressed currents and the deformed currents $(\HR, \MJ_{-t})$: 
\begin{align}\label{Hrelation}
	\begin{aligned}
		&\tilde{\mathcal{H}}_R(\hat{v})=\frac{\HR(t,y)}{1-\lambda \tilde{\MJ}_+(t,y)+\frac{\lambda^2 k}{4\pi }\MH_R(t,y)},
		\quad \tilde{\mathcal{J}}_-(\hat{v})=\frac{\MJ_{-t}(t,y)}{1-\lambda \tilde{\MJ}_+(t,y)+\frac{\lambda^2 k}{4\pi }\MH_R(t,y)}
	\end{aligned}
\end{align}
where  $y-t+\delta \hat{v}$ is replaced by $\hat{v}$.
Using the relation between the physical stress tensor and the undeformed stress tensor \eqref{TransHJ},  we get another relation between the  {dressed} currents and the undeformed currents, 
\begin{align}\label{H0relationA}
	\begin{aligned}
		&\tilde{\mathcal{H}}_R(\hat{v})=\frac{\HR^{(0)}(t,y)}{\z(1-\lambda \tilde{\MJ}_+(t,y)+\frac{\lambda^2 k}{4\pi }\MH_R(t,y)\y)^2}=({\partial \hat v\over \partial y})^{-2} \HR^{(0)}(t,y),\\
		&\tilde{\mathcal{J}}_-(\hat{v})=\frac{\MJ_-^{(0)}(t,y)}{1-\lambda \tilde{\MJ}_+(t,y)+\frac{\lambda^2 k}{4\pi }\MH_R(t,y)}=({\partial \hat v\over \partial y})^{-1} \MJ_{-}^{(0)}(t,y)
	\end{aligned}
\end{align}

\item Applying $f(y')=\delta(y-y')$, we can express the dressed currents in terms of the deformed currents
\begin{align}
    \tilde{\MH}_R(v)=\int dy'\delta(v-\hat v')\MH_R(t,y'),\quad \tilde{\MJ}_-(v)=\int dy'\delta(v-\hat{v}')\MJ_{-t}(t,y')
\end{align}
where $ \hat v'=y'-t+\delta \hat v(t,y')$. 
\item 
Applying $f(y')=G(y+\delta\hat{v}-y')$ for $\HR$, we have 
\begin{equation}\label{Ginverse}
    \begin{aligned}
        &\int dy' G(y+\delta\hat{v}(t,y)-y')\tilde{\MH}_R(y'-t)=\bar{\eta}_0(t,y)
    \end{aligned}
\end{equation}
This provides another derivation of the inverse relation between the coordinates and non-local coordinates:
by rewriting $\delta \hat v$ in terms of the dressed currents and moving to the left hand side of  \eqref{nonlocal}.
\end{itemize}
\section{Comments on the right-moving currents}\label{comment}
The construction of the dressed operator presented in Section \ref{sec:dress} has so far been purely classical. This derivation is expected to hold for any primary operator in the undeformed theory that satisfies the Poisson bracket relation in \eqref{h0O0}. Furthermore, in the absence of a classical central charge, the undeformed symmetry currents themselves satisfy \eqref{h0O0}. Having established this classical framework, we now examine two key questions: first, whether the dressed operator \eqref{Flowedoperator} is consistent with the dressed currents defined in \eqref{dressedcureents}; and second, whether these deformed currents can be identified with the physical operators introduced here. To address these questions, we apply
\begin{align}\mathcal O^{(0)}_{0,2,0}=\mathcal H_R^{(0)}, \quad \mathcal O^{(0)}_{0,1,0}=\mathcal \MJ_{-}^{(0)}.\end{align} 
It is clear that the relation between the {dressed} and undeformed symmetry currents \eqref{H0relation} takes the same form as \eqref{O0relation}, implying that the two ways of constructing the {dressed} symmetry currents are indeed compatible.

A careful comparison between \eqref{TransHJ} and \eqref{physicalOp} shows that the deformed stress tensor does not coincide with the physical operator, but the two are related via
\begin{align}\label{o20}
\mathcal O_{0,2,0}=\frac{\mathcal H_R}{1-\lambda \tilde{\MJ}_+}\end{align}
We also have the relation between $\MJ_{-t}$ and physical operator
\begin{equation}\label{o10}
    \MO_{0,1,0}=\frac{\MJ_{-t}}{1-\lambda \tilde{\MJ}_+}
\end{equation}
Substituting this into \eqref{Flowedoperator},
\begin{equation}\label{secondHL}
\begin{aligned}
&\tilde{\mathcal{O}}_{0,2,0} (u,v)\equiv \sum_{n=0}^\infty   \frac{1}{n!} \partial_v^{n}\left((-\delta\hat{v})^{n}\frac{\mathcal H_R}{1-\lambda \tilde{\MJ}_+}\right)\\
&\tilde{\mathcal{O}}_{0,1,0} (u,v)\equiv \sum_{n=0}^\infty   \frac{1}{n!} \partial_v^{n}\left((-\delta\hat{v})^{n}\frac{\MJ_{-t}}{1-\lambda \tilde{\MJ}_+}\right)
\end{aligned}
\end{equation}
Although the above expression appears distinct from \eqref{dressedcureents}, and a direct verification of their equivalence is difficult, we can proceed indirectly. Substituting \eqref{o20} and \eqref{o10} into \eqref{relation}, we find that the dressed $(0,2)$  and $(0,1)$ operators with zero charge in the nonlocal coordinates are given by
\begin{equation}
\begin{aligned}
    & \tilde{\mathcal{O}}_{0,2, 0} (u,\hat{v})={\MH_R(u,v)\over 1-\lambda \tilde{\MJ}_+(u) +\frac{\lambda^2k}{4\pi}\HR}= \tilde{\MH}_R(\hat{v})\\
    & \tilde{\mathcal{O}}_{0,1, 0} (u,\hat{v})={\MJ_{-t}(u,v)\over 1-\lambda \tilde{\MJ}_+(u) +\frac{\lambda^2k}{4\pi}\HR}= \tilde{\MJ}_-(\hat{v})
\end{aligned}
\end{equation}
This is consistent with \eqref{H0relation} derived from \eqref{dressedcureents}, thereby providing a consistency check that definitively shows \eqref{secondHL} and \eqref{H0relation} describe the same dressed currents.

\section{Comparison to earlier results}\label{sec:compare}
In this section, we compare our physical operator with an operator introduced in \cite{Guica:2021fkv}, and show that the two types of operators are different both in conceptual and technical perspectives. This is the main reason the correlation function reported in this paper is different from that of  \cite{Guica:2021fkv}.

In the following, we first summarize the definition of the operator from \cite{Guica:2021fkv}, hereafter termed the G-operator. This is to be distinguished from the physical operator introduced in this work \eqref{Ophys-m-E}, which we denote as the CDS-operator, constructed according to the strategy developed for $T\bar T$-deformed CFTs in \cite{Chen:2025jzb}.

\subsection{A brief review of previous work}
Let us first summarize the definition and main properties of the G-operator  \cite{Guica:2021fkv}.
In \cite{Guica:2021fkv}, symmetry generators and  operators are defined on the cylinder, and then an infinite radius limit is taken to obtain results on the plane. 
The dressed symmetry generators and dressed operators, also denoted by tilded variables, are solutions to the flow equation on the cylinder, analogous to \eqref{flowedeq}. 
Unlike the case on the plane, the flow equation of the deformed Hamiltonian is not just a canonical transformation, but also contains a diagonal piece \eqref{XY} which is responsible to the flow of the eigenvalues. As a result, the dressed symmetry generators, referred to as the spectral-flowed generators, do not contain either the left-moving nor the right-moving energy.
Nevertheless, a set of unflowed generators can be constructed from the flowed ones, with the properties that they are i) conserved and ii) the zero modes are physical generators of the theory. The unflowed generators are related to the flowed ones in a form that resembles a spectral flow transformation, but with a field-dependent spectral-flow parameter. 
The classical version of the unflowed generators is given by \footnote{Here $J_n$ is $P_n$ in (3.10)-(3.11) of \cite{Georgescu:2024ppd}. Our $q$ is $\tilde q$ in \cite{Guica:2021fkv} and \cite{Georgescu:2024ppd}. }, 
    \begin{align}\label{spectralflow}
   RQ_n &= R \tilde{Q}_n + \lambda H_R \tilde{P}_n + \frac{\lambda^2 k H_R^2}{4} \delta_{n,0}, \quad  J_n = \tilde{J}_n + \frac{\lambda k H_R}{2} \delta_{n,0}
\\
\label{spectralflowR}
    R_v \bar{Q}_n &= R \tilde{\bar Q}_n + \lambda H_R \tilde{{\bar P}}_n + \frac{\lambda^2 k H_R^2}{4} \delta_{n,0},\quad \bar  J_n = \tilde{\bar J}_n + \frac{\lambda k H_R}{2} \delta_{n,0}, \nonumber \\
    R_v&=R-\lambda(J_0-\bar J_0), \quad  {Q}_0=H_L ,\quad  \bar{Q}_0=H_R
\end{align}
where the tilded variables are the dressed (spectral-flowed) symmetry generators, and $Q_n,\,J_n$ etc are the unflowed generators. 
The quantum version is obtained by
\begin{align}\label{L2Q}
&   R Q_n \rightarrow L_n, \quad R\tilde Q_n \rightarrow\tilde L_n,
  \quad  R_v \bar Q_n \rightarrow  {\bar L}_n , \quad R \tilde {\bar Q}_n \rightarrow  \tilde {\bar L}_n  ,
\end{align} and normal ordering is assumed. 
The generators are related by (3.17) in \cite{Guica:2021fkv}
\begin{align}\label{spectralflowq}
    L_n=\tilde{L}_n+\lambda : H_R \tilde J_n :+{\lambda^2 k H_R^2  \over 4}\delta_{n,0},\quad J_n=\tilde J_n+{\lambda k H_R \over 2} \delta_{n,0} \\
    {\bar L}_n=\tilde{{\bar L} }_n+\lambda :H_R {\tilde {\bar J }}_n:+{\lambda^2 k H_R^2  \over 4}\delta_{n,0},\quad {\bar J}_n=\tilde {\bar J}_n+{\lambda k H_R \over 2} \delta_{n,0}\label{spectralflowqR}
\end{align} 

Acting \eqref{spectralflow} on an eigenstate of the energy and charges, the eigenvalues of flowed and unflowed generators satisfy a relation which is equivalent to the deformed spectrum on the cylinder.
The relation \eqref{spectralflow} resembles the spectral flow transformation in a CFT, but with a field-dependent spectral flow operator $\lambda H_R$. Consider a primary state under the left-moving Virasoro-Kac-Moody generators $\tilde L_n,\tilde J$, with fixed right-moving energy $E_R=\bar p$, 
The relation \eqref{L2Q} then suggests the following shift of the left-moving conformal weight
\begin{align}
    h_\lambda= h+\lambda \bar p q+{\lambda^2 \bar p ^2 k\over 4}, \label{shifthL}
\end{align}
where we have defined $h_\lambda=R E_L+{c\over 12},\quad  h=R {\tilde{E}}_L+{c\over 12}$, similar to the case of a CFT. Note that the above relation is true for the left moving sector. We have not required any  property of the state under the right-moving generators other than being an eigenstate of $H_R$. Since $H_R$ commutes with the entire left-moving sector, its eigenvalue can be viewed as a constant spectral flow parameter and the usual spectral flow rule can be directly applied. In fact, so far the relation \eqref{shifthL} is only a rewriting of the spectrum in a way that looks like a shift in the conformal shift. But the existence of a local left-moving conformal symmetry is a strong indication that it is possible to find an operator that behaves like a primary operator with the shifted weight $h_\lambda$. 
Starting from a primary operator $\tilde {\mathcal O}(p,\bar p)$ under the dressed Virasoro algebra $\tilde L_m, {\bar {\tilde L} }_m$ with conformal weight $(h,\bar h)$, 
the authors of \cite{Guica:2021fkv} define a type of operator with the following properties:  
\begin{itemize}
    \item[(1)] It is a primary operator of the left-moving Virasora-Kac-Moody algebra with conformal weight $h_\lambda$, 
    \begin{align}\label{c1}
       C_1:\quad [ L_n,{\mathcal O}^G(p,\bar p)]&=[n(h_\lambda-1)+p R]{\mathcal O}^G(p-{n\over R},\bar p)     
    \end{align}
    \item  
and shifted $U(1)$ charge, 
\begin{align}\label{c2}
    C_2:\quad  [J_n,{\mathcal O}^G(p,\bar p)]&=(q+{k\over 2}\lambda \bar p){\mathcal O}^G(p-{n\over R},\bar p)
\end{align}
    \item[(2)]It corresponds to an eigenstate of the  right-moving Hamiltonian  \begin{align}  C_3: \quad [H_R,{\mathcal O}^G(p,\bar p)]=\bar p {\mathcal O}^G(p,\bar p).\label{c3}\end{align}
\end{itemize}
The condition $C_2$ and $C_3$ imply that the operator has fixed charge under the chiral-current $\tilde{J}_\mu$,  
\begin{align}
  C_2':\quad  [\tilde J_0, {\mathcal O}^G(p,\bar p)]=q{\mathcal O}^G(p,\bar p)\label{c2p}
\end{align}
Using the relation \eqref{L2Q}, we learn that 
\begin{align}
    \tilde{\bar L}_0=H_R (R-\lambda \tilde J_0-{\lambda^2 k\over 4}H_R)
\end{align}
which together with conditions \eqref{c2}\eqref{c3} implies that 
\begin{align}
   C_3':\quad  [\tilde {\bar L}_0, \, {\mathcal O}^G(p,\bar p)]=[\bar p (R+\lambda q+{\lambda^2 k\over 4}\bar p)-\lambda (q+k \lambda \bar p/2) H_R-\lambda  \bar{p} {\tilde J}_0]{\mathcal O}^G(p,\bar p)\label{c4}
\end{align}
Neither relation $C_2'$ nor $C_3'$ is an additional condition, but a consequence of the above conditions \eqref{c1}-\eqref{c3}. We note that the G-operator carries a fixed charge $q$ under the chiral current, and a shifted charge $q_\lambda$ under the quasi-chiral current. This property is different from that of the spectrum, where the quasi-chiral charges are fixed.

To explicitly construct the above primary operators,  a specific ansatz was made in equation (3.19) of \cite{Guica:2021fkv}, which we reproduce here, 
\begin{align}\label{ansatz}
     \mathcal{O}^G(p,\bar p)=&\int dwd\bar{w}\,e^{-pw-\bar{p}\bar{w}}e^{A_{\mathcal{O}}w+B_{\mathcal{O}}\bar{w}}
e^{\eta_{\mathcal{O}}\sum_{n=1}^\infty\frac{1}{n}(e^{nw}\tilde{J}_{-n}+e^{n\bar{w}}\tilde{\bar{J}}_{-n})}\\
&\tilde{\mathcal{O}}(w,\bar{w}) e^{-\eta_{\mathcal{O}} \sum_{n=1}^\infty\frac{1}{n}(e^{-nw}\tilde{J}_{n}+e^{-n\bar{w}}\tilde{\bar{J}}_{n})}\nonumber
\end{align}
where $w,\bar w$ are complex worldsheet coordinates, and $A_{\mathcal O}, \, B_{\mathcal O}$ are operators to be determined. They are assumed to be a linear combination of the charges $J_0,\, {\bar J}_0$ and right-moving energy $H_R$, and also depend on the conserved quantum numbers of the operator $\mathcal O$. Plugging the ansatz to the Ward identities \eqref{c1}\eqref{c2}\eqref{c3}, $A_{\mathcal O}, \, B_{\mathcal O}$ are fixed in \cite{Guica:2021fkv}.

From the ansatz \eqref{ansatz}, it is already clear that the resulting operator is different from the our definition of the physical operator \eqref{Ophys-m-E}. The most obvious distinction is in the right-moving sector: Our physical operators depend on the right-moving dressed stress tensor, but not on the right-moving current, whereas the above ansatz depends the right-moving current $\tilde {\bar {J}}$ and only the zero mode of the right-moving stress tensor $H_R$. This is the main technical reason we get a different correlation function from that of \cite{Guica:2021fkv}.

\subsubsection{The plane limit}
To better understand the distinction between the G-operator and the CDS-operator, let us now take the plane limit. To warm up, let us first consider the expansion of the conformal Killing vectors on the cylinder in terms of those on the plane,
\begin{align}
    l^{cl}_n=-Re^{nw\over R}\partial_w= \sum_{m=0}^\infty{R^{1-m} n^{m}\over (m)!} l_{m-1}^{pl}, \quad  l_{n}^{pl}=-w^{n+1}\partial_w \label{cl2pl}
\end{align} 
where we use the superscript $cl$ for quantities on the cylinder, and $pl$ for those on the plane. 
The plane generators $l_{m-1}^{pl}$ can be read off from the coefficient of $n^m$ in the above expansion. Alternatively, we can expand $l_n^{cl},\, n=0,\pm1, \cdots \pm l$ up to the order of $l+1$, and solve $l^{pl}_{m},\, m=-1,\cdots l$ in terms of $l^{cl}_{n}$. This will allow us to find explicit expressions of plane generators in terms of the cylinder ones for all $l^{pl}_{m},\, m=-1,\cdots l$. 
Similarly, the current can be expanded as
\begin{align}
    j^{cl}_n= \sum_{m=0}{R^{-m} n^{m}\over m!} j_{m}^{pl}, \label{cl2pl}
\end{align}
Plugging into the spectral flow relation \eqref{spectralflow},
we can read the following relation between plane generators from the coefficient of $n^m$ \begin{align}
    L_{m-1}^{pl}-{\tilde L}_{m-1}^{pl}=\lim_{R\to \infty}\lambda{H_R\over  R} {\tilde J}_m^{pl}= 0,   \quad   { J}_m^{pl}={\tilde J}_m^{pl}+{\lambda k H_R\over2} , \quad m=0,1,\cdots
\end{align}
From the relation above, it follows that in the plane limit, the two generators related by spectral flow on the cylinder reduce to the same generator on the plane.
This statement holds for both the left-moving sector and the right-moving sector.
More explicitly, let us start with the dressed charges on the cylinder in Table 1 of \cite{Georgescu:2024ppd},
\begin{equation}
    \begin{aligned}
   & \tilde{Q}_n=\int^R_0d\sigma e^{\frac{in u}{R}}\z[\mathcal{H}_L-\frac{\lambda H_R}{R}(\MJ_++\frac{\lambda k}{2}\mathcal{H}_R)+\frac{\lambda^2kH_R^2}{4R^2}\y]\\
   & \tilde{\bar{Q}}_n=\int^R_0d\sigma e^{\frac{in \hat{v}}{R_v}}\z[\frac{R_v}{R}\mathcal{H}_R-\frac{\lambda H_R}{R}(\MJ_-+\frac{\lambda k}{2}\mathcal{H}_R)+\frac{\lambda^2kH_R^2}{4R}\z(\frac{1}{R}-\frac{\lambda\hat{\phi}'}{R_v}\y)\y].
\end{aligned}
\end{equation}
Expand the $e^{\frac{inu}{R}}$ and $e^{\frac{in\hat{v}}{R_v}}$ and collect the terms of the same $u^{m+1}$ and $\hat{v}^{m+1}$, we can see 
\begin{equation}
\begin{aligned}
    &   \tilde{L}^{pl}_{m}\equiv\lim_{R\to\infty}\lim_{n\to0}\frac{\p}{\p n^{m+1}}R^{m+1}\tilde{Q}_n=\int^\infty_0 d\sigma u^{m+1}\mathcal{H}_L\\
    &\tilde{\bar{L}}^{pl}_{m}\equiv\lim_{R\to\infty}\lim_{n\to0}\frac{\p}{\p n^{m+1}}R^{m+1}\tilde{\bar{Q}}_n=\int^\infty_0 d\sigma \hat{v}^{m+1}\mathcal{H}_R
\end{aligned}
\end{equation}
The symmetry generators obtained from the large-radius limit of the cylinder reproduce our results \eqref{Charges1} and \eqref{Charges2}, up to the specific definition of the nonlocal coordinate $\hat{v}$. This discrepancy in the nonlocal coordinate arises from our choice of $U(1)$ currents and does not affect the conceptual conclusion. We have therefore shown that our direct construction of symmetry generators on the plane is consistent with the large-radius limit of their cylinder counterparts.

We now consider the plane limit of the Ward identities. By applying the same expansion procedure used for the symmetry generators, we expand the Ward identities \eqref{c1}-\eqref{c4} in powers of $1/R$, yielding the following relations on the plane:
\begin{align}\label{csplane}
       C_1:\quad &[ L_n,{\mathcal O}^G(p,\bar p)]=[ {\tilde L}_n,{\mathcal O}^G(p,\bar p)]=(-1)^{n}[(n+1)(h_\lambda-1)\partial p^n- p\partial p^{n+1}]{\mathcal O}^G(p,\bar p)\\
       C_2:\quad &[J_0,{\mathcal O}^G(p,\bar p)]=(q+{k\over 2}\lambda \bar p){\mathcal O}^G(p,\bar p)\label{quasi-chiral}\\
       C_2':\quad & [\tilde J_0, {\mathcal O}^G(p,\bar p)]=q{\mathcal O}^G(p,\bar p)\label{chiralq}\\
      C_3, \,C_3':\quad &[H_R,{\mathcal O}^G(p,\bar p)]=\bar p {\mathcal O}^G(p,\bar p)
    \end{align}
In the right-moving sector, the two conditions \eqref{c3} and \eqref{c4} coalesce on the plane into the single condition that the operator must have a definite right-moving momentum $\bar p$. In the left-moving sector, both the flowed and unflowed Virasoro generators on the cylinder reduce to the same ones on the plane. This is consistent with the fact that the physical left-moving stress tensor already satisfies the flow equation \eqref{flowedeq}, which does not happen on the cylinder due to the presence of the diagonal term \eqref{XY}.
Note that the two currents $J_n$ and $\tilde{J}_n$ remain distinct, and the G-operator still has fixed chiral charge $q$ under the chiral current. The charge property of the G-operator is different from that of the spectrum, which has fixed charge under $J_0$ instead.  On the other hand, the charges of the CDS-operator agree with those of the spectrum, as can be seen from Table 2. 
 
 It turns out that the commutator between the operator \eqref{ansatz} 
 with the right-moving Virasoro-Kac-Moody also takes the same form of a primary operator after taking the plane limit.
\begin{align}
 &[ {\bar L}^{pl}_n,{\mathcal O}^G(p,\bar p)]=[ {\tilde {\bar  L}}^{pl}_n,{\mathcal O}^G(p,\bar p)]=(-1)^{n}[(n+1)({\bar h}^G_\lambda-1)\partial {\bar p}^n- \bar p\partial {\bar p}^{n+1}]{\mathcal O}^G(p,\bar p)\label{r1}\\
  &[ {\bar J}_0, {\mathcal O}^G(p,\bar p)]=(\bar q+{\lambda k \bar p\over 2}) {\mathcal O}^G(p,\bar p)\label{r2}
\\
 &{\bar h}^G_\lambda=\bar h+\lambda \bar p\bar q+{\lambda^2 k\bar p^2\over4 }\label{weightR}
\end{align}
The above Ward identity arises as a consequence of the specific choice of ansatz \eqref{ansatz} made in \cite{Guica:2021fkv}.  It is precisely the choice of ansatz \eqref{ansatz} that makes the resulting operator have primary-like behavior under the right-moving Virasoro–Kac–Moody symmetries, thereby imposing strong constraints on the correlation functions.
As shown in \cite{Georgescu:2024ppd}, the correlation function of the operator \eqref{ansatz} on the plane takes the same form as the CFT result in momentum space, but with shifted conformal weights $h^G_\lambda = h_\lambda$ and $\bar h^G_\lambda$ given by \eqref{shifthL} and \eqref{weightR}. Comparing with our result \eqref{Gn}, a crucial difference emerges: in our case, the right-moving weight depends on the left-moving charge $q$, similarly to the left-moving weight. In \cite{Guica:2021fkv}, however, the shifted right-moving weight depends on $\bar q$ instead. This discrepancy stems from the fact that the two papers consider different operators.

\subsection{Comparison}

We now compare the G-operator introduced in \cite{Guica:2021fkv,Georgescu:2024ppd} with the CDS physical operator defined in this paper \eqref{Ophys-m-E}. Both operators are constructed from dressed operators and currents, and both transform as primary operators under the left-moving Virasoro algebra. Despite these  similarities, the two types of operators differ significantly, both conceptually and technically.

Conceptually, this difference arises from the distinct guiding principles used to identify a suitable operator basis. As summarized in the previous section, the G-operator was constructed in \cite{Guica:2021fkv} by imposing the Ward identities \eqref{c1}–\eqref{c4}, proposing an ansatz of the form \eqref{ansatz}, and solving the resulting constraints. The specific form of the ansatz was motivated by insights from the $J\wedge \bar J$ deformation, together with the a posteriori that the resulting solution also transforms as a primary operator under the right-moving pseudo-conformal symmetry on the plane.
The latter property is useful for constraining correlators, although it does not follow from any first-principles argument.

In contrast, the CDS-operator is constructed by requiring the following two properties:
i) At the classical level, it is pointwise defined in the sense that it depends only on the undeformed operator, currents, and stress tensor at a given point. In particular, no integrated quantities appear in the relation \eqref{physicalOp}.
ii) It can be expressed entirely in terms of dressed operators and dressed symmetry operators in a manageable way. 
Since the dressed quantities fully respect the symmetries of the undeformed CFT, the CDS-type operator has the advantage of being amenable to study within an auxiliary CFT framework, where Ward identities and correlation functions are readily applicable.
Notably, no specific symmetry properties are imposed a priori, although the resulting operator turns out to be primary under the left-moving conformal symmetry. It remains an open question whether the CDS operators are the unique ones satisfying the two properties above; however, they are the simplest examples we have found so far. For instance, attempting to express the undeformed operator directly in terms of dressed operators leads to a highly complicated expression whose utility for explicit computations is unclear.

At the technical level, the primary distinction lies in the construction of the two operators. The ansatz \eqref{ansatz} for the G-operator is built from $\tilde O$, $\tilde {\mathcal J}$, and $\tilde {\bar {\mathcal J}}$, whereas the CDS operator \eqref{Ophys-m-L} is obtained by dressing $\tilde O$ with a Wilson-line-like operator that involves the left-moving chiral current $\tilde {\mathcal J}$ and the dressed right-moving stress tensor $\tilde {\mathcal H}_R$.
Both operators are primary under the left-moving Virasoro–Kac–Moody algebra \eqref{csplane} and share the same shifted conformal weight $h_\lambda$. However, their charges under the chiral current differ. As shown in \eqref{shiftL}, the CDS operator carries a shifted chiral charge that is consistent with the spectrum. In contrast, the G-operator has a fixed charge under the chiral current \eqref{chiralq}, but exhibits a shifted charge under the quasi-chiral current \eqref{quasi-chiral}.
In the right-moving sector, the distinction between the two operators becomes more obvious. The G-operator transforms as a primary operator under the right-moving Virasoro–Kac–Moody algebra, as shown in \eqref{r1}, whereas the CDS operator does not.
When considering two-point correlation functions, the leading logarithmic contribution of the CDS operator takes a form similar to that of the G-operator. In both cases, the structure mirrors that of a CFT correlator in momentum space, but with different shifted conformal weights. While the shifted weights in the left-moving sector coincide, they differ in the right-moving sector, as can be seen from \eqref{weightR} and \eqref{Gn}.

\bibliographystyle{JHEP}
\bibliography{ref.bib}

\providecommand{\href}[2]{#2}\begingroup\raggedright\begin{thebibliography}{10}

\bibitem{Chen:2025jzb}
L.~Chen, Z.~Du, K.~Liu and W.~Song, \emph{{Symmetries and operators in $T\bar{T}$ deformed CFTs}},  \href{https://arxiv.org/abs/2507.08588}{{\ttfamily 2507.08588}}.

\bibitem{Strominger:1996sh}
A.~Strominger and C.~Vafa, \emph{{Microscopic origin of the Bekenstein-Hawking entropy}}, \href{https://doi.org/10.1016/0370-2693(96)00345-0}{\emph{Phys. Lett. B} {\bfseries 379} (1996) 99} [\href{https://arxiv.org/abs/hep-th/9601029}{{\ttfamily hep-th/9601029}}].

\bibitem{Brown:1986nw}
J.~D. Brown and M.~Henneaux, \emph{{Central Charges in the Canonical Realization of Asymptotic Symmetries: An Example from Three-Dimensional Gravity}}, \href{https://doi.org/10.1007/BF01211590}{\emph{Commun. Math. Phys.} {\bfseries 104} (1986) 207}.

\bibitem{Strominger:1997eq}
A.~Strominger, \emph{{Black hole entropy from near horizon microstates}}, \href{https://doi.org/10.1088/1126-6708/1998/02/009}{\emph{JHEP} {\bfseries 02} (1998) 009} [\href{https://arxiv.org/abs/hep-th/9712251}{{\ttfamily hep-th/9712251}}].

\bibitem{Maldacena:1997re}
J.~M. Maldacena, \emph{{The Large $N$ limit of superconformal field theories and supergravity}}, \href{https://doi.org/10.4310/ATMP.1998.v2.n2.a1}{\emph{Adv. Theor. Math. Phys.} {\bfseries 2} (1998) 231} [\href{https://arxiv.org/abs/hep-th/9711200}{{\ttfamily hep-th/9711200}}].

\bibitem{Gubser:1998bc}
S.~S. Gubser, I.~R. Klebanov and A.~M. Polyakov, \emph{{Gauge theory correlators from noncritical string theory}}, \href{https://doi.org/10.1016/S0370-2693(98)00377-3}{\emph{Phys. Lett. B} {\bfseries 428} (1998) 105} [\href{https://arxiv.org/abs/hep-th/9802109}{{\ttfamily hep-th/9802109}}].

\bibitem{Witten:1998qj}
E.~Witten, \emph{{Anti de Sitter space and holography}}, \href{https://doi.org/10.4310/ATMP.1998.v2.n2.a2}{\emph{Adv. Theor. Math. Phys.} {\bfseries 2} (1998) 253} [\href{https://arxiv.org/abs/hep-th/9802150}{{\ttfamily hep-th/9802150}}].

\bibitem{tHooft:1993dmi}
G.~'t~Hooft, \emph{{Dimensional reduction in quantum gravity}}, {\emph{Conf. Proc. C} {\bfseries 930308} (1993) 284} [\href{https://arxiv.org/abs/gr-qc/9310026}{{\ttfamily gr-qc/9310026}}].

\bibitem{Susskind:1994vu}
L.~Susskind, \emph{{The World as a hologram}}, \href{https://doi.org/10.1063/1.531249}{\emph{J. Math. Phys.} {\bfseries 36} (1995) 6377} [\href{https://arxiv.org/abs/hep-th/9409089}{{\ttfamily hep-th/9409089}}].

\bibitem{Ryu:2006bv}
S.~Ryu and T.~Takayanagi, \emph{{Holographic derivation of entanglement entropy from AdS/CFT}}, \href{https://doi.org/10.1103/PhysRevLett.96.181602}{\emph{Phys. Rev. Lett.} {\bfseries 96} (2006) 181602} [\href{https://arxiv.org/abs/hep-th/0603001}{{\ttfamily hep-th/0603001}}].

\bibitem{Hubeny:2007xt}
V.~E. Hubeny, M.~Rangamani and T.~Takayanagi, \emph{{A Covariant holographic entanglement entropy proposal}}, \href{https://doi.org/10.1088/1126-6708/2007/07/062}{\emph{JHEP} {\bfseries 07} (2007) 062} [\href{https://arxiv.org/abs/0705.0016}{{\ttfamily 0705.0016}}].

\bibitem{Almheiri:2019psf}
A.~Almheiri, N.~Engelhardt, D.~Marolf and H.~Maxfield, \emph{{The entropy of bulk quantum fields and the entanglement wedge of an evaporating black hole}}, \href{https://doi.org/10.1007/JHEP12(2019)063}{\emph{JHEP} {\bfseries 12} (2019) 063} [\href{https://arxiv.org/abs/1905.08762}{{\ttfamily 1905.08762}}].

\bibitem{Penington:2019kki}
G.~Penington, S.~H. Shenker, D.~Stanford and Z.~Yang, \emph{{Replica wormholes and the black hole interior}}, \href{https://doi.org/10.1007/JHEP03(2022)205}{\emph{JHEP} {\bfseries 03} (2022) 205} [\href{https://arxiv.org/abs/1911.11977}{{\ttfamily 1911.11977}}].

\bibitem{Almheiri:2019qdq}
A.~Almheiri, T.~Hartman, J.~Maldacena, E.~Shaghoulian and A.~Tajdini, \emph{{Replica Wormholes and the Entropy of Hawking Radiation}}, \href{https://doi.org/10.1007/JHEP05(2020)013}{\emph{JHEP} {\bfseries 05} (2020) 013} [\href{https://arxiv.org/abs/1911.12333}{{\ttfamily 1911.12333}}].

\bibitem{Guica:2008mu}
M.~Guica, T.~Hartman, W.~Song and A.~Strominger, \emph{{The Kerr/CFT Correspondence}}, \href{https://doi.org/10.1103/PhysRevD.80.124008}{\emph{Phys. Rev. D} {\bfseries 80} (2009) 124008} [\href{https://arxiv.org/abs/0809.4266}{{\ttfamily 0809.4266}}].

\bibitem{Bredberg:2009pv}
I.~Bredberg, T.~Hartman, W.~Song and A.~Strominger, \emph{{Black Hole Superradiance From Kerr/CFT}}, \href{https://doi.org/10.1007/JHEP04(2010)019}{\emph{JHEP} {\bfseries 04} (2010) 019} [\href{https://arxiv.org/abs/0907.3477}{{\ttfamily 0907.3477}}].

\bibitem{Castro:2010fd}
A.~Castro, A.~Maloney and A.~Strominger, \emph{{Hidden Conformal Symmetry of the Kerr Black Hole}}, \href{https://doi.org/10.1103/PhysRevD.82.024008}{\emph{Phys. Rev. D} {\bfseries 82} (2010) 024008} [\href{https://arxiv.org/abs/1004.0996}{{\ttfamily 1004.0996}}].

\bibitem{Haco:2018ske}
S.~Haco, S.~W. Hawking, M.~J. Perry and A.~Strominger, \emph{{Black Hole Entropy and Soft Hair}}, \href{https://doi.org/10.1007/JHEP12(2018)098}{\emph{JHEP} {\bfseries 12} (2018) 098} [\href{https://arxiv.org/abs/1810.01847}{{\ttfamily 1810.01847}}].

\bibitem{Haco:2019ggi}
S.~Haco, M.~J. Perry and A.~Strominger, \emph{{Kerr-Newman Black Hole Entropy and Soft Hair}},  \href{https://arxiv.org/abs/1902.02247}{{\ttfamily 1902.02247}}.

\bibitem{Bredberg:2011hp}
I.~Bredberg, C.~Keeler, V.~Lysov and A.~Strominger, \emph{{Cargese Lectures on the Kerr/CFT Correspondence}}, \href{https://doi.org/10.1016/j.nuclphysbps.2011.04.155}{\emph{Nucl. Phys. B Proc. Suppl.} {\bfseries 216} (2011) 194} [\href{https://arxiv.org/abs/1103.2355}{{\ttfamily 1103.2355}}].

\bibitem{Compere:2012jk}
G.~Comp{\`e}re, \emph{{The Kerr/CFT Correspondence and its Extensions}}, \href{https://doi.org/10.12942/lrr-2012-11}{\emph{Living Rev. Rel.} {\bfseries 15} (2012) 11} [\href{https://arxiv.org/abs/1203.3561}{{\ttfamily 1203.3561}}].

\bibitem{Anninos:2008fx}
D.~Anninos, W.~Li, M.~Padi, W.~Song and A.~Strominger, \emph{{Warped AdS(3) Black Holes}}, \href{https://doi.org/10.1088/1126-6708/2009/03/130}{\emph{JHEP} {\bfseries 03} (2009) 130} [\href{https://arxiv.org/abs/0807.3040}{{\ttfamily 0807.3040}}].

\bibitem{Compere:2009zj}
G.~Compere and S.~Detournay, \emph{{Boundary conditions for spacelike and timelike warped $AdS_{3}$ spaces in topologically massive gravity}}, \href{https://doi.org/10.1088/1126-6708/2009/08/092}{\emph{JHEP} {\bfseries 08} (2009) 092} [\href{https://arxiv.org/abs/0906.1243}{{\ttfamily 0906.1243}}].

\bibitem{Compere:2014bia}
G.~Comp{\`e}re, M.~Guica and M.~J. Rodriguez, \emph{{Two Virasoro symmetries in stringy warped AdS$_{3}$}}, \href{https://doi.org/10.1007/JHEP12(2014)012}{\emph{JHEP} {\bfseries 12} (2014) 012} [\href{https://arxiv.org/abs/1407.7871}{{\ttfamily 1407.7871}}].

\bibitem{Song:2016gtd}
W.~Song, Q.~Wen and J.~Xu, \emph{{Modifications to Holographic Entanglement Entropy in Warped CFT}}, \href{https://doi.org/10.1007/JHEP02(2017)067}{\emph{JHEP} {\bfseries 02} (2017) 067} [\href{https://arxiv.org/abs/1610.00727}{{\ttfamily 1610.00727}}].

\bibitem{Bardeen:1999px}
J.~M. Bardeen and G.~T. Horowitz, \emph{{The Extreme Kerr throat geometry: A Vacuum analog of AdS(2) x S**2}}, \href{https://doi.org/10.1103/PhysRevD.60.104030}{\emph{Phys. Rev. D} {\bfseries 60} (1999) 104030} [\href{https://arxiv.org/abs/hep-th/9905099}{{\ttfamily hep-th/9905099}}].

\bibitem{Guica:2010ej}
M.~Guica and A.~Strominger, \emph{{Microscopic Realization of the Kerr/CFT Correspondence}}, \href{https://doi.org/10.1007/JHEP02(2011)010}{\emph{JHEP} {\bfseries 02} (2011) 010} [\href{https://arxiv.org/abs/1009.5039}{{\ttfamily 1009.5039}}].

\bibitem{Compere:2010uk}
G.~Compere, W.~Song and A.~Virmani, \emph{{Microscopics of Extremal Kerr from Spinning M5 Branes}}, \href{https://doi.org/10.1007/JHEP10(2011)087}{\emph{JHEP} {\bfseries 10} (2011) 087} [\href{https://arxiv.org/abs/1010.0685}{{\ttfamily 1010.0685}}].

\bibitem{El-Showk:2011euy}
S.~El-Showk and M.~Guica, \emph{{Kerr/CFT, dipole theories and nonrelativistic CFTs}}, \href{https://doi.org/10.1007/JHEP12(2012)009}{\emph{JHEP} {\bfseries 12} (2012) 009} [\href{https://arxiv.org/abs/1108.6091}{{\ttfamily 1108.6091}}].

\bibitem{Song:2011sr}
W.~Song and A.~Strominger, \emph{{Warped AdS3/Dipole-CFT Duality}}, \href{https://doi.org/10.1007/JHEP05(2012)120}{\emph{JHEP} {\bfseries 05} (2012) 120} [\href{https://arxiv.org/abs/1109.0544}{{\ttfamily 1109.0544}}].

\bibitem{Guica:2017lia}
M.~Guica, \emph{{An integrable Lorentz-breaking deformation of two-dimensional CFTs}}, \href{https://doi.org/10.21468/SciPostPhys.5.5.048}{\emph{SciPost Phys.} {\bfseries 5} (2018) 048} [\href{https://arxiv.org/abs/1710.08415}{{\ttfamily 1710.08415}}].

\bibitem{Zamolodchikov:2004ce}
A.~B. Zamolodchikov, \emph{{Expectation value of composite field T anti-T in two-dimensional quantum field theory}},  \href{https://arxiv.org/abs/hep-th/0401146}{{\ttfamily hep-th/0401146}}.

\bibitem{Smirnov:2016lqw}
F.~A. Smirnov and A.~B. Zamolodchikov, \emph{{On space of integrable quantum field theories}}, \href{https://doi.org/10.1016/j.nuclphysb.2016.12.014}{\emph{Nucl. Phys. B} {\bfseries 915} (2017) 363} [\href{https://arxiv.org/abs/1608.05499}{{\ttfamily 1608.05499}}].

\bibitem{Cavaglia:2016oda}
A.~Cavagli\`a, S.~Negro, I.~M. Sz\'ecs\'enyi and R.~Tateo, \emph{{$T \bar{T}$-deformed 2D Quantum Field Theories}}, \href{https://doi.org/10.1007/JHEP10(2016)112}{\emph{JHEP} {\bfseries 10} (2016) 112} [\href{https://arxiv.org/abs/1608.05534}{{\ttfamily 1608.05534}}].

\bibitem{Chakraborty:2018vja}
S.~Chakraborty, A.~Giveon and D.~Kutasov, \emph{{$ J\overline{T} $ deformed CFT$_{2}$ and string theory}}, \href{https://doi.org/10.1007/JHEP10(2018)057}{\emph{JHEP} {\bfseries 10} (2018) 057} [\href{https://arxiv.org/abs/1806.09667}{{\ttfamily 1806.09667}}].

\bibitem{Aharony:2018ics}
O.~Aharony, S.~Datta, A.~Giveon, Y.~Jiang and D.~Kutasov, \emph{{Modular covariance and uniqueness of $J\bar{T}$ deformed CFTs}}, \href{https://doi.org/10.1007/JHEP01(2019)085}{\emph{JHEP} {\bfseries 01} (2019) 085} [\href{https://arxiv.org/abs/1808.08978}{{\ttfamily 1808.08978}}].

\bibitem{Anous:2019osb}
T.~Anous and M.~Guica, \emph{{A general definition of $JT_a$ -- deformed QFTs}}, \href{https://doi.org/10.21468/SciPostPhys.10.4.096}{\emph{SciPost Phys.} {\bfseries 10} (2021) 096} [\href{https://arxiv.org/abs/1911.02031}{{\ttfamily 1911.02031}}].

\bibitem{Bzowski:2018pcy}
A.~Bzowski and M.~Guica, \emph{{The holographic interpretation of $J \bar T$-deformed CFTs}}, \href{https://doi.org/10.1007/JHEP01(2019)198}{\emph{JHEP} {\bfseries 01} (2019) 198} [\href{https://arxiv.org/abs/1803.09753}{{\ttfamily 1803.09753}}].

\bibitem{Witten:2001ua}
E.~Witten, \emph{{Multitrace operators, boundary conditions, and AdS / CFT correspondence}},  \href{https://arxiv.org/abs/hep-th/0112258}{{\ttfamily hep-th/0112258}}.

\bibitem{Gubser:2002zh}
S.~S. Gubser and I.~Mitra, \emph{{Double trace operators and one loop vacuum energy in AdS / CFT}}, \href{https://doi.org/10.1103/PhysRevD.67.064018}{\emph{Phys. Rev. D} {\bfseries 67} (2003) 064018} [\href{https://arxiv.org/abs/hep-th/0210093}{{\ttfamily hep-th/0210093}}].

\bibitem{Gubser:2002vv}
S.~S. Gubser and I.~R. Klebanov, \emph{{A Universal result on central charges in the presence of double trace deformations}}, \href{https://doi.org/10.1016/S0550-3213(03)00056-7}{\emph{Nucl. Phys. B} {\bfseries 656} (2003) 23} [\href{https://arxiv.org/abs/hep-th/0212138}{{\ttfamily hep-th/0212138}}].

\bibitem{Apolo:2018qpq}
L.~Apolo and W.~Song, \emph{{Strings on warped AdS$_{3}$ via $ \mathrm{T}\bar{\mathrm{J}} $ deformations}}, \href{https://doi.org/10.1007/JHEP10(2018)165}{\emph{JHEP} {\bfseries 10} (2018) 165} [\href{https://arxiv.org/abs/1806.10127}{{\ttfamily 1806.10127}}].

\bibitem{Lunin:2005jy}
O.~Lunin and J.~M. Maldacena, \emph{{Deforming field theories with U(1) x U(1) global symmetry and their gravity duals}}, \href{https://doi.org/10.1088/1126-6708/2005/05/033}{\emph{JHEP} {\bfseries 05} (2005) 033} [\href{https://arxiv.org/abs/hep-th/0502086}{{\ttfamily hep-th/0502086}}].

\bibitem{Maldacena:2008wh}
J.~Maldacena, D.~Martelli and Y.~Tachikawa, \emph{{Comments on string theory backgrounds with non-relativistic conformal symmetry}}, \href{https://doi.org/10.1088/1126-6708/2008/10/072}{\emph{JHEP} {\bfseries 10} (2008) 072} [\href{https://arxiv.org/abs/0807.1100}{{\ttfamily 0807.1100}}].

\bibitem{Araujo:2018rho}
T.~Araujo, E.~O. Colg\'ain, Y.~Sakatani, M.~M. Sheikh-Jabbari and H.~Yavartanoo, \emph{{Holographic integration of $T \bar{T}$ \textbackslash{}\& $J \bar{T}$ via $O(d,d)$}}, \href{https://doi.org/10.1007/JHEP03(2019)168}{\emph{JHEP} {\bfseries 03} (2019) 168} [\href{https://arxiv.org/abs/1811.03050}{{\ttfamily 1811.03050}}].

\bibitem{Borsato:2018spz}
R.~Borsato and L.~Wulff, \emph{{Marginal deformations of WZW models and the classical Yang\textendash{}Baxter equation}}, \href{https://doi.org/10.1088/1751-8121/ab1b9c}{\emph{J. Phys. A} {\bfseries 52} (2019) 225401} [\href{https://arxiv.org/abs/1812.07287}{{\ttfamily 1812.07287}}].

\bibitem{Apolo:2019zai}
L.~Apolo, S.~Detournay and W.~Song, \emph{{TsT, $T\bar{T}$ and black strings}}, \href{https://doi.org/10.1007/JHEP06(2020)109}{\emph{JHEP} {\bfseries 06} (2020) 109} [\href{https://arxiv.org/abs/1911.12359}{{\ttfamily 1911.12359}}].

\bibitem{Apolo:2021wcn}
L.~Apolo and W.~Song, \emph{{TsT, black holes, and $ T\overline{T} $ + $ J\overline{T} $ + $ T\overline{J} $}}, \href{https://doi.org/10.1007/JHEP04(2022)177}{\emph{JHEP} {\bfseries 04} (2022) 177} [\href{https://arxiv.org/abs/2111.02243}{{\ttfamily 2111.02243}}].

\bibitem{Guica:2020eab}
M.~Guica, \emph{{Symmetries versus the spectrum of $ J\bar T$-deformed CFTs}}, \href{https://doi.org/10.21468/SciPostPhys.10.3.065}{\emph{SciPost Phys.} {\bfseries 10} (2021) 065} [\href{https://arxiv.org/abs/2012.15806}{{\ttfamily 2012.15806}}].

\bibitem{Chakraborty:2023wel}
S.~B. Chakraborty, S.~Georgescu and M.~Guica, \emph{{States, symmetries and correlators of $T\bar{T}$ and $ J\bar{T} $ symmetric orbifolds}}, \href{https://doi.org/10.21468/SciPostPhys.16.1.011}{\emph{SciPost Phys.} {\bfseries 16} (2024) 011} [\href{https://arxiv.org/abs/2306.16454}{{\ttfamily 2306.16454}}].

\bibitem{Georgescu:2024ppd}
S.~Georgescu and M.~Guica, \emph{{Classical $J\bar T$ symmetries -- three ways -- and a precision holography check}},  \href{https://arxiv.org/abs/2412.21176}{{\ttfamily 2412.21176}}.

\bibitem{Georgescu:2025jlx}
S.~Georgescu, \emph{{Non-linear asymptotic symmetries in warped AdS$_3$ holography}},  \href{https://arxiv.org/abs/2507.00144}{{\ttfamily 2507.00144}}.

\bibitem{Kruthoff:2020hsi}
J.~Kruthoff and O.~Parrikar, \emph{{On the flow of states under $T\overline{T}$}},  \href{https://arxiv.org/abs/2006.03054}{{\ttfamily 2006.03054}}.

\bibitem{Hirano:2025alr}
S.~Hirano and V.~Raj, \emph{{Nonperturbative effects in $T\bar{T}$-deformed conformal field theories: A toy model for Planckian physics}},  \href{https://arxiv.org/abs/2507.16262}{{\ttfamily 2507.16262}}.

\bibitem{Guica:2019vnb}
M.~Guica, \emph{{On correlation functions in $J\bar T$-deformed CFTs}}, \href{https://doi.org/10.1088/1751-8121/ab0ef3}{\emph{J. Phys. A} {\bfseries 52} (2019) 184003} [\href{https://arxiv.org/abs/1902.01434}{{\ttfamily 1902.01434}}].

\bibitem{Guica:2021fkv}
M.~Guica, \emph{{A definition of primary operators in $J\bar T$-deformed CFTs}}, \href{https://doi.org/10.21468/SciPostPhys.13.3.045}{\emph{SciPost Phys.} {\bfseries 13} (2022) 045} [\href{https://arxiv.org/abs/2112.14736}{{\ttfamily 2112.14736}}].

\bibitem{Georgescu:2022iyx}
S.~Georgescu and M.~Guica, \emph{{Infinite $\mathrm{T\bar T}$-like symmetries of compactified LST}}, \href{https://doi.org/10.21468/SciPostPhys.16.1.006}{\emph{SciPost Phys.} {\bfseries 16} (2024) 006} [\href{https://arxiv.org/abs/2212.09768}{{\ttfamily 2212.09768}}].

\bibitem{ACSSW}
L.~Apolo, W.~Cui, H.~Shu, W.~Song and J.~Wang, \emph{{Correlation functions in the $TsT/J\bar{T}$ correspondence and beyond, to appear}}, .

\bibitem{Cui:2023jrb}
W.~Cui, H.~Shu, W.~Song and J.~Wang, \emph{{Correlation functions in the ${\text{TsT}}/T\overline{T }$ correspondence}}, \href{https://doi.org/10.1007/JHEP04(2024)017}{\emph{JHEP} {\bfseries 04} (2024) 017} [\href{https://arxiv.org/abs/2304.04684}{{\ttfamily 2304.04684}}].

\bibitem{Aharony:2023dod}
O.~Aharony and N.~Barel, \emph{{Correlation functions in $ \textrm{T}\overline{\textrm{T}} $-deformed Conformal Field Theories}}, \href{https://doi.org/10.1007/JHEP08(2023)035}{\emph{JHEP} {\bfseries 08} (2023) 035} [\href{https://arxiv.org/abs/2304.14091}{{\ttfamily 2304.14091}}].

\bibitem{Guica:2020uhm}
M.~Guica and R.~Monten, \emph{{Infinite pseudo-conformal symmetries of classical $T \bar T$, $J \bar T $ and $J T_a$ - deformed CFTs}}, \href{https://doi.org/10.21468/SciPostPhys.11.4.078}{\emph{SciPost Phys.} {\bfseries 11} (2021) 078} [\href{https://arxiv.org/abs/2011.05445}{{\ttfamily 2011.05445}}].

\bibitem{Dubovsky:2012wk}
S.~Dubovsky, R.~Flauger and V.~Gorbenko, \emph{{Solving the Simplest Theory of Quantum Gravity}}, \href{https://doi.org/10.1007/JHEP09(2012)133}{\emph{JHEP} {\bfseries 09} (2012) 133} [\href{https://arxiv.org/abs/1205.6805}{{\ttfamily 1205.6805}}].

\bibitem{Azeyanagi:2012zd}
T.~Azeyanagi, D.~M. Hofman, W.~Song and A.~Strominger, \emph{{The Spectrum of Strings on Warped AdS$_3 \times$ S$^3$}}, \href{https://doi.org/10.1007/JHEP04(2013)078}{\emph{JHEP} {\bfseries 04} (2013) 078} [\href{https://arxiv.org/abs/1207.5050}{{\ttfamily 1207.5050}}].

\bibitem{Cardy:2019qao}
J.~Cardy, \emph{{$T\bar T$ deformation of correlation functions}}, \href{https://doi.org/10.1007/JHEP12(2019)160}{\emph{JHEP} {\bfseries 12} (2019) 160} [\href{https://arxiv.org/abs/1907.03394}{{\ttfamily 1907.03394}}].

\bibitem{Bautista:2019qxj}
T.~Bautista and H.~Godazgar, \emph{{Lorentzian CFT 3-point functions in momentum space}}, \href{https://doi.org/10.1007/JHEP01(2020)142}{\emph{JHEP} {\bfseries 01} (2020) 142} [\href{https://arxiv.org/abs/1908.04733}{{\ttfamily 1908.04733}}].

\bibitem{Seiberg:1999vs}
N.~Seiberg and E.~Witten, \emph{{String theory and noncommutative geometry}}, \href{https://doi.org/10.1088/1126-6708/1999/09/032}{\emph{JHEP} {\bfseries 09} (1999) 032} [\href{https://arxiv.org/abs/hep-th/9908142}{{\ttfamily hep-th/9908142}}].

\bibitem{Marino:2012zq}
M.~Mari{\~n}o, \emph{{Lectures on non-perturbative effects in large $N$ gauge theories, matrix models and strings}}, \href{https://doi.org/10.1002/prop.201400005}{\emph{Fortsch. Phys.} {\bfseries 62} (2014) 455} [\href{https://arxiv.org/abs/1206.6272}{{\ttfamily 1206.6272}}].

\bibitem{Benjamin:2023nts}
N.~Benjamin, S.~Collier, J.~Kruthoff, H.~Verlinde and M.~Zhang, \emph{{S-duality in $ T\overline{T} $-deformed CFT}}, \href{https://doi.org/10.1007/JHEP05(2023)140}{\emph{JHEP} {\bfseries 05} (2023) 140} [\href{https://arxiv.org/abs/2302.09677}{{\ttfamily 2302.09677}}].

\bibitem{Dei:2024sct}
A.~Dei, B.~Knighton, K.~Naderi and S.~Sethi, \emph{{Tensionless AdS$_{3}$/CFT$_{2}$ and single trace $ T\overline{T} $}}, \href{https://doi.org/10.1007/JHEP11(2024)145}{\emph{JHEP} {\bfseries 11} (2024) 145} [\href{https://arxiv.org/abs/2408.00823}{{\ttfamily 2408.00823}}].

\bibitem{Gu:2025tpy}
J.~Gu, Y.~Jiang and H.~Wang, \emph{{Resurgent properties of TT{\textasciimacron}-deformed conformal field theories}}, \href{https://doi.org/10.1103/dvwg-9jcy}{\emph{Phys. Rev. D} {\bfseries 112} (2025) 045007} [\href{https://arxiv.org/abs/2503.19350}{{\ttfamily 2503.19350}}].

\end{thebibliography}\endgroup
\end{document}